\documentclass[11pt]{article}

\usepackage{amssymb,amsmath,amsfonts,amssymb}
\usepackage{graphics,graphicx,color,tikz}
\usepackage{empheq}

\def\@abssec#1{\vspace{.05in}\footnotesize \parindent .2in
{\bf #1. }\ignorespaces}
\graphicspath{{/EPSF/}{Figures/}}
\DeclareGraphicsExtensions{.eps}
\usepackage[margin=0.94in]{geometry}

\newtheorem{theorem}{Theorem}[section]
\newtheorem{lemma}[theorem]{Lemma}
\newtheorem{proposition}[theorem]{Proposition}
\newtheorem{corollary}[theorem]{Corollary}

\def \Rm {\mathbb R}
\def \Nm {\mathbb N}
\def \Cm {\mathbb C}

\def \Mm {\mathbb M}

\def \Zm {\mathbb Z}
\def \Sm {\mathbb S}
\newcommand{\eps}{\varepsilon}

\newcommand{\dsum}{\displaystyle\sum}
\newcommand{\dint}{\displaystyle\int}

\newcommand{\aver}[1]{\langle {#1} \rangle}

\newcommand{\mA}{\mathcal A}

\newcommand{\mC}{\mathcal C}
\newcommand{\mF}{\mathcal F}
\newcommand{\mH}{\mathcal H}
\newcommand{\mI}{\mathcal I}
\newcommand{\mK}{\mathcal K}

\newcommand{\mP}{\mathcal P}

\newcommand{\mS}{\mathcal S} \newcommand{\mT}{\mathcal T}
\newcommand{\mU}{\mathcal U}

\newcommand{\rF}{{\rm F}}

\newcommand{\mD}{\mathfrak D}
\newcommand{\mJ}{{\mathfrak J}}
\newcommand{\fa}{{\mathfrak a}}

\newcommand{\fm}{\mathfrak m}

 \newcommand{\fS}{{\mathfrak S}}

\newcommand{\ns}{{\rm N/S}}
\newcommand{\BD}{{\rm BD}}

\newcommand{\cout}[1]{}

\newcommand{\ind}{{\rm Index\,}}
\newcommand{\sgn}[1]{\,{\rm sign}(#1)}

\newcommand{\ow}{{\rm Op}^w}

\newcommand{\Tr}{{\rm Tr}}

\newcommand{\Ran}{{\rm Ran}}
\newcommand{\trr}{{\rm tr}}

\newcommand{\pup}{PUP_{\Ran P}}
\newcommand{\ipup}{\ind\,PUP_{\Ran P}}

\newcommand{\rP}{{\rm P}}

 \renewcommand{\arraystretch}{1.5}

\title{Continuous Topological Insulators\\ Classification and Bulk Edge Correspondence}

\author{Guillaume Bal  \thanks{Departments of Statistics and Mathematics and CCAM, University of Chicago, Chicago, IL 60637;
    {\tt guillaumebal@uchicago.edu}}} %\thanks{University of Chicago; {\tt guillaumebal@uchicago.edu}}}

\begin{document}
 
\maketitle

%\tableofcontents

\begin{abstract}
   This paper reviews recent results on the classification of partial differential operators modeling bulk and interface topological insulators in Euclidean spaces. Our main objective is the mathematical analysis of the unusual, robust-to-perturbations, asymmetric transport that necessarily appears at interfaces separating topological insulators in different phases. The central element of the analysis is an interface-current-observable describing this asymmetry. We show that this observable may be computed explicitly by spectral flow when the interface Hamiltonian is explicitly diagonalizable. 
   
We review the classification of bulk phases for Landau and Dirac operators and provide a general classification of elliptic interface pseudo-differential operators by means of domain walls and a corresponding bulk-difference invariant (BDI). The BDI is simple to compute by the  Fedosov-H\"ormander formula implementing in a Euclidean setting an Atiyah-Singer index theory. A generalized bulk-edge correspondence then states that the interface current observable and the BDI agree on elliptic operators, whereas this is not necessarily the case for non-elliptic operators. 
\end{abstract}


 
\renewcommand{\thefootnote}{\fnsymbol{footnote}}
\renewcommand{\thefootnote}{\arabic{footnote}}

\renewcommand{\arraystretch}{1.1}

%\begin{keywords}
%\end{keywords}

%\begin{AMS}
%\end{AMS}

%\pagestyle{myheadings}
%\thispagestyle{plain}

%%%%%%%%%%%%%%%%%%%%%%
%%% BEGINNING TEXT %%%
%%%%%%%%%%%%%%%%%%%%%%

%
%%
\section{Introduction}
\label{sec:intro}
Since the first experiments displaying the quantum Hall effect \cite{klitzing1980new,tsui1982two} and their topological interpretation \cite{PhysRevLett.61.2015,laughlin1981quantized,PhysRevLett.49.405}, the analysis of topological phase of matter has received continuous interest in the engineering, physics, and mathematics literatures. We refer the interested reader to the monographs \cite{asboth2016short,bernevig2013topological,moessner2021topological,shen2012topological} for general introductions to the broad field of topological insulators and topological phases of matter. This review focuses on the topological classification and operator properties of continuous single particle Hamiltonians, which are adapted to the description of large-scale, macroscopic transport features in topologically non-trivial systems.

A topological insulator is an insulating system in a certain energy (or frequency) range. As a standalone object it is not particularly interesting: excitations generated within that energy range dissipate locally without exhibiting any large-scale transport.  A central property of topological insulators occurs at interfaces separating two insulators in different topological phases. Heuristically, the difference of topological phases may be considered as anomalous. This is compensated by another anomalous behavior near the interface, where large-scale, asymmetric, transport is guaranteed. This large-scale transport is topological in nature and hence robust to continuous deformations and as such immune to (in principle arbitrary large) amounts of perturbations. For one-dimensional interfaces separating two-dimensional insulators, this unusual robust transport comes as a direct obstruction to the Anderson localization we expect to observe for topologically trivial materials. In many instances, one of the phases is vacuum but this is not necessary and will not be the case in the systems we consider.

Non-trivial topological phases of matter are by and large wave phenomena and therefore find applications in many areas of physical sciences where Hamiltonian dynamics are reasonably accurate. These include electronics as we saw with the quantum Hall effect, but also moir\'e structures in multilayer two-dimensional materials \cite{bistritzer2011moire,watson2023bistritzer}, topological photonics \cite{hafezi2013imaging,PhysRevLett.100.013904,lee2019elliptic,lin2022mathematical,lu2014topological,silveirinha2015chern,slobozhanyuk2017three}, geophysical shallow water models \cite{delplace2017topological,souslov2019topological} as well as cold plasma models \cite{fu2021topological,parker2020topological,qin2023topological,frazier2025topological}, to name a few. This review considers differential models that find applications in the aforementioned areas. These models act on functions of the Euclidean space $\Rm^d$ where $d$ is spatial dimension. Our main focus is on $d=2$, arguably the most relevant dimension in applications.
\\[2mm]
{\bf Bulk phases.} A large fraction of the literature on topological insulators focuses on the definition and computation of bulk invariants. An important observation is the universality of the topological classification of bulk Hamiltonians satisfying prescribed symmetries, such as time-reversal, parity, and chiral symmetries \cite{bernevig2013topological,kitaev2009periodic}. Eventually, the receptacles for classification are either $\Zm$, in which case classification may be obtained as the index of a Fredholm operator, or $\Zm_2$ (materials are either trivial or non-trivial with the first realization proposed in \cite{PhysRevLett.95.146802}), in which case the $\Zm_2$ invariant may sometimes be written as the mod 2 index of an odd Fredholm operator \cite{atiyah1969index,schulz2015z2}.   For information on the well-developed operator algebraic K-theoretic descriptions of bulk and interface invariants, we refer the reader to, e.g., \cite{kellendonk2017c,prodan2016bulk,thiang2016k}.

This paper focuses on $\Zm$ indices for the complex classes A (no symmetry) and AIII (chiral symmetry).  Such operators necessarily break time-reversal symmetry \cite{avron1994,bernevig2013topological,prodan2016bulk}. Section \ref{sec:bulk} considers the definition of a bulk invariant for the Landau operator, which is a partial differential model of the integer quantum Hall effect. We follow the derivation in \cite{avron1994} and review the necessary material on Fredholm operators, Fredholm modules, indices of pairs of projections, Fedosov formula, and Chern numbers, to define the Hall conductivity and show that each Landau level is topologically nontrivial. A similar construction is then applied to a regularized two-dimensional Dirac equation following the presentation in \cite{bal2019continuous}. Dirac operators are then analyzed in all spatial dimensions, alternatively in class A and class AIII in section \ref{sec:Diracnd}. We also present stability results of the invariant against perturbations leveraging the power of the functional-calculus Helffer-Sj\"ostrand formula.

The Dirac operator displays a peculiarity not seen for discrete models or systems with a well-defined Brillouin zone. Because dual variables live in the non-compact $\xi\in(\Rm^d)^*=\Rm^d$, bulk phases cannot be defined naturally when the Hamiltonian's behavior as $|\xi|\to\infty$ depends on the direction $\xi/|\xi|$. While this is a well-recognized issue \cite{silveirinha2015chern,shen2012topological} that can be remedied with by appropriate regularization, we show in later sections that regularization is not necessary when considering interface Hamiltonians.
\\[2mm]
{\bf Interface invariants.} Interface Hamiltonians model a transition from one bulk Hamiltonian to another one. These transitions are typically modeled by spatially varying coefficients, such as a mass term in a Dirac system or a Coriolis force parameter in a shallow water equation. The central object in our analysis is the interface current observable $\sigma_I[H]$ defined in \eqref{eq:sigmaI} in section \ref{sec:IH} below. This observable, defined as the expectation of a current operator against a density of states confined to the vicinity of the interface, is used in one form or another in all analyses of interface transport in various scenarios \cite{elbau2002equality,elgart2005equality,kellendonk2004quantization,kotani2014quantization,prodan2016bulk}.

We first recall in section \ref{sec:SF} a simple result relating the interface current $\sigma_I$ to the spectral flow associated to invariant Hamiltonians with respect to translations along the interface. This shows that $\sigma_I$ may be easily computed provided one has a full spectral decomposition of $H$. Such a decomposition is rarely available, and this justifies the remainder of this review.  Still, we apply it in section \ref{sec:appliSF} to a number of problems including the Landau operator, the magnetic Dirac operator, and the shallow water model.
\\[2mm]
{\bf Classification of elliptic operators by domain walls.} Section \ref{sec:elliptic} introduces classes of pseudo-differential operators (PDO) for which the interface current observable $\sigma_I[H]$ is well defined. We recall the functional calculus of \cite{bony2013characterization,dimassi1999spectral,grigis1994microlocal,zworski2022semiclassical} for elliptic  PDOs in section \ref{sec:Weyl}, and show in section \ref{sec:classIC} that $2\pi\sigma_I\in\Zm$ is quantized  following \cite{quinn2024approximations}.

We then move on to a general simple classification for elliptic PDOs in section \ref{sec:classDW} that applies equally to bulk Hamiltonians, interface Hamiltonians, and so-called higher-order topological insulator (HOTI) Hamiltonians, as well as possibly non-Hermitian Hamiltonians. This classification is based on twisting a Hamiltonian $H_k$ confined in $k$ out of $d$ spatial directions by domain walls in $(x_{k+1},\ldots,x_d)$. A resulting Fredholm operator has index given explicitly by a Fedosov-H\"ormander formula \eqref{eq:FH}, implementing in a Euclidean  setting an Atiyah-Singer result \cite{atiyah1968index}. We follow the construction in \cite{bal2023topological} and refer to \cite{bott1978some,callias1978axial,faure2023manifestation,gesztesy2016callias} for other topological applications of the formula. 

The invariant is recast in section \ref{sec:BDI}  as a {\em bulk-difference} invariant \cite{bal2022topological}  in terms of the properties of the bulk insulators the interface separates. This bulk-difference invariant still makes sense when no invariants may be naturally defined for each bulk, intuitively indicating that phase differences are more generally defined than differences of phases.
\\[2mm]
{\bf The Bulk-edge correspondence (BEC).} The role of the bulk edge correspondence, treated in section \ref{sec:BEC}, is to show that the two aforementioned classifications, one based on $\sigma_I[H]$ and one based on the Fedosov-H\"ormander formula \eqref{eq:FH}, agree for self-adjoint elliptic operators. 

While many operators, such as the ubiquitous Dirac operators, are elliptic, the Landau operator, the magnetic Dirac, and the shallow water operator are examples of {\em non} elliptic operators. While the BEC applies to Landau and magnetic Dirac operator in a different form, it does not apply directly to the shallow water model. 

The BEC is a pillar in the understanding of topological phases of matter \cite{asboth2016short,bernevig2013topological,moessner2021topological,shen2012topological}. The derivation of the BEC recalled in this review bears some similarities with \cite{essin2011bulk,faure2023manifestation,volovik2009universe}. Operator algebra K-theoretic techniques may be used to obtain general BEC, see, e.g., \cite{bourne2017k,bourne2018chern,prodan2016bulk}. A general bulk-edge correspondence applies to discrete Hamiltonians \cite{elbau2002equality,elgart2005equality}. The bulk-edge correspondence has also been established for second-order differential equations with microscopic periodic structures; see \cite{drouot2021microlocal,drouot2019bulk,drouot2020edge,gontier2023edge}.  

We focus on a bulk edge correspondence for elliptic PDO operators, first in the two-dimensional setting in Theorem \ref{thm:TCC} and next in arbitrary spatial dimensions in Theorem \ref{thm:tcc}. We also consider a number of applications of the BEC and of its possible violations for non-elliptic operators. In the setting of shallow water equations with possibly discontinuous Coriolis force parameter, the BEC is in fact shown not to apply as demonstrated in Fig. \ref{fig:shallow} below and analyzed in more detail in \cite{bal2024topological}. 
\\[2mm]
{\bf Additional references.}  There is a huge literature on methods to estimate Chern numbers, winding numbers and other topological degrees of maps, for single particle Hamiltonians as described here as well as interacting systems \cite{bal2023topological,bernevig2013topological,moessner2021topological,prodan2016bulk,quinn2024approximations}. 

We do not address how topological insulators may be realized in practice.  Some systems generate non-trivial topologies by shining light onto electronic structures, leading to the field of Floquet topological insulators \cite{cayssol2013floquet,rechtsman2013photonic}; see also \cite{bal2022multiscale}.  In Topological Anderson insulators, different topologies may be generated by means of highly oscillatory media \cite{groth2009theory,li2009topological}. A mathematical justification by homogenization theory and resolvent estimates was recently proposed in \cite{bal2024bistopological}.

This review focuses on qualitative topological classifications. Quantitative and geometric descriptions of the asymmetric transport are also relevant and influenced by the nontrivial topology \cite{1751-8121-41-40-405203,fulga2012scattering}. While scattering theory is well-developed for perturbations of spatially homogeneous unperturbed operators \cite{H-II-SP-83,RS-79-III,yafaev1992mathematical}, very little is known for topological insulators, where scattering occurs only along the non-constrained spatial dimensions. \cite{buchendorfer2006scattering,frohlich2000extended,hislop2008edge} develop scattering theories for the Landau operator while the Dirac operator with linear domain wall is considered in \cite{chen2025scattering}. Scattering theory displays how $2\pi\sigma_I\not=0$ acts as a quantized obstruction to Anderson localization \cite{bal2019topological,bal2024mathbb}. Scattering theory also provides robust means to compute interface transport numerically Klein-Gordon and Dirac operators \cite{bal2023integral,bal2025integral,bal2023asymmetric}. Numerous computational tools to estimate invariants have been developed in, e.g., \cite{Loring,LS19,LS20,MPLW,Prodan,quinn2024approximations}. 

A complementary quantitative understanding of $H$ may be obtained from (finite time) solutions to the evolution (Schr\"odinger) equation $(D_t+H)u=0$. In particular, it is fruitful to look at the scattering-free semiclassical regime where the typical scale of the propagating wavepacket $u$ is small compared to the variations of the macroscopic coefficients. It turns out that the semiclassical analysis is somewhat anomalous since wavepackets in such settings are confined near an interface; see \cite{bal2024semiclassical,bal2023magnetic,bal2023edge,D22}.

\medskip\noindent {\bf Notation.}
We say that a function is a {\em switch} function $f\in \fS[a,b;c,d]$ when $f:\Rm\to\Rm$ is a bounded function with $f(x)=a$ for $x\leq c$ and $f(x)=b$ for $x>d$. We denote by $\fS[a,b]$ the union of the sets $\fS[a,b;c,d]$ for all $c\leq d$. We also denote $\fS=\fS[0,1]$. The set of smooth switch functions will be denoted by $C^\infty \fS[\cdot]$.

\section{Bulk Hamiltonians}
\label{sec:bulk}

The two simplest examples of gapped differential Hamiltonians with non-trivial topological features are the Landau operator and the Dirac operator. 
\\[2mm]
{\bf The Landau (or magnetic Schr\"odinger) operator} is given in two space dimensions by
\begin{gather} \label{eq:2dmagSch}
   H = (D-A)^2 + V, \quad D=(D_x,D_y)^t=(\frac 1i\partial_x,\frac1i\partial_y)^t,\quad A=(A_x,A_y)^t
\end{gather}
where $\partial_x A_y-\partial_y A_x = B$ is magnetic field and $V$ is electric potential while $(D-A)^2=(D-A)\cdot(D-A)$. In the Landau gauge, $(A_x=0,A_y=Bx)$ generates a constant magnetic field $B$. When $B>0$ is constant and $V=0$, the above operator admits an explicit diagonalization on $\mH=L^2(\Rm^2;\Cm)$. The spectrum is pure point, with a countable sequence of infinitely degenerate eigenvalues $E_n=(2n-1)B$ for $n\geq1$ called the Landau levels. 

Two features of this operator are: (i) an infinite number of spectral gaps, i.e., energies $E$ such that $E_n<E<E_{n+1}$; and (ii) eigenspaces associated to $E_n$, and given as the range of an orthogonal projector $P_n$, are topologically non-trivial. Moreover, the topological invariant that will be assigned to $P_n$ will be shown to equal $1$ and be robust against perturbations of $A$ and $V$ in \eqref{eq:2dmagSch}. The Landau operator is a model for the integer quantum Hall effect \cite{avron1994,bellissard1994noncommutative}.
\\[2mm]
{\bf The massive Dirac operator} is given in two space dimensions by
\begin{equation}\label{eq:2dDirac}
  H = D\cdot\sigma + (m + \eta \Delta)\sigma_3 = \begin{pmatrix}m + \eta \Delta & D_x-iD_y \\ D_x + i D_y & -(m + \eta \Delta) \end{pmatrix},
\end{equation}
where $D\cdot\sigma=D_x\sigma_1+D_y\sigma_2$, $m(x,y)$ is a mass term, $\eta$ is a constant, $-\Delta=D_x^2+D_y^2$, and $\sigma_j$ for $j=1,2,3$ are the standard Pauli matrices. Strictly speaking, $H$ is a (first order) Dirac operator when $\eta=0$ but we will continue to refer to operators such as $H$ above as Dirac operators.  The regularizing term $\eta\Delta$ is added mostly out of mathematical convenience to obtain a simple model for which bulk invariants may be defined unambiguously in section \ref{sec:topoDirac}. This is in contrast to the case of interface Hamiltonians considered later in the paper for which invariants are well-defined even when $\eta=0$.  The operator $H$ is an unbounded self-adjoint operator on $\mH=L^2(\Rm^2;\Cm^2)$. The only property of the Pauli matrices we use is the anti-commuting property
$ \sigma_i\sigma_j+\sigma_j\sigma_i = 2 \delta_{ij} I_2$
where $I_2$ is identity on $\Cm^2$. This property naturally leads to 
\[
  H^2 = [-\Delta + (m+\eta \Delta)^2 ] I_2
\]
when $m$ is constant. When $\eta=0$, we thus observe that $H$ has a spectral gap in $(-|m|,|m|)$. This gap persists when $1-2|\eta m|\geq0$, which we assume. Moreover, we also verify that $H$ has purely absolutely continuous spectrum with two branches represented in the Fourier domain by 
\[
 \Rm^2 \ni \xi \mapsto E_\pm(\xi) = \pm \sqrt{|\xi|^2 + (m-\eta|\xi|^2)^2}.
\]

The main two features of this operator are: (i) a unique spectral gap for energies $E$ such that $-|m|<E<|m|$ (when $1-2|\eta m|\geq0$); and (ii) the orthogonal projectors $\Pi_\pm$ onto the branches with positive/negative energies are topologically non-trivial. The invariants associated to $\Pi_\pm$ will be shown to equal $\pm\frac12(\sgn{m}+\sgn{\eta})$ and to be robust against perturbations. Since $\Pi_++\Pi_-=I$ (identity on $\mH$), these invariants have to sum to $0$. 
Dirac operators are ubiquitous in topological phases of matter as the simplest two-band model with non-trivial topology \cite{Fefferman2016,fefferman2012honeycomb,lee2019elliptic,shen2012topological}.
\subsection{Fredholm operators and modules and pairs of projections.}
A natural method to devise topological classifications and assign topological invariants to objects such as $P_n$ or $\Pi_\pm$ above is to construct Fredholm operators and compute their index. 
\\[2mm]
{\bf Fredholm operators.} Let $\mH$ be a Hilbert space and $T$ a linear bounded operator on $\mH$. The operator $T$ is said to be Fredholm if its kernel and co-kernel are finite-dimensional. This implies that the range of $T$ is closed and that both kernels in the direct sums $\mH={\rm Ran } T \oplus {\rm Ker } T^*={\rm Ran } T^* \oplus {\rm Ker } T$ are finite dimensional, where $T^*$ is the adjoint operator to $T$. The index of $T$ is then defined as
\begin{equation*} %\label{eq:index}
 \ind\,T = {\rm dim}\, {\rm Ker }\, T \ - \ {\rm dim}\, {\rm Ker } \,T^*.
\end{equation*}
When $T$ maps $\Rm^m$ to $\Rm^n$ (an $n\times m$ matrix), then the index equals $m-n$, indicating its stability. For any $K$ compact on $\mH$, then $\ind\, (T+K)=\ind\, T$. Fredholm operators are an open set in the uniform topology so that any continuous family of Fredholm operators $[0,1]\ni t\mapsto T_t$ is such that  $\ind\, T_t$ is independent of $t$.  Finally, the Atkinson criterion states that $F$ is Fredholm if and only if it is left and right invertible up to (possibly different) compact operators, i.e., there is an operator $S$ such that $I-ST$ and $I-TS$ are compact operators (and $S$ may be chosen so that they are in fact finite rank). Then $\ind\,T + \ind\,S=0$. The index of $T$ vanishes when the latter two compact operators may be chosen to be the same. See, e.g., \cite[Chapter 19]{H-III-SP-94}.
\\[2mm]
{\bf Fredholm modules.} This standard tool to produce Fredholm operators plays an important role in the classification of elliptic operators on compact manifolds \cite{atiyah1970global} and generalizations in non-commutative geometry \cite{connes1994noncommutative} (see also \cite{carey2011spectral} for a review on the related notion of spectral triples).  

For $\mH$ a Hilbert space and $\mA$ an algebra of bounded operators on $\mH$, we say that $(\mH,F)$ is a Fredholm module over $\mA$ if $F$ is a bounded self-adjoint operator on $\mH$ such that $F^2=I$ and such that $[F,f]=Ff-fF$ is compact for all $f\in \mA$. In the absence of additional symmetry, we call the Fredholm module {\bf odd} and verify that $\mP=\frac12(I+F)$ is then an orthogonal projector.

If in addition, there exists an operator $\gamma$ such that $\gamma F+F\gamma=0$, then we call the Fredholm module {\bf even}. This symmetry allows one to decompose the Hilbert space in such a way that $F$ may be written for a unitary operator $\mU$ as 
\begin{equation}\label{eq:FmU}
  F =  \begin{pmatrix} 0 & \mU^* \\ \mU & 0 \end{pmatrix}.
\end{equation}

The odd/even Fredholm modules are used to classify Hamiltonians in (effective) odd/even spatial dimensions as we will see. We will come back to what we mean by `effective' here. In applications to topological insulators, the operator $F$ and the corresponding projector $\mP$ or unitary $\mU$ in even/odd dimensions take the form of multiplication by matrix-valued functions of the spatial coordinates. In particular, $\mP(x)$ is a Heaviside function in dimension one while $\mU(x,y)=(x+iy)/|x+iy|$ is a function with unit winding number about the origin. These objects become matrix valued in higher spatial dimensions. 

The operator $F$ is then used to {\em test} the topology of the elements in the algebra $\mA$. In (effective) even dimension, the tested objects in $\mA$ are projectors $P=P(H)$ associated to a Hamiltonian $H$ of interest, such as $P_m$ or $\Pi_\pm$ above. Under the assumption that $[\mU,P]$ is compact, we observe that $P \mU P$ is a Fredholm operator when restricted to the range of $P$. Indeed
\[
 P\mU P P \mU^* P = P \mU P \mU^* P = P P \mU \mU^* P + K = P+K
\]
with $K$ compact so that $P \mU^* P$ is a right inverse of $P\mU P$ modulo a compact operator as well as a left inverse (modulo compact) by the same principle. The Atkinson criterion then implies that $P\mU P$ and $P \mU^* P$ are Fredholm operators with opposite indices. The index of $P\mU P$ is the {\em topological invariant} associated to $H$. 

In (effective) odd dimension, the objects of interest in $\mA$ are unitary operators $U=U(H)$ associated to a Hamiltonian $H$. We then verify as above that $\mP U \mP$ is a Fredholm operator on the range of $\mP$. The index of $\mP U \mP$ is then the topological invariant associated to $H$. 

Our next objective is to write the index as the trace of an appropriate operator.
\\[2mm]
{\bf Trace-class operators.}  Let $K$ be a compact operator on $\mH$ and let $\lambda_j$ be the sequence of its singular values, i.e., eigenvalues of $(K^*K)^{\frac12}$. The Schatten ideal $\mI_p$ is the space of compact operators such that $\lambda_j\in l^p(\Nm)$. Then $\|(\lambda_j)\|_p$ is a norm for $K\in \mI_p$ turning $\mI_p$ into a Banach space. The space $\mI_1$ is the space of {\em trace-class} operators while $\mI_2$ is called the space of Hilbert-Schmidt operators. For two operators in $\mI_p$ and $\mI_q$, the product lies in $\mI_{r}$ with $\frac1p+\frac 1q=\frac1r$ with all $p,q,r\in[1,\infty]$. In particular $K\in \mI_p$ implies that $K^p\in \mI_1$ for $p\in\Nm^*$.
\\[2mm]
{\bf Fedosov formula.} Also referred to as the Calder\'on-Fedosov formula, it states that the index of a Fredholm operator may often be written as the trace of a trace-class operator. Let $T$ and $S$ be two operators such that $R_1=I-ST$ and $R_2=I-TS$ are compact operators. Then $T$ and $S$ are Fredholm as we saw. Let us further assume that $R_1^n$ and $R_2^n$ are trace-class for some $n\geq1$. Then we have the Fedosov formula:
\begin{equation}\label{eq:Fedosov}
  \ind\,T = {\rm Tr}\, R_1^n - {\rm Tr}\, R_2^n = -\ind\,S.
\end{equation}
For the canonical example of the right shift $S$ on $l^2(\Nm)$, we obtain that $\ind S = \Tr\,[S,S^*]=-\Tr\,\Pi_0 =-1$ with $\Pi_0$ the orthogonal projector onto the $0$th component. Neither $SS^*$ nor $S^*S$ are trace-class but the commutator $[S,S^*]$ is with non-vanishing trace.

The above formula applies to elements in a Fredholm module as follows. Let $P$ be an orthogonal projector and $U$ a unitary operator on $\mH$.  Assume that $[P,U]$ is compact so that $PUP$ and $PU^*P$ are Fredholm operators on the range of $P$. Let $R_1=P-ST=P-PU^*PUP$ and $R_2=P-TS=P-PUPU^*P$  (with $P$ identified with identity on the range of $P$). Assuming that $R_1^n$ and $R_2^n$ are trace-class, then
\[ 
   \ind\, PU^*P = {\rm Tr}\, R_1^n - {\rm Tr}\, R_2^n = -\ind\, PUP.
\]
{\bf Index of pairs of projectors.} Let $P$ and $Q$ be orthogonal projectors on $\mH$. Then \cite{avron1994} defines the index of a pair of projections $P$ and $Q$ as the excess of the dimension of the range of $P$ compared to that of $Q$. When they are finite-rank, then $\ind\, (P,Q)\equiv \dim P - \dim Q = \Tr (P-Q)$ computes the difference of dimensions, showing the topologically invariant nature of the object. This is generalized for $P$ and $Q$ such that $P-Q$ is compact as 
\begin{equation*}%\label{eq:indexpair}
  \ind\, (P,Q) := \dim {\rm Ker} (P-Q-1) - \dim {\rm Ker} (P-Q+1).
\end{equation*}
The relation to the Fredholm module constructions is elucidated by the choice $Q=UPU^*$. The index of the pair may also be written as the trace of an appropriate operator \cite{avron1994}. When $(P-Q)^{2n+1}$ is trace-class for some $n\geq0$, then for all $m\geq n$, we have
\begin{equation}\label{eq:indextrace}
  \ind\,(P,Q)  = {\rm Tr}\, (P-Q)^{2m+1}.
\end{equation}
Under the same hypotheses with now $Q=UPU^*$, we then have that
$\ind\,(P,Q)   = - \ind\,PUP$.

The above formulas show that topological indices may be computed as a trace provided that appropriate powers of $[P,U]U^*$ are trace-class. This is the structure used to define and compute topological invariants for two-dimensional Landau and Dirac operators.

\subsection{Topology associated to Landau levels}
We come back to the Landau operator in \eqref{eq:2dmagSch} with constant magnetic field $B$ and vanishing $V$ and follow \cite{avron1994}. Recall that $E_n=(2n-1)B$ are the Landau levels and consider an energy level $E\in (E_m,E_{m+1})$. We then define by spectral calculus the orthogonal projector
\begin{equation}\label{eq:projneg}
 P = P[H] := \chi(H - E) = \chi_{\eps}(H - E)
\end{equation}
where $\chi(h)=1$ for $h<0$ and $\chi(h)=0$ for $h\geq0$ whereas $\chi_\eps$ is a smooth function such that $\chi_\eps(h)=\chi(h)$ for $|h|\geq\eps$. Since $[E-\eps,E+\eps]$ is in a spectral gap of $H$ for $\eps>0$ sufficiently small, the above equality holds by spectral calculus and is independent of the choice of $E\in (E_m,E_{m+1})$.

Let $u(x)$ be a complex valued function with $|u(x)|=1$ differentiable away from $x=0$ such that 
\[
  |u(x+y)-u(x)| \leq C \frac{|x|}{|y|}.
\]
Let $U$ be the unitary operator on $L^2(\Rm^2)$ of point-wise multiplication by $u(x)$. 
The winding number $N(U)$ of $u$ about $x=0$ may then be computed. We find for instance that $N(u)=n$ for $u=(x_1+ix_2)^n/|x_1+ix_2|^n$. The choice $\mU=U$ and then $F$ as in \eqref{eq:FmU} generates an even Fredholm module (in even spatial dimension $d=2$). Then we have
\begin{theorem}[\cite{avron1994}]\label{thm:Landau}
 The operator $PUP$ is Fredholm on the range of $P$ and, moreover, 
 \[
   \ind\, PUP = -m N(U).
 \]
\end{theorem}
The index is thus given as the product of the number of Landau levels captured by $P$ by the winding number of $U$.  If $P_m$ is the projector onto the $m$th Landau level, then the above result states that $\ind\,P_mUP_m = -N(U)$ for each level $m$.

We provide some steps of the derivation. The first step is an analysis of the Schwartz kernel $p(x,y)$ of the projector $P$, in the sense that for $f\in \mH=L^2(\Rm^2;\Cm)$, then $(Pf)(x) = \int_{\Rm^2} p(x,y) f(y) dy$.

Using a Combes-Thomas estimate, we prove that 
\begin{equation}\label{eq:CTeta}
  |p(x,y)| \leq \frac{C}{1+|x-y|^\eta},
\end{equation}
for some $\eta>2$. The estimate holds for smooth compactly supported perturbations of $A$ and $V$ following Combes-Thomas estimates in \cite{avron1994,germinet2003operator} for smooth functional $P=\chi_\eps(H-E)$. 

The second step is to show that $(P-Q)^3$ is trace-class for $Q=UPU^*$ so that \eqref{eq:indextrace} applies. This is a consequence of the following criterion \cite{GOFFENG2012357,russo1977hausdorff}:
\begin{lemma}[Russo's criterion]\label{lem:russo}
 Assume $T$ bounded on $L^2(\Rm^d)$ for $d\geq1$ with integral kernel $t(x,y)$ while $t^*(x,y)$ is the kernel of the adjoint operator $T^*$. Let $p>2$ with conjugate $q=\frac{p}{p-1}$. Assume $\|t\|_{q,p}$ and $\|t^*\|_{q,p}$ are bounded, where
 \begin{equation} \label{eq:Russo}
    \|t\|_{q,p} = \Big( \dint_{\Rm^d} \Big( \dint_{\Rm^d} |t(x,y)|^q dx \Big)^\frac pq  dy\Big) ^{\frac 1p}. 
 \end{equation} 
  Then $T\in\mI_p$ with $\|T\|_p\leq \|t\|_{q,p}^{\frac12}\|t^*\|_{q,p}^{\frac12}$.  For $n\geq p$ an integer, $T^n$ is trace-class with $\|T^n\|_1\leq \|t\|_{q,p}^{\frac12}\|t^*\|_{q,p}^{\frac12}$. Moreover, if $T^n$ has Schwartz kernel $k(x,y)$, then $\Tr\, T^n = \int_{\Rm^d} k(x,x) dx$.
\end{lemma}
We apply this criterion to $T=P-Q$ with Schwartz kernel $p(x,y)(1-u(x)/u(y))$ as done in \cite{avron1994} to obtain that \eqref{eq:Russo} holds, that $(P-Q)^3$ is trace-class and that following \eqref{eq:indextrace} we have
\[
 -\ind\, PUP =  \dint_{\Rm^6} p(x,y)p(y,z)p(z,x) \big(1-\frac{u(x)}{u(y)}\big) \big(1-\frac{u(y)}{u(z)}\big) \big(1-\frac{u(z)}{u(x)}\big)\,dxdydz.
\]
We remark that a non-trivial index requires $p$ to be complex valued. When $p$ is real valued, for instance because it comes from a time-reversal symmetric operator, the above integral is purely imaginary and hence vanishes. Indeed, the term involving the cyclic product of $u(\cdot)$ is odd under complex conjugation. A non-trivial magnetic field, which breaks time-reversal symmetry, is then necessary to obtain both spectral gaps and nontrivial topological invariants.

While the computation of this integral remains a formidable task in general, it simplifies when $B$ is constant and $V=0$ by using an invariance of the Landau operator with respect to magnetic translations. Let $a\in\Rm^2$ and $T_af(x)=f(x-a)$ the unitary shift operator. We verify that in the Landau gauge, $T_a A(x) =  A(x)-\nabla \Lambda_a(x)$ for $\Lambda_a(x)=Ba_1x_2$, using coordinates $x=(x_1,x_2)$ and $a=(a_1,a_2)$. Define $U_a(x)=e^{-i\Lambda_a(x)}$ and observe the conjugation:
\[
  e^{i\Lambda_a} D e^{-i\Lambda_a} = D -\nabla \Lambda_a(x)\quad \mbox{ so that } \ \ T_a(D-A)^2 T_{-a} = U_a^* (D-A)^2 U_a,
\]
implying the invariance $(U_aT_a)g(H)(U_aT_a)^*=g(H)$ for spectral functionals of the unperturbed Landau operator. This symmetry is inherited by $P$ so that 
\begin{equation}\label{eq:magsym}
  p(x,y)p(y,z)p(z,x) = p(0,y-x)p(y-x,z-x)p(z-x,0).
\end{equation}

\paragraph{Geometric identity.}  We observe that a complete separation in the roles of $u$ and $p$ may be achieved when the following formula holds
\begin{equation}\label{eq:connesformula} 
  \dint_{\Rm^2}   (1-\frac{u(x-a)}{u(x-b)}) (1-\frac{u(x-b)}{u(x-c)}) (1-\frac{u(x-c)}{u(x-a)})dx =2\pi i N(U) {\rm Area}(a,b,c).
\end{equation}
Here $(a,b,c)$ are three vectors in $\Rm^2$ and the area is that of the rectangle with ${\rm Area}(a,b,c)=a\wedge b\wedge c+c\wedge a$ twice the oriented area of the triangle with vertices $a$, $b$, and $c$. This central geometric identity due to Connes is derived in detail in \cite[Lemma 4.4]{avron1994} and generalized to higher dimensions in \cite{prodan2016bulk};  see also \cite{bal2019continuous}. A combination of the above equalities shows that
\[
  \ind\, PUP = -2\pi i N(U) \dint_{\Rm^4} p(0,x) p(x,y) p(y,0) x\wedge y dx dy
\]
with $x\wedge y=x_1y_2-x_2y_1$. The computation of the above integral may then be obtained from explicit expressions for the projectors $P_m$ onto Landau levels. Following for instance \cite{avron1994}, we obtain that $\ind\, P_m U P_m=-N(U)$. We presented the result in Theorem \ref{thm:Landau} for the Landau operator. The theorem applies so long as the bound \eqref{eq:CTeta} and the symmetry assumption \eqref{eq:magsym} hold.

The above construction provides a topological characterization of spectral gaps of $H$ by means of the index of $PUP$.  This characterization has a physical interpretation by a Laughlin argument and is related by a Kubo formula to a Hall conductivity; see \cite{avron1994,bernevig2013topological,frohlich2000extended,laughlin1981quantized}.  The quantum Hall effect was also analyzed in great detail in \cite{bellissard1994noncommutative} using K-theoretic tools. 
\subsection{Topology associated to Dirac bands.}
\label{sec:topoDirac}
We now move to the two-dimensional Dirac operator in \eqref{eq:2dDirac}. As for the Landau operator, the Dirac operator breaks time-reversal symmetry as soon as the component in front of $\sigma_3$ does not vanish. Time reversal symmetry for such operators is implemented by a anti-unitary operator $\mT=i\sigma_2 \mK$ with $\mK$ complex conjugation. We verify that $\mT^*H\mT=H$ only when the component in front of $\sigma_3$ vanishes. Breaking time-reversal symmetry is a common feature that is necessary for indices of Fredholm operators of the form $PUP$ to be non-trivial \cite{avron1994,bernevig2013topological}.

The operator $H=\mF^{-1} \hat H(\xi)\mF$ with $\mF$ two-dimensional Fourier transform and $\hat H(\xi)=\xi\cdot\sigma + (m-\eta|\xi|^2)^2\sigma_3$ so that $\hat H^2(\xi)\geq \eta^2|\xi|^4+m^2$. Thus $H$ is self-adjoint as an unbounded operator on $\mH=L^2(\Rm^2;\Cm^2)$ with domain $\mD(H)=\mF^{-1}(\eta^2|\xi|^4+m^2)^{-1}\hat H \mF \mH$ as we verify that $(H\pm i)\mD(H)=\mH$. We also verify that the perturbed operator $H+V$ for $V$ multiplication by $V(x)$ smooth and compactly supported, say, is self-adjoint with domain $\mD(H)$ as well. 

Since $H$ has a spectral gap between $(-|m|,|m|)$, we may also define the projector $\Pi_-=P[H]$ as in \eqref{eq:projneg}. We will call $P$ that projector to simplify notation. We denote by $u(x)=\frac{x_1+ix_2}{|x_1+ix_2|}$ so that for $U$ the unitary operator of multiplication by $u(x)$, we have $N(U)=1$ for concreteness. Then:
\begin{theorem}[\cite{bal2019continuous}]
 The operator $PUP$ is Fredholm on the range of $P$ and, moreover, 
 \begin{equation} \label{eq:indexDirac}
   -\ind\, PUP = \frac12 \big(\sgn{m} + \sgn{\eta}\big).
 \end{equation}
 \end{theorem}
The derivation parallels that for the Landau operator with a few modifications. The main difference is that the Schwartz kernel of $P$ is no longer as smooth as it was for the Landau operator. In fact, it is the reason why $\eta=0$ is singular, as displayed explicitly in the above formula, which makes no sense when $\eta=0$.

The first step is to prove that $(P-Q)^3$ is trace-class with $Q=UPU^*$. This is obtained by applying Russo's criterion in Lemma \ref{lem:russo}. Writing $P-Q=(P-B)-U(P-B)U^*$ for $B=\frac12(I+\sgn{\eta}\sigma_3)$ so that $UBU^*=B$, and defining $R=P-B$, we find that $R$ has a Schwartz kernel $r=r(x-y)$ by invariance of $H$ under spatial translations whose Fourier transform is given by
\[
  \hat r(\xi) =  -\frac12 \frac{\xi\cdot\sigma}{|\xi,m|} - \frac12 (\sgn{\eta}+\frac{m-\eta|\xi|^2}{|\xi,m|}) \sigma_3,\quad |\xi,m|:=(|\xi|^2+(m-\eta|\xi|^2)^2)^{\frac12}.
\]
Asymptotic expansions for large $|\xi|$ and bounds on $\aver{x}^\alpha |p(x)|$ for any $\alpha$ with $\aver{x}=\sqrt{1+|x|^2}$ yields that the Schwartz kernel of $P-Q$ satisfies the bound:
\[
   |p(x-y)-q(x,y)| = |p(x-y) (1-\frac{u(x)}{u(y)})|  \leq C_\beta \min\big(1,\frac{|x-y|}{|y|} \big) \frac{1}{|x-y| \aver{x-y}^\beta}
\]
for any $\beta\in \Nm$. This is sufficient to apply Russo's criterion \cite[Lemma 3.3]{bal2019continuous} when $|\eta|\not=0$. Writing the trace of $(P-Q)^3$ explicitly and using Connes' geometric identity \eqref{eq:connesformula} then leads to
\[
 {\rm Tr}\,(P-Q)^3 = -2\pi i \dint_{\Rm^4} {\rm tr} \, p(-x) p (x-z) p(z) \ x\wedge(x-z)\ dxdz.
\]
{\bf Chern number.} The operator $P=\chi(H)=\mF^{-1} \hat P(\xi) \mF$ with $\hat P(\xi)$ the Fourier transform of $p(x)$. The above formula thus states, using formulas such as $\mF (-ix_j)\mF^{-1}=\partial_{\xi_j}$, that $-\ind\, PUP = {\rm Tr}\,(P-Q)^3 = {\rm Ch}[P]$ where we have defined the Chern number:
\begin{equation} \label{eq:chernnumber}
  {\rm Ch}[P] = \frac{i}{2\pi} \dint_{\Rm^2} {\rm tr}\, \hat P [\partial_1 \hat P,\partial_2 \hat P] d\xi =  \frac{i}{2\pi} \dint_{\Rm^2} {\rm tr}\,  \hat P d\hat P\wedge d\hat P.
\end{equation}
Here, $d\hat P=\partial_1 \hat P d\xi_1 + \partial_2 \hat P d\xi_2$. The last expression shows the invariance of the integral against changes of variables. Upon mapping $\Rm^2$ onto the (Riemann) sphere by stereographic projection and with the induced notation $\hat P(\xi)= R(\theta)$, then 
$
  {\rm Ch}[P] = \frac{i}{2\pi} \int_{\Sm^2} {\rm tr}\,  R dR\wedge dR.
$
Since $\hat P(\xi)$ converges to $\frac12(I-\sigma_3)$ as $|\xi|\to\infty$ (because $\eta\not=0$), then $\theta\mapsto R(\theta)$ is continuous. The above expression is therefore the Chern number associated to the vector bundle over the sphere $\Sm^2$ with fiber at $\theta$ the range of the projector $R(\theta)$. We thus know that ${\rm Ch}[P]$ is an integer and standard algebraic manipulations \cite{bal2019continuous} imply that 
\[
 {\rm Ch}[P] = \dfrac{1}{4\pi} \dint_{\Rm^2} \dfrac{m+\eta |k|^2}{(|k|^2+(m-\eta|k|^2)^2)^{\frac32}} dk =\dfrac{1}{2}\dint_0^\infty \!\! \dfrac{ (m+\eta r^2)r}{(r^2+(m-\eta r^2)^2)^{\frac32}}dr = \frac12 \big(\sgn{m} + \sgn{\eta}\big).
\]

This provides a classification that depends on the choice of sign of $\eta\not=0$, which is difficult to justify from a physical point of view. We note that \eqref{eq:chernnumber} may also be computed when $\eta=0$ to yield $\frac12 \sgn{m}$. This cannot be the Chern number associated to any vector bundle on any compact manifold and reflects the fact that $\hat P(\xi)$ no longer converges to a unique limit as $\xi\to\infty$. The resulting projector $R(\theta)$ is therefore not continuous at the south pole (assuming this is where $\infty$ is mapped to by the stereographic projection). This also shows that $(P-Q)^{2n+1}$ cannot be trace-class when $\eta=0$. The regularization is necessary to apply the Fedosov formula \eqref{eq:indextrace}.

Note however that the transition from an operator with positive mass term $m>0$ to one with negative mass term $m<0$ corresponds to a change of topological invariant given by $\frac12 \big(\sgn{m} + \sgn{\eta}\big) - \frac12 \big(\sgn{-m} + \sgn{\eta}\big) =\sgn{m}$ independent of the regularization $\eta$. This reflects the fact that we may more generally define topological phase differences (defined also when $\eta=0$) rather than absolute phases (defined only for $\eta\not=0$). This observation will be leveraged to the definition of bulk-difference invariants (rather than difference of bulk invariants) when we consider interface invariants in the next section.

\subsection{Dirac operators in arbitrary spatial dimensions}
\label{sec:Diracnd}
The assignment of operators of the form $PUP$ generalizes to other Hamiltonians acting on functions in Euclidean space. A roadmap is to look at trace-class properties of powers $(P-UPU^*)^{2n+1}$. We saw that $n=1$ was appropriate for two-dimensional operators. The paper \cite{bal2019continuous} considers the extension of \eqref{eq:2dDirac} to other spatial dimensions. We briefly summarize the main results.
\\[2mm]
{\bf Dirac operator in one dimension.} The Dirac operator $H=D\sigma_1$ satisfies the chiral symmetry $\sigma_3 H+H\sigma_3=0$ (see \eqref{eq:FmU} or \eqref{eq:chirsymDirac} below), with $D=D_x$ an unbounded self-adjoint operator on $\mH=L^2(\Rm)$ with domain $H^1(\Rm)$.
%The Dirac operator $H=D=D_x$ with domain $H^1(\Rm)$ is a self-adjoint unbounded operator on $\mH=L^2(\Rm)$. 
It may be classified as follows. Let $\mP(x)=\chi(x- x_0)$ be a Heaviside function associated to the odd Fredholm module with $F=\sgn{x-x_0}$. Let $U(h)=e^{2\pi i\varphi(h)}=1+W(h)$ be a unitary function with $\varphi\in\fS$ and $\varphi'(h)$ and $W(h)$ smooth and compactly supported to simplify. We then construct the unitary operator $U(D)=\mF^{-1} \hat U(\xi) \mF$ with $\mF$ one-dimensional Fourier transform. We verify that $[\mP,U]$ is a compact operator on $\mH$ and $\mP-U\mP U^*$ is trace-class. Thus, $\mP U \mP$ is a Fredholm operator on the range of $\mP$ and 
\[
  {\rm Tr}\, [\mP,U]U^* = -\ind\, \mP U\mP_{{\rm Ran} \mP} = \int_{\Rm^2} (p(x)-p(y)) u(x-y) u^*(y-x) dxdy,
\]
where $p(x)\delta(x-y)$ and $u(x-y)$ are the Schwartz kernels of $\mP$ and $U$, respectively.
This is computed, using the one-dimensional simple geometric identity $\int_{\Rm} [p(x)-p(x-X)] dx =X$, as:
\[\begin{array}{ll}
  \dint_{\Rm^2} (p(x)-p(y))u(x-y)u^*(y-x) dxdy 
   = \dint_{\Rm} X u(X)u^*(-X) dX = \frac{i}{2\pi} \dint_{\Rm} \partial_\xi \hat u(\xi) \hat u^*(\xi) d\xi =-1.
   %= \frac{-1}{2\pi i} \dint_{\Rm} \hat u^*(\xi) d \hat u(\xi) = -w_1[\hat u],
\end{array}\]
Thus, $\ind\,\mP U\mP_{{\rm Ran} \mP} =1$, reflecting right-moving transport associated to $H=D$.
\\[2mm]
{\bf Geometric algebra representation.} 
In order to describe Dirac operators in higher dimensions and their classification, we need to generalize the Pauli matrices used in the representation of the complex Clifford algebra $Cl(\Cm^2)$.   Let $d=2\kappa$ or $d=2\kappa-1$ for $\kappa\geq1$. For $\kappa=1$, the matrices $\gamma_2^{1,2,3}$ are $\sigma_{1,2,3}$, respectively. We verify that $\sigma_3=(-i)^\kappa\sigma_1\sigma_2$ for $\kappa=1$ and recall that $\sigma_i\sigma_j+\sigma_j\sigma_i=2\delta_{ij}$. %We need to generalize this structure to higher dimensions. 
A standard construction in higher dimensions is as follows. 

Let $d=2\kappa$ and assume constructed the square matrices $\gamma_d^j$ of size $2^\kappa$ for $1\leq j\leq d+1$ with 
\begin{equation}\label{eq:constCA}
\gamma_d^{d+1}=(-i)^\kappa\gamma_d^1\ldots\gamma_d^d,\qquad \gamma_d^j\gamma_d^k+\gamma_d^k\gamma_d^j=2\delta_{jk}\quad \mbox{ for all } \quad1\leq j,k\leq d+1.
\end{equation}
The construction in dimensions $2\kappa+2$ and $2\kappa+1$ is then
\[
  \gamma_{d+2}^j=\gamma_{d+1}^j=\sigma_1\otimes \gamma_d^j,\quad 1\leq j\leq d+1,\qquad \gamma_{d+2}^{d+2}=\gamma_{d+1}^{d+2}=\sigma_2\otimes I_{2^\kappa}
\]
followed by the chiral symmetry matrix $\gamma_{d+2}^{d+3}=(-i)^{\kappa+1}\gamma_{d+2}^1\ldots\gamma_{d+2}^{d+2} = \sigma_3\otimes I_{2^\kappa}$.  \eqref{eq:constCA} holds in even dimension while in odd dimension $d=2\kappa-1$, it is replaced by the constraint
\begin{equation}\label{eq:constCAodd}
\gamma_d^{d+2}=(-i)^\kappa\gamma_d^1\ldots\gamma_d^{d+1},\qquad \gamma_d^j\gamma_d^k+\gamma_d^k\gamma_d^j=2\delta_{jk}\quad \mbox{ for all } \quad1\leq j,k\leq d+1.
\end{equation}
As a convenient notation, we introduce the following vectors of $\gamma$ matrices
\[
  \Gamma_{d} = (\gamma_d^1,\ldots,\gamma_d^{d+1}),  \qquad \gamma_d=(\gamma_d^1,\ldots,\gamma_d^{d})
\]
both for even and odd dimensions $d$. When $d=2\kappa$, then $\gamma$ is the collection of generators of the $2^\kappa$ dimensional Clifford algebra $Cl(\Cm^d)$. The Dirac bulk Hamiltonian in dimension $d$ is then defined as
\begin{equation}\label{eq:Diracdim}
  H= h_{d} \cdot \Gamma_d = D\cdot\gamma_d+ m_\eta \gamma_{d}^{d+1},\qquad h_d=(D_1 , \ldots, D_d, m_\eta),\quad m_\eta = m+\eta\Delta.
\end{equation}
This expression applies to both even and odd dimensions $d$. When $d=2\kappa+1$ is odd, then the above Hamiltonian anti-commutes with the chiral matrix $\gamma_d^{d+2}$ (symmetry class AIII).  
\\[2mm]
{\bf Topological classification in even dimensions.}
Let $H$ be as in \eqref{eq:Diracdim} and define the idempotent 
\[
 P = P[H] = \frac I2 - \frac 12 \sgn{H}.
\]
Note that $H$ is gapped at $0$ since $H^2=-\Delta + (m+\eta\Delta)^2$ is gapped in $(0,m^2)$ for $\eta$ sufficiently small.

We construct an even Fredholm module to test the topology of $P[H]$ defined as
\begin{equation}\label{eq:fmodeven}
   F = \frac{x\cdot\gamma_d}{|x\cdot\gamma_d|}.
\end{equation}
Since every matrix $\gamma^j_d$ in $\gamma_d$ anti-commutes with $\gamma^{d+1}$, they may be written as
\[
  \gamma^j = \begin{pmatrix} 0 & (\check\gamma^j_d)^* \\ \check\gamma^j_d & 0 \end{pmatrix} \quad \mbox{ so that } \quad F =  \begin{pmatrix} 0 & \mU^* \\ \mU & 0 \end{pmatrix}\quad \mbox{ with } \quad 
  \mU(x) = \frac{x\cdot\check\gamma_d}{|x\cdot\check\gamma_d|}
\]
where $\check\gamma_d = (\check\gamma_d^1,\ldots,\check\gamma_d^{d})$.
Thus $F$ satisfies \eqref{eq:FmU}. We verify that $\mU(x)=\frac{x_1+ix_2}{|x_1+ix_2|}$ when $d=2$. 

The operator $x\cdot\gamma_d$ is often referred to as a Dirac operator as a first-order differential operator in the dual variables $\xi$ to $x$. We thus have the collusion of two notions of Dirac operators, the physical one in \eqref{eq:Diracdim} and the classifying one in \eqref{eq:fmodeven} used in the construction of the Fredholm module.

The operator $H$ acts on $L^2(\Rm^d)\otimes \Cm^{2^\kappa}$ while the unitary operator $\mU(x)$ acts on $L^2(\Rm^d)\otimes \Cm^{2^{\kappa-1}}$. We define the operators $\tilde P=P\otimes I_{2^{\kappa-1}}$ and $\tilde \mU=I_{2^{\kappa}}\otimes \mU$ on the tensor product $L^2(\Rm^d)\otimes \Cm^{2^\kappa}\otimes \Cm^{2^{\kappa-1}}\cong L^2(\Rm^d)\otimes \Cm^{2^{d-1}}$.  Define the {\em even Chern number} as
\begin{equation}\label{eq:evenChern}
{\rm Ch}_d[P] = \dfrac{i^\frac d2}{(2\pi)^{\frac d2}(\frac d2)!} \dint_{\Rm^d} \trr \ \hat P (d\hat P)^{\wedge d} .
\end{equation}
\begin{theorem}[\cite{bal2019continuous}]\label{thn:Diraceven}
 The operator $\tilde P \tilde U \tilde P$ is Fredholm on the range of $\tilde P$ and 
 \begin{equation}\label{eq:bulkevenindex}
	-\ind\,{\tilde P\tilde \mU\tilde P} 
	=(-1)^{\frac d2+1} {\rm Ch}_d[P]
	=\frac12 (\sgn{m}+\sgn{\eta}).
\end{equation} \end{theorem}
The proof follows a similar structure to the case $d=2$. Russo's criterion \eqref{eq:Russo} applies for $p=d+1$ allowing us to write the index as the trace of $(\tilde P-\tilde Q)^{d+1}$ where $\tilde Q=\tilde\mU \tilde P \tilde \mU^*$ and hence as the integral along of the diagonal of its Schwartz kernel. 
Connes' geometric identity \eqref{eq:connesformula} is replaced by its $d-$dimensional equivalent \cite{bal2019continuous,prodan2016bulk}:
\[
 	\dint_{\Rm^d} \trr \ \mU(x)(\mU^*(x)-\mU^*(x+y_d))\ldots (\mU^*(x+y_1)-\mU^*(x)) dx =\dfrac{(2\pi i)^{\frac d2}}{\frac d2!} {\rm Det}(y_d,\ldots,y_1) I.
\]
Writing the Schwartz kernel of $[\tilde P,\tilde \mU]$ as $p(x-y)\otimes (\mU(x)-\mU(y))$ and using the above identity shows after some algebra that $-\ind\, {\tilde P\tilde \mU\tilde P} = (-1)^{\frac d2+1} {\rm Ch}_d[P]$. By stereographic projection onto the one-point compactification $\Sm^d\cong\Rm^d\cup\infty$, we observe that the above integral is a bona fide (even) Chern number of a line bundle over a sphere. Since $\hat P$ is written in a Clifford algebra representation, the computation of ${\rm Ch}_d[P]$ may be carried out explicitly by expressing it as the topological degree of an appropriate map on $\Sm^d$; see \cite{bal2019continuous,prodan2016bulk} for details that lead to the derivation of \eqref{eq:bulkevenindex}.

Note that ${\rm Ch}_{d=2}[P]$ is often referred to as the {\em first} Chern number (and often denoted by $c_1$), while ${\rm Ch}_{d=4}[P]$ is called the {\em second} Chern number (and often denoted by $c_2$), and so on. We follow the convention in \cite{carey2011spectral,prodan2016bulk}.
\\[2mm]
{\bf Topological classification in odd dimensions.}
In odd spatial dimensions $d=2\kappa+1$, the roles of the projectors and unitaries are reversed. The operator $H$ in \eqref{eq:Diracdim} satisfies a chiral symmetry $\gamma_d^{d+2}H+H\gamma_d^{d+2}=0$ (and hence belongs to operators of complex class AIII). We may therefore define by functional calculus:
\begin{equation}\label{eq:chirsymDirac}
    \dfrac{H}{|H|} = \left(\begin{matrix} 0&U^*\\U&0 \end{matrix}\right), \qquad U=\frac{h_1\gamma^1_d+\ldots h_d\gamma^d_d + i h_{d+1}}{|H|},
\end{equation}
where $h_d$ is defined in \eqref{eq:Diracdim} and $U$ is a unitary operator. %, although not a Hermitian one.

The odd Fredholm module is defined as
\begin{equation}\label{eq:PUodd}
F=\sgn{x\cdot\gamma_d} =\sgn{\sum_{j=1}^d x_j \gamma^j_d},\qquad \mP=\frac12(I+F).
\end{equation}
This is again the sign function of a Dirac operator $x\cdot\gamma_d$ written in spatial variables. It generalizes the Heaviside function $\mP(x)=\frac12(1+\sgn{x})$ used above in dimension $d=1$.

We form the operator $I_{2^{\kappa+1}}\otimes \mP\ U[H]\otimes I_{2^\kappa}\ I_{2^{\kappa+1}}\otimes \mP$. These operators are defined on  $L^2(\Rm^d)\otimes\Cm^{2^{\kappa+1}}\otimes \Cm^{2^{\kappa}}$. We define $\tilde\mP=I_{2^{\kappa+1}}\otimes\mP$ and $\tilde U=U[H]\otimes I_{2^\kappa}$.  Define the {\em odd winding number} (a.k.a. {\em odd Chern number}):
\begin{equation}\label{eq:oddwinding}
 {\rm W}_d[U] = \frac{1}{2^d d!!} \big(  \frac i\pi \big)^{\frac {d+1}2} \dint_{\Rm^d} {\rm tr}\,(\hat U^{-1}d\hat U)^{\wedge d}.
 \end{equation}
\begin{theorem}
\label{thm:bulkodd}
The operator $\tilde\mP \tilde U\tilde\mP$ is Fredholm on Ran$\tilde \mP$. Moreover, 
\begin{equation}\label{eq:bulkoddindex}
	-\ind\,{\tilde\mP \tilde U\tilde\mP} 
	=-W_d[U] 
	=\frac12 (\sgn{m}+\sgn{\eta}).
\end{equation}
\end{theorem}
The proof is similar to that of the even dimensional setting; see \cite{bal2019continuous}. A minor difference compared to the even case is that Russo's lemma \ref{lem:russo} applies only for $p>d$. Since the resulting bound is independent of $p>d$, we obtain that  $\tilde\mP - \tilde U\tilde\mP \tilde U^*\in \mI_d$ so the index is indeed written in terms of the Schwartz kernel of $(\tilde\mP - \tilde U\tilde\mP \tilde U^*)^d$.

The Dirac operators in even dimensions are prototypical examples of two-band operators of dimension $d$ in the complex class A. The Dirac operators in odd dimension are prototypical examples of two-band operators of dimension $d$ in the complex class AIII ({\em chiral} symmetry) \cite{bernevig2013topological,kitaev2009periodic}.
\subsection{Stability under perturbations and Helffer-Sj\"ostrand formula}
\label{sec:stabbulk}
The topological classifications and computations of topological invariants were obtained for unperturbed operators so far, with constant magnetic field for the Landau operator and constant mass term for the Dirac operator. We now show that the invariants are stable against perturbations. Heuristically, the topological classification depends on the operator's behavior as the Fourier variables $\xi\to\infty$ so that spatially local perturbations should not modify the topological index. 
\\[2mm]
{\bf Odd Fredholm modules.}
Consider the odd-dimensional case. Let $U=U(H)$ be a unitary operator constructed by spectral calculus and $P=P(x)$ a projector obtained from an odd Fredholm module. We know that $P(x)U(H)P(x)$ is a Fredholm operator as soon as $[P,U]$ is compact. 

Assume that $H$ is perturbed to $H_V=H+V$ for $V$ a Hermitian perturbation such that $H+V$ is self-adjoint on $\mD(H+V)=\mD(H)\subset\mH$. We would like to show that $P(x)U(H+V)P(x)$ remains a Fredholm operator with the same index as that of $P(x)U(H)P(x)$. A sufficient criterion ensuring the stability of the index is that $U(H+V)-U(H)$ is compact for then $P(x)U(H+V)P(x)-P(x)U(H)P(x)$ on the range of $P$ is compact as well.

A classical method to analyze the operator $U(H+V)-U(H)$ is to use the Helffer-Sj\"ostrand formula, which relates functional calculus to resolvent integrals; see \cite[Chapter 8]{dimassi1999spectral} and \cite{davies1995spectral}.  Let $f\in C^\infty_0(\Rm)$ be a smooth function vanishing at infinity. Then we have the existence of a smooth {\em almost analytic extension} $\tilde f(z)$ for $z \in \Cm$ such that 
\begin{equation}\label{eq:aee}
  \tilde f(\lambda)=f(\lambda),\quad \lambda\in \Rm, \qquad |\bar\partial \tilde f(z) | \leq C_N |\Im z|^N, \ \ \forall N \in \Nm,
\end{equation}
where for $z=\lambda+i\mu$, $\bar\partial = \frac12 (\partial_\lambda + i \partial_\mu)$ is the Cauchy-Riemann operator. The extension, which is not unique, may be chosen smooth with compact support (arbitrarily close to the real axis) in $\Cm$.% and smooth when $f$ is itself compactly supported.

This extension allows us to describe the functional calculus in terms of resolvent operators of $H$ an unbounded self-adjoint operator by the {\em Helffer-Sj\"ostrand formula}
\begin{equation}\label{eq:HS}
  f(H) = \dfrac{-1}{\pi} \dint_{\Cm} \bar\partial \tilde f(z) (z-H)^{-1} d^2z
\end{equation}
where $d^2z=d\lambda d\mu$ is Lebesgue measure on $\Cm$. Thus, writing $U=I+W$ with $W\in C^\infty_c(\Rm)$, we have
\begin{equation}\label{eq:diffU}
   U(H+V) - U(H) = \dfrac{-1}{\pi} \dint_{Z} \bar\partial \tilde W(z)  (z-H-V)^{-1} V (z-H)^{-1} d^2z,
\end{equation}
for $Z$ any compact domain including the support of $\tilde W$.
We then have the following stability result:
\begin{proposition}\label{prop:compactperturb}
  Let $H$ and $H+V$ be self-adjoint operators such that for each $\eps$, there is $V_\eps$ such that $\|(V-V_\eps)(z-H)^{-1}\|\leq \eps |\Im z|^{-1} $ and $V_\eps(z-H)^{-1}$ is compact in $\mI_p$ for some $p<\infty$ with $\|V_\eps(z-H)^{-1}\|_p \leq C_\eps |\Im z|^{-q}$ uniformly for $z\in Z$ for some $q\in \Nm$. 
    
  Then $U(H+V)-U(H)$ is compact and $P(x) U(H+V) P(x)$ is a Fredholm operator on the range of $P$ and
   $\ind\,P(x) U(H+V) P(x) = \ind \, P(x) U(H) P(x)$. 
\end{proposition}
This is a direct consequence of \eqref{eq:diffU} and \eqref{eq:aee}. 
\\[2mm]

{\bf Even Fredholm modules.} In this setting, the operator of interest is of the form $P(H)U(x)P(H)$. The ranges of $P(H)$ and $P(H+V)$ do not need to coincide. We therefore consider the Fredholm operator $T(H)=P(H)U(x)P(H) + (I-P(H))$ which is clearly Fredholm on (the whole of) $\mH$.  We then obtain that $T(H+V)$ is also Fredholm on $\mH$ provided for instance that $P(H+V)-P(H)$ is a compact operator. 

To obtain the latter, we cannot apply the Helffer-Sj\"ostrand formula directly since $h\mapsto P(h)$ is neither smooth nor compactly supported.  We observe however that $P(h)=\frac12(1-\sgn{h})=P(h/|h|)=P\circ \Phi(h)$ depends only on the sign of $h$ and not its magnitude so that the above equality holds for any bounded function $\Phi$ such that $\sgn{h}\Phi(h)>0$ for $h\not=0$. We also assume that $H$ is gapped near zero, i.e., there is an interval $\Delta\ni 0$ such that $\Delta\cap {\rm Spec}(H)=\emptyset$. Let us assume that $V$ is such that $\Delta\cap {\rm Spec}(H+V)=\emptyset$, for instance because $V$ is sufficiently small in the uniform sense or by some compactness argument. Then $P(H+V)=P\circ \Phi(H+V)= P_\eta\circ \Phi(H+V)$ for $\eta$ sufficiently small where $P_\eta$ is smooth and equal to $P(h)$ for $|h|\geq\eta$. If we assume $\Phi\in C^\infty(\Rm)$ with $\sgn{h}\Phi(h)>0$ for $h\not=0$ decaying sufficiently fast at infinity, then we may apply \eqref{eq:HS} to $P_\eta\circ \Phi$ \cite[Theorem 2.3.1]{davies1995spectral}. For $V$ sufficiently small and compactly supported and $H$ either the Landau operator or the Dirac operator in even dimension, we deduce as in Proposition \ref{prop:compactperturb} that $T(H+V)$ is Fredholm on $\mH$ and that the index is independent of $V$.

\section{Interface Hamiltonian}
\label{sec:IH}
By interface Hamiltonian $H=H_I$, we mean a self-adjoint (pseudo-)differential operator with symbol transitioning from a bulk Hamiltonian as $y=L\gg1$ to another bulk Hamiltonian as $y=-L\ll-1$ (in two-space dimensions parametrized by $(x,y)$). The prototypical example is the Dirac operator with a domain wall generated by a spatially varying mass term $m(y)$:
\begin{equation}\label{eq:HIDirac}
  H_I = D_x \sigma_3 - D_y \sigma_2 + m(y) \sigma_1
\end{equation}
with $m(y)\geq m_0>0$ for $y\geq L\gg1$, say, while $m(y)\leq-m_0$ for $y=\leq-L\ll1$.

We interpret the interface $y=0$ as separating two insulators in different topological phases. This topological difference is compensated by a rather anomalous behavior along the interface: transport is asymmetric, with more signal propagating in one direction than the other. This excess is moreover quantized as we will see.  For the above Dirac operator, we observe that
\[
  \phi(x,y;\xi) = e^{ix\xi} \psi_0(y), \quad \psi_0(y) =e^{-M(y)}  \begin{pmatrix} 0 \\ 1 \end{pmatrix},\qquad M(y)=\int_0^y m(z) dz
\]
is a solution of $(H_I+\xi)\phi(x,y;\xi)=0$. We assume that $m(y)\geq m_0>0$ for $y\geq L\ll1$ and $m(y)\leq -m_0$ for $y<-L$ so that $e^{-M(y)}\in L^2(\Rm)$. This corresponds to a whole branch of absolutely continuous spectrum of $H_I$ with dispersion relation $E(\xi)=-\xi$ and group velocity $E'(\xi)=-1$. This branch thus models propagating modes with constant speed $-1$ along the $x$-axis. This may be further confirmed by the construction of the time dependent wave-packet
\[
  \psi(t,x,y) = \int_{\Rm} \hat u_0(\xi)  e^{-iE(\xi)t}  \phi(x,y;\xi) \frac{d\xi}{2\pi} = \int_{\Rm} \hat u_0(\xi) e^{i(x+t)\xi} \frac{d\xi}{2\pi} \psi_0(y)= u_0(x+t)\psi_0(y),
\]
a left-moving solution of the evolution Schr\"odinger equation $(D_t+H_I)\psi(t,x,y)=0$ with initial condition $\psi(0,x,y)=u_0(x)\psi_0(y)$.

\begin{figure}[htbp]
\begin{center}
\includegraphics[width=6cm]{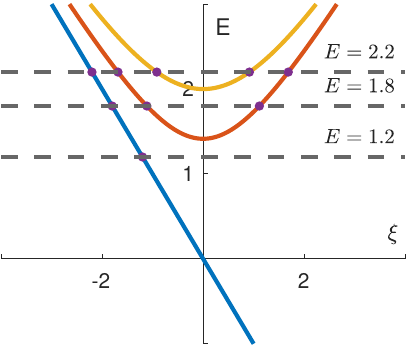}  \hspace{.9cm}  \includegraphics[width=6cm]{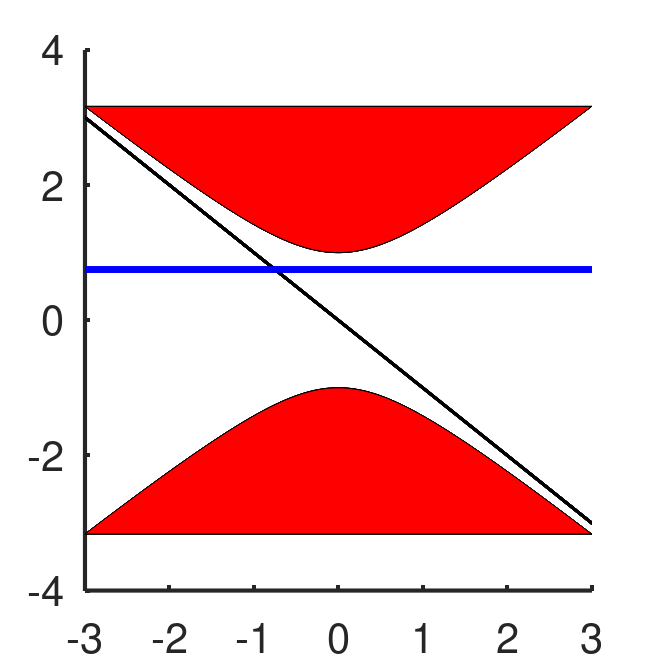} 
\end{center}
\caption{Spectral of Dirac interface Hamiltonian with: Left: mass term $m(y)=y$ (only positive energies displayed for clearer presentation); Right: mass term $m(y)=\sgn{y}$ with a unique branch of edge spectrum while filled (red) parts correspond to bulk spectrum.}
\label{fig:Dirac}
\end{figure}

The fact that $\psi_0(y)$ above is independent of $\xi$ is not generic. Neither is the fact that the group velocity is constant (and hence the mode non-dispersive). The above calculation presents the simplest example of an edge state propagating in an asymmetric manner. It is characterized by a spectral branch negatively crossing an interval $\Delta\ni 0$ sufficient small. A topological classification of interface Hamiltonians may in fact be obtained by counting the number of branches of spectrum crossing the energy level $0$. For $H_I$ that number would equal $-1$. See Fig. \ref{fig:Dirac} for a spectral decomposition of $H_I$ with $m(y)=y$ and $m(y)=\sgn{y}$.
\subsection{Interface current observable}
A robust and general way to quantify the interface asymmetry is to consider the following interface current observable. Let $P(x)\in\fS$ be a switch function. Then $i[H_I,P]$ may be interpreted physically as a current operator modeling current across the region (near a vertical line) where $P$ transitions from $0$ to $1$. Let $\varphi\in\fS$ be another switch function such that $\varphi'$ is supported in the spectral gap of each bulk insulator in $y\gg L$ and $y \ll -L$. For the above Dirac example, this means that $\varphi'$ is supported in $(-m_0,m_0)$, i.e., $\varphi\in \fS[0,1;-m_0,m_0]$. The role of $\varphi'$ is to filter out energies that are not in the bulk spectral gaps and hence can propagate into the (no longer insulating) bulks.

The expectation value of the current observable $i[H_I,P]$ for a density of states $\varphi'(H_I)$ is then:
\begin{equation}\label{eq:sigmaI}
  \sigma_I [H_I] = {\rm Tr}\, i[H_I,P] \varphi'(H_I)
\end{equation}
assuming that $i[H_I,P] \varphi'(H_I)$ is a trace-class operator. Here, ${\rm Tr}$ is the standard trace on the Hilbert space $\mH$, which is $L^2(\Rm^2;\Cm^2)$ for the above Dirac operator. The terminology $\sigma_I$ stems from the electronic setting, where $\sigma_I$ may be interpreted as a conductivity. We will refer to it as an interface (current) observable. We will take this observable as the physical object describing asymmetric transport along an interface, which at the moment is the $x-$axis.

\medskip

 The analysis of \eqref{eq:sigmaI} is the starting point of many mathematical studies of interface Hamiltonians and of their interplay with bulk Hamiltonians. An interface observable of the form $ \tilde \sigma_I := {\rm Tr}_v \, i[H,X] \varphi'(H)$ was introduced in  \cite{schulz2000simultaneous} following an analysis of edge modes in \cite{halperin1982quantized}. Here $X$ is the position operator of multiplication by the spatial variable $x$ and ${\rm Tr}_v$ is a trace per volume. Such an object is shown to be well defined for Landau operators with random coefficients that are ergodic and stationary. This interface observable is also the starting point of algebraic (operator K-theoretic) analyses of edge effects in continuous and discrete settings \cite{Avila2013,bourne2017k,prodan2016bulk}. For inhomogeneous perturbations, $\tilde \sigma_I$ is not necessarily stable and \eqref{eq:sigmaI} should be preferred. The current observable \eqref{eq:sigmaI} and its generalizations  is also central in the works \cite{elbau2002equality,elgart2005equality} on discrete Hamiltonians and in the analysis of second-order Hamiltonians with modulated locally periodic coefficients \cite{drouot2021microlocal}.%%%

One of the main objectives of this review is: (i) to show that $2\pi\sigma_I\in\Zm$ is quantized and hence stable against continuous deformations for large classes of interface Hamiltonians $H_I$; and (ii) to compute this invariant in general settings. 

\subsection{Interface observable and spectral flow.}
\label{sec:SF}

The computation of $\sigma_I$ is in general a difficult task. It may be computed by means of spectral flows when $H\equiv H_I$, assumed to be unbounded self-adjoint on $\mH$, is invariant by translation along the $x-$axis. In the latter case, $\mF H \mF^{-1} = \int^{\oplus}_{\Rm} \hat H(\xi) d\xi$, with $\mF$ one-dimensional Fourier transform. Let $\Delta$ be an open interval including the support of $\varphi'$, $\Phi$ a smooth function supported on $\Delta$, and assume that $H$ admits the following spectral decomposition
\begin{equation}\label{eq:spectralH}
 \mF \Phi(H)  \mF^{-1} = \sum_{j\in J} \dint^\oplus_{\Rm} \Phi \circ E_j(\xi) \Pi_j(\xi)
\end{equation}
where $\xi\mapsto E_j(\xi)$ are branches of spectrum of $H$ for $1\leq j\leq J<\infty$ and $\Pi_j(\xi)$ are generalized projectors with Schwartz kernel $\Pi_j(\xi;x,x') = \tilde \Pi_j(\xi) \frac{1}{2\pi} e^{i(x-x')\xi}$ while 
$\tilde\Pi_j(\xi)$ may be taken as a rank-one projector without loss of generality.  We assume that $\xi \mapsto E_j(\xi)$ are smooth branches crossing the interval $\Delta=(\alpha,\beta)$ on a compact domain. We define $E_j^\pm$ as equal to $\alpha$ or $\beta$ depending on whether $E_j(\xi)$ leaves $\Delta$ through $\alpha$ or $\beta$ as $\xi\to\pm\infty$. Then we have  (see e.g., \cite{bal2022topological,combes2005edge,quinn2024approximations} for different forms):
\begin{lemma} \label{lem:sftrace}
  Let $\Phi(h)$ and $\Psi(h)$ be in $C^\infty_c(\Delta)$ and $H$ as above. Assume that $i[\Psi(H),P] \Phi'(H)$ is trace-class and with trace computed as an integral along the diagonal of its Schwartz kernel.  Then:
  \[
    {\rm Tr}\, 2\pi i[\Psi(H),P] \Phi(H) = \dsum_j \dint_\Rm \partial_\xi ( \Psi\circ E_j(\xi)) \Phi\circ E_j(\xi) d\xi.
  \]
\end{lemma}
\begin{corollary}[Spectral flow]
  Under the hypotheses of the above Lemma and $\varphi\in  \fS[0,1;\alpha,\beta]$, we have:
  \[
    2\pi \sigma_I = \sum_{j}  \int_{\Rm} d (\varphi \circ E_j)(\xi) =\dsum_j \varphi(E^+_j)-\varphi(E^-_j) \in \Zm.
  \]
\end{corollary}
By construction, we find that $\varphi(E^+_j)-\varphi(E^-_j)=1$ ($-1$) when the branch $\xi\to E_j(\xi)$ crosses the interval $\Delta=(\alpha,\beta)$ from $\alpha$ to $\beta$  (from $\beta$ to $\alpha$), while $\varphi(E^+_j)-\varphi(E^-_j)=0$ otherwise.

The derivation of the corollary stems from choosing $\Phi=\varphi'$ and $\Psi(h)=h$ on the support of $\varphi'$ and using the cyclicity of the trace to verify that ${\rm Tr} \, 2\pi i[\Psi(H),P] \Phi(H)={\rm Tr} \, 2\pi i[H,P] \Phi(H)$. The main steps of the derivation of Lemma \ref{lem:sftrace}, applied in different forms in \cite{bal2022topological,combes2005edge,quinn2024approximations}, are as follows.  Using \eqref{eq:spectralH}, we deduce the following expression for the Schwartz kernels in the $x$-variables:
\begin{align*}
 \begin{array}{rcl}
    2\pi i [\Psi(H),P ] (x,x')
     &=& \dsum_j \dint_{\Rm} 2\pi i \Psi\circ E_j(\xi)  \tilde \Pi_j(\xi) \frac{1}{2\pi} (P(x')-P(x)) e^{i(x-x')\xi} d\xi,\\
     \Phi(H) (x',x'') &=&  \dsum_k \dint_{\Rm}\Phi\circ E_k(\xi')  \tilde \Pi_k(\xi') \frac{1}{2\pi} e^{i(x'-x'')\xi'} d\xi'.
 \end{array}
\end{align*}
Using $\int_{\Rm} (P(x+z)-P(x)) dx=z$ and the notation ${\rm Tr'}$ for trace on $L^2_\xi$, we find
\begin{align*}
 \begin{array}{rl}
    {\rm Tr}\,2\pi i [\Psi(H),P ]  \Phi(H) = {\rm Tr'}\,\dsum_{j,k} \dint_{\Rm^3} \Psi\circ E_j(\xi) \Phi\circ E_k(\xi')  \tilde \Pi_j(\xi)\tilde \Pi_k(\xi')\frac{-iz}{2\pi}  e^{iz(\xi-\xi')}   d\xi d\xi' dz.
         \end{array}
\end{align*}
Integrating out the variable $z$ and using the relations $\partial_\xi \tilde \Pi_j \tilde\Pi_k + \partial_\xi \tilde \Pi_k \tilde\Pi_j  =0$ and  ${\rm Tr'} \, \partial_\xi \tilde \Pi_j \tilde\Pi_j=0$ with $\Tr' \tilde \Pi_j=1$ gives the result. For the Dirac operator, $2\pi\sigma_I[H_I]=-1$, consistent with Figure \ref{fig:Dirac} \cite{bal2022topological}.

Let $\rP$ be a Heaviside function and $U(H)=e^{i2\pi \varphi(H)}$ a unitary operator. Define $W=U-I$. Then, similarly, we have \cite{bal2022topological,quinn2024approximations}:
\begin{corollary}[Spectral flow and Index] \label{cor:SF}
  Under the hypotheses of the above Lemma, and assuming $[U(H),\rP]$ is trace-class with vanishing trace, then
  \begin{equation}\label{eq:SF}
    \ind\, \rP U \rP  = \sum_{j}  w_1[e^{i 2\pi \varphi \circ E_j}]=  2\pi \sigma_I \in \Zm,\quad w_1[f]=\frac{1}{2\pi i} \dint_{\Rm} f^*(\xi) df(\xi).
  \end{equation}
\end{corollary}
Here, $w_1$ is the winding number of a function from $\Sm^1$ to $\Sm^1$. The right equality is deduced from the above lemma by choosing $\Phi(h)=U^*(h)=I+V^*(h)$ and $\Psi(h)=\frac{1}{2\pi i} U(h)= \frac{1}{2\pi i} (I+W(h))$ knowing that $W(h)\in C^\infty_c(\Delta)$. For each branch, we verify that $w_1[e^{i 2\pi \varphi \circ E_j}]=\varphi(E^+_j)-\varphi(E^-_j)$. The left equality comes from the fact that since $[U,\rP]$ is compact, then $\rP U \rP$ restricted to ${\rm Ran}\rP$ is a Fredholm operator whose index is given using the Fedosov formula by ${\rm Tr}\,[U,\rP]U^*$.

\subsection{Application to continuous operators.}
\label{sec:appliSF}

The formula \eqref{eq:SF} allows one to compute interface invariants when enough information on the spectral decomposition of the interface Hamiltonian $H_I$ is available. 
\\[2mm]
{\bf Landau operator.}
Consider the confined Landau operator $H=(D-A)^2+V$. \cite{combes2005edge} analyzes the setting $H=D_x^2+(D_y-Bx)^2+V+W$ for $B>0$ where $V$ is confining and $W$ is a perturbation. Let $\varphi'$ be supported in $(B(2N-1),B(2N+1)$ inside the $N$th spectral gap of the Landau operator.  Assume $W=0$ and $V=V(x)$ with $V(x)=0$ for $x\geq0$ and $V(x)$ greater than $B(2N+1)$ for $x<-1$. Then $V(x)$ acts as a confining potential near $x=0$, implying that $N$ Landau levels are confined near the interface $x=0$. As a result, the operator $H=D_x^2+(D_y-Bx)^2+V$ is invariant with respect to translations in $y$. Using the above spectral flow calculation we have in \cite[Proposition 1]{combes2005edge} that 
\[2\pi \Tr\, i[H,P(y)]\varphi'(H)=N,\]
and, moreover, the latter is stable against small perturbations $W$ \cite{combes2005edge}.

In \cite{dombrowski2011quantization}, the Landau operator is analyzed for a spatially varying magnetic field such that $B(x,y)=B(x)$ converges to $B_\pm$ as $x\to\pm\infty$. Let $\varphi'$ be supported in a gap for both Hamiltonians, i.e., supported in an interval meeting no point of the form $B_+(2n+1)$ or $B_-(2n+1)$. The transition from $B_-$ to $B_+$ for an energy level in the support of $\varphi'$ therefore crosses $\Zm\ni N=\sgn{B_-}n_--\sgn{B_+} n_+$ Landau levels. This generates a number of $N$ edge modes and \cite[Theorem 2.2]{dombrowski2011quantization} indeed shows using spectral flow calculations that $2\pi \Tr i[H,P(y)]\varphi'(H)=N$ in that case as well. The latter result is also shown to be stable against perturbations \cite{dombrowski2011quantization}. See also the related approach based on the St\v reda formula in \cite{cornean2024orbital}.
\\[2mm]
{\bf Magnetic Dirac operator.} Consider the magnetic Dirac operator $H = D_x \sigma_1 + (D_y - A_2(x)) \sigma_2 + m(x) \sigma_3 + V(x)$ where $A_2 (x) = x B (x)$ and $B \in \fS [B_-, B_+]$, $m \in \fS [m_-, m_+]$, and $ V \in \fS [V_-, V_+]$ for constants $B_\pm, m_\pm, V_\pm \in \mathbb{R}$ with $B_\pm \ne 0$. These three domain walls include the two domain walls we just considered for the Landau operator plus the standard domain wall in the mass term as in \eqref{eq:HIDirac}. The fact that $B_\pm \ne 0$ shows that the spectrum of the limiting bulk operators is purely composed of Landau levels. As $(B,V,m)$ vary, a number of Landau levels are crossed and this generates an equal number of edge modes.  Let $\varphi'$ be supported in a spectral gap common to both bulk insulators. Let $\alpha$ be an energy level in the support of $\varphi'$. Then we have the following result
\begin{theorem}[{\cite[Theorem 2.1]{quinn2024asymmetric}}]\label{thm:sf0}
We have $2\pi\sigma_I[H] =  I (H_-;\alpha) -I(H_+;\alpha)$, where
\begin{equation}\label{eq:I}
    I (H_\pm; \alpha) = \sgn {B_\pm}\sgn {\alpha - V_\pm - m_\pm\sgn {B_\pm}} \Big(N(H_\pm;\alpha) + \frac{1}{2}\Big)
\end{equation} 
for $N(H_\pm;\alpha)=0$ when $|\alpha-V_\pm|< \sqrt{2|B_\pm| + m_\pm^2}$ and $N(H_\pm;\alpha)=k$ when $\sqrt{2 k |B_\pm| + m_\pm^2} < |\alpha-V_\pm| < \sqrt{2 (k+1) |B_\pm| + m_\pm^2}$ for $k \in \mathbb{N}_+$.\end{theorem}
The above result is obtained by spectral flow \cite{quinn2024asymmetric}. The main difficulty is to understand the behavior of the branches of absolutely continuous spectrum as the dual parameter $\zeta\to\infty$. The structure of the Dirac operator allows us to still show that branches of spectrum are simple and also stable against perturbations; see \cite[Theorems 3.2-3.4]{quinn2024asymmetric}.
\\[2mm]
{\bf Shallow water Hamiltonian.}
The simplest model of atmospheric transport is the following linearized system of shallow water equations
\begin{equation}\label{eq:H}
  H = D_x\gamma_1+D_y\gamma_4 - f(y) \gamma_7 = \begin{pmatrix} 0 & D_x & D_y \\ D_x & 0 & if(y) \\ D_y & -if(y) & 0\end{pmatrix},\quad \psi =  \begin{pmatrix} \eta  \\ u \\ v \end{pmatrix},
\end{equation}
where $H$ acts on vector-valued functions of the above form $\psi$ and $\gamma_{1,4,7}$ are Gell-Mann matrices. Here, $(x,y)\in\Rm^2$ are spatial coordinates, $f(y)$ is a Coriolis force parameter, and mass transport is modeled by $\eta(x,y)$ the height of an atmospheric or oceanic layer, $u(x,y)$ its horizontal velocity, and $v(x,y)$ its vertical velocity. See \cite{delplace2017topological} for a derivation of this model from Boussinesq primitive equations.

When $f$ is constant, the operator then $H\equiv H_B$ may be diagonalized as
\begin{equation}\label{eq:hatHB}
  H_B = \mF^{-1} \hat H_B \mF,\qquad  (\xi,\zeta)\mapsto \hat H_B(\xi,\zeta) = 
  \xi\gamma_1+ \zeta\gamma_4-f\gamma_7
\end{equation}
with three branches of absolutely continuous spectrum parametrized by
\[
  (\xi,\zeta) \mapsto  E_0(\xi,\zeta)=0,\qquad (\xi,\zeta) \mapsto  E_\pm(\xi,\zeta)=\pm \sqrt{\xi^2+\zeta^2+f^2}.
\]
When $f\not=0$, we thus have two spectral gaps in $(-|f|,0)$ and $(0,|f|)$. The Coriolis force parameter is positive in the northern hemisphere and negative in the southern hemisphere. A reasonable model is in fact given by $f(y)=\beta y$ in a $\beta-$plane model \cite{delplace2017topological,matsuno1966quasi}. Let $\mF_{x\to\xi}$ be the Fourier transform in the first variable only. Then, we have the partial diagonalization 
\begin{equation}\label{eq:hatH}
  H_I = \mF_{\xi\to x}^{-1} \hat H \mF_{x\to\xi} ,\qquad \Rm\ni \xi \mapsto \hat H(\xi) =  \begin{pmatrix} 0 & \xi& D_y \\ \xi & 0 & if(y) \\ D_y& -if(y)& 0\end{pmatrix}.
\end{equation}
\begin{figure}[htbp]
  \begin{center}
  \includegraphics[width=7cm]{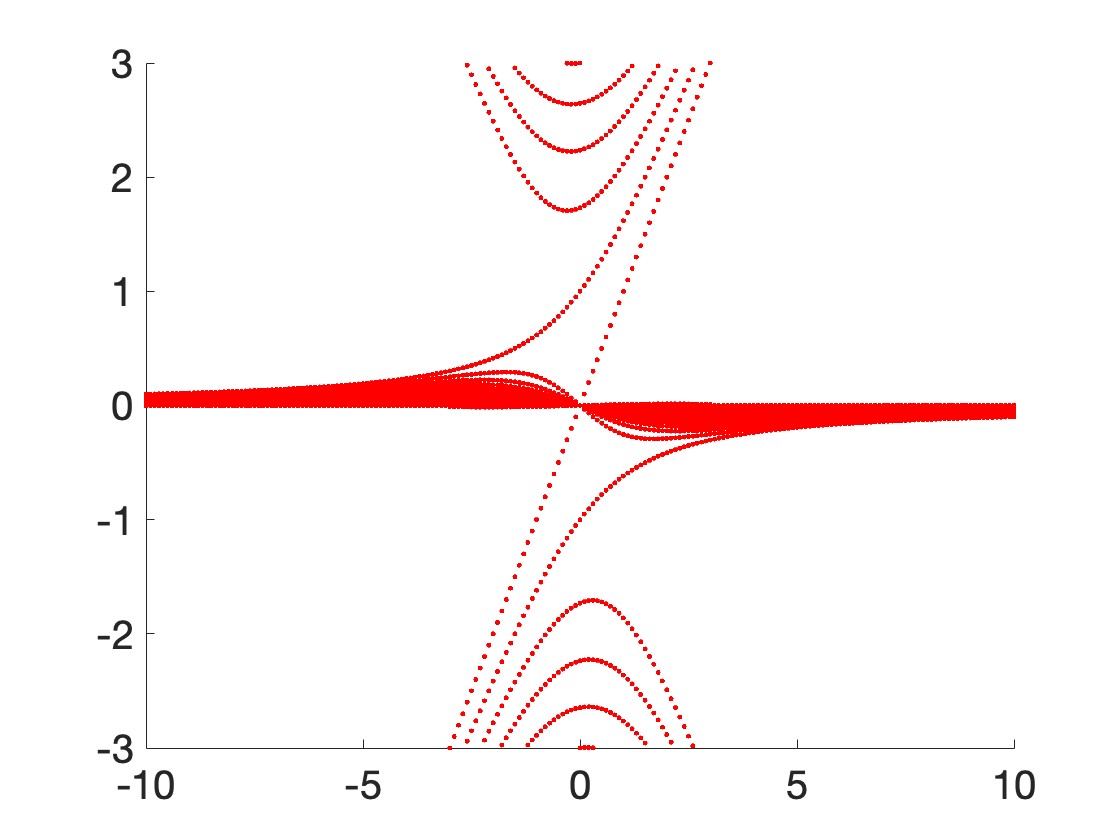}  \hspace{1cm} \includegraphics[width=7cm]{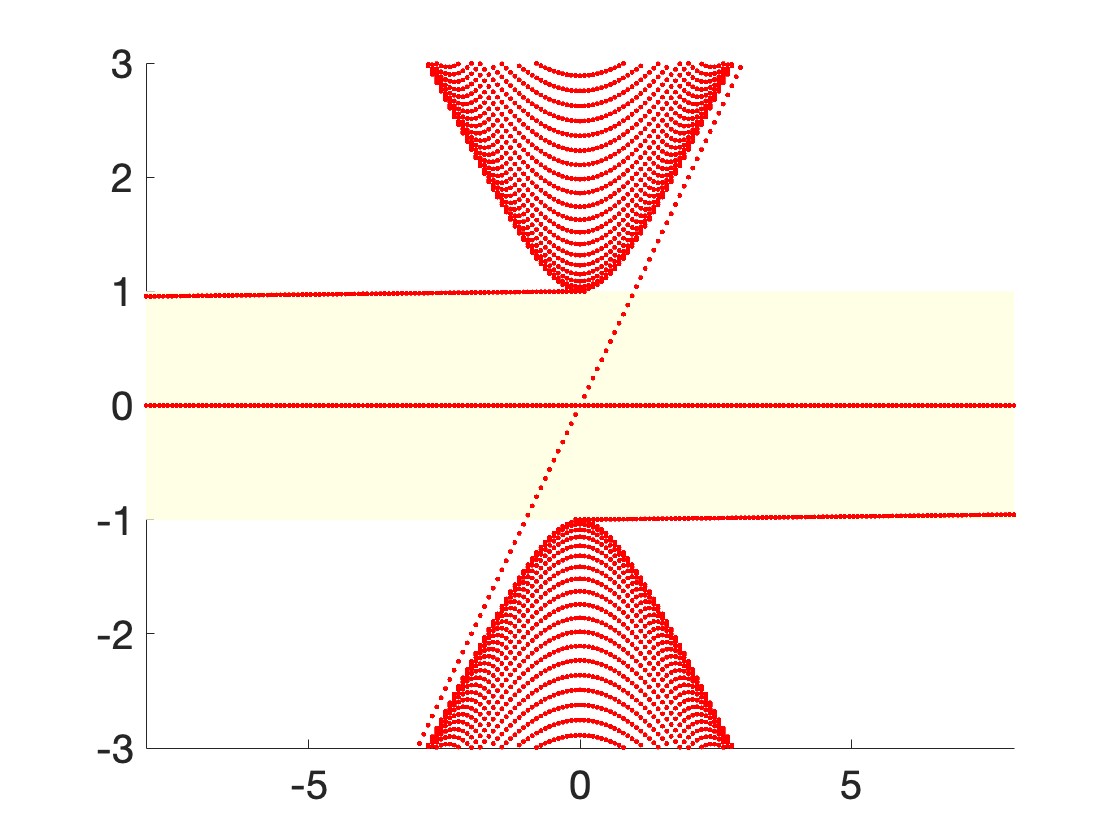}
  \end{center}
  \caption{Left: spectrum of $H$ when $f(y)=y$ with a spectral flow equal to $2$ as dictated by the BEC. Right: spectrum of $H$ when $f(y)=\sgn{y}$.} 
  \label{fig:shallow}
\end{figure}
Assume $\varphi'$ is supported in $(a,b)$ with $0<a<b$ and to simplify $|f(y)|\to\infty$ as $|y|\to\infty$. Then only a finite number of branches of $\hat H(\xi)$ cross the support of $\varphi'$ and hence $\sigma_I[H_I]$ is well defined. \cite[Theorems 2.1 \& 2.2]{bal2024topological} show that $2\pi\sigma_I[H_I]=2$ when $f'(y)$ is bounded; for instance when $f(y)=y$ as shown on the right panel of Fig.\ref{fig:shallow}. However, as soon as $f(y)$ admits jumps, then under additional restrictions on the support $\varphi'$, we have $2\pi \sigma_I  = 2 - \mJ_L(E)+ \mJ_R(E)$ where $E$ is in the support of $\varphi'$ and $\mJ_L(E)$ is the number of positive jumps with half-jump values above $E$ while $\mJ_R(E)$ is the number of negative jumps with half-jump (absolute) values above $E$. When $f(y)=\sgn{y}$ for $|y|\leq L\gg1$ and $\varphi'$ is supported in $(0,1)$, then $2\pi \sigma_I  = 1$ as shown on the left panel of Fig.\ref{fig:shallow}. The derivation of this result is based on a qualitative analysis of branches of absolutely continuous spectrum for which no explicit expression is available.  Unlike the cases of Landau or Dirac operators \cite{combes2005edge,dombrowski2011quantization,quinn2024asymmetric}, no complete stability results exist (or are even expected to exist) for $\sigma_I[H+V]$ with $V$ a compactly supported perturbation; see \cite{bal2022topological,quinn2024approximations} for some theoretical and computational results in this direction.  This surprising result acts as a violation of the bulk-edge correspondence, which states that the number of edge modes should be related to the bulk properties of the two insulators joined at an interface irrespective on the way these two insulators are joined. Jumps in the Coriolis force parameter introduce singularities that modify the flow of the spectral branches.

\medskip

This paper focuses on the spectral flow of the family of Hamiltonians $\xi\mapsto\hat H(\xi)$ with $\xi$ the dual variable to one of the spatial variables of the problem. We may more generally consider families of Hamiltonians $I\ni \mu\mapsto H(\mu)$ with for instance $I$ a bounded interval. See, e.g., \cite{faure2023manifestation}, for a notion of spectral flow as the external parameter $\mu$ varies. This spectral flow is then linked to a winding number in the variables $(\mu,x,\xi)$ with $(x,\xi)$ the phase space variables of the system. This winding number is then computed using the same Fedosov-H\"ormander formula that we will be using in \eqref{eq:FH} below. The methodology applies to the symbol of Hamiltonians rather than the Hamiltonians itself. It is semiclassical in nature and cannot capture violations to the bulk-edge correspondence as identified above in Fig. \ref{fig:shallow}.

\section{Classification of elliptic pseudo-differential operators}
\label{sec:elliptic}

Arguably the most striking feature of topological systems is the asymmetric transport displayed along interfaces separating insulating bulks and characterized by the current interface observable $\sigma_I$. This section reviews a general classification by domain walls, which is then related to $\sigma_I$ in the next section.

%We review settings where $\sigma_I$ is indeed defined and may be computed relatively easily. 
%
\subsection{Weyl quantization and symbol classes.}
\label{sec:Weyl}

We focus on elliptic pseudo-differential operators and first recall relevant notions of pseudo-differential calculus from \cite{dimassi1999spectral,grigis1994microlocal,zworski2022semiclassical}. We first recall our convention for the $d-$dimensional Fourier transform:
\begin{equation} \label{eq:FT}
     \hat f(\xi) = \mF_{x\to \xi} f (\xi) = \dint_{\Rm^d} e^{-ix\cdot\xi} f(x) dx,\ \ 
     f(x) = \mF^{-1}_{\xi \to x} \hat f(x) = \dfrac{1}{(2\pi)^d} \dint_{\Rm^d} e^{ix\cdot\xi} \hat f(\xi) d\xi.
\end{equation}
A differential operator $P(D)$ for $P$ a polynomial may be written as $P(D)=\mF^{-1} P(\xi) \mF$. More generally, any differential operator with not necessarily constant coefficients may be written using their {\em Weyl quantization} and defined by \cite{grigis1994microlocal}
\begin{equation}\label{eq:weylquantization}
  (\ow a) f(x)  = \dint_{\Rm^{2d}} \dfrac{e^{i(x-y) \cdot\xi}}{(2\pi)^d}  a(\frac{x+y}2,\xi) f(y)  d\xi dy.
\end{equation}
The properties of an operator $H=\ow a$ are then described via those of its symbol $a$. 
\\[2mm]
{\bf Class of symbols $S^m$.} We assume $a(x,\xi) \in C^\infty(\Rm^d\times\Rm^d)\otimes {\mathbb M}_n(\Cm)$.  We say that $a\in S^m$ when for each multi-index $\alpha=(\alpha_1,\ldots,\alpha_d)\in \Nm^d$ and $\beta=(\beta_1,\ldots,\beta_d)\in \Nm^d$, we have the following bound on the semi-norm
\begin{equation}\label{eq:Smbound}
  | \partial^\alpha_\xi \partial^\beta_x a |(x,\xi) \leq C_{\alpha,\beta} \aver{\xi}^{m-|\alpha|}
\end{equation}
for some constant $C_{\alpha,\beta}$. Here $|\alpha|=\alpha_1+\ldots+\alpha_d$, and $|\cdot |$ is a norm on ${\mathbb M}_n(\Cm)$.

When $a(x,\xi)$ is the symbol of a partial differential operator of order $m$ with smooth (bounded) coefficients, then indeed $a\in S^m$. The class of operators associated to symbols in $S^m$ by \eqref{eq:weylquantization} is called $\ow S^m$. Any partial differential operator of order $m$ with (uniformly) smooth (bounded) coefficients is in $\ow S^m$. In particular, the Dirac operator with smooth coefficients (and $\eta=0$) is in $S^1$ while the Laplace operator (and the modified Dirac operator involving the term $\eta\Delta$) is in $S^2$.

We define by $S^\infty$ the union of all $S^m$ over $m\in \Nm$ and by $S^{-\infty}$ the intersection of all $S^m$ over $m\in \Nm$. We observe that $S^\infty$ is a graded algebra in the sense that if $a\in S^m$ and $b\in S^n$, then $ab$ and $ba$ are in $S^{m+n}$. We also have an algebra of operators $\ow S^\infty$. The class of symbols $S^m$ may be assigned a Fr\'echet topology by taking the best constants $C_{\alpha,\beta}$ above \cite{dimassi1999spectral,grigis1994microlocal}.
\\[2mm]
{\bf Order functions and composition calculus.} An {\em order function} $\fm:\Rm^{2d}\to [0,\infty)$ is such that there exists constants $C_0$ and $N_0$ such that for all $X=(x,\xi)\in \Rm^{2d}$ and all $Y\in\Rm^{2d}$,  we have 
$\fm(X) \leq C_0 \aver{X-Y}^{N_0} \fm(Y)$,
where $\aver{X}=\sqrt{1+|X|^2}$. Examples of order functions include $\aver{X}^p$ for $p\in \Rm$ and $X_+={\rm max}(X,0)$ (defined element-wise on each coordinate).  When $\fm_1$ and $\fm_2$ are order functions, then so is $\fm_1\fm_2$.  Associated to an order function is a class of symbols $S(\fm)$ defined as all $a(x,\xi) \in C^\infty(\Rm^d\times\Rm^d)\otimes {\mathbb M}_n(\Cm)$ such that for all $\alpha\in \Nm^{2d}$,
\begin{equation}\label{eq:Sofmbound}
  |\partial^\alpha a (X)| \leq C_\alpha \fm(X) \quad \forall X\in \Rm^{2d}.
\end{equation}
Choosing the best constants $C_\alpha$ provides a Fr\'echet topology for $S(\fm)$. We use the convenient notation $S^{-\infty}(\fm)=\cap _{m\in \Zm} S((\fm)^m)$. 
The operators constructed by \eqref{eq:weylquantization} from $a\in S(\fm)$ are denoted by $\ow S(\fm)$. Such operators, as well as those in $\ow S^m$, are called {\em pseudo-differential operators} (PDO). These families include partial differential operators but are much larger and have better composition and invertibility properties. Let $a \in S(\fm_1)$ and $b\in S(\fm_2)$. Then the composed operator $\ow a \ow b\in \ow S(\fm_1\fm_2)$ and there is $c\in S(\fm_1\fm_2)$ such that $\ow a \ow b=\ow c$. There is also an explicit formula for $c$ \cite{dimassi1999spectral}.  A similar result holds for $a\in S^m$ and $b\in S^n$ with $c\in S^{m+n}$.
\\[2mm]
{\bf Ellipticity, Self-adjointness, Resolvents and Functional Calculus.} We recall functional calculus results \cite{bony2013characterization,dimassi1999spectral}.  Assume $a\in S(1)$ or $a\in S^m$ a Hermitian symbol (so that $\ow a$ is self-adjoint) 
and $f\in C^\infty_0(\Rm)$. We wish to know when $f(\ow a)$ is itself a PDO. The Helffer-Sj\"ostrand formula \eqref{eq:HS} shows that $f(\ow a)$ may be written in terms of the resolvent operator $(z-\ow a)^{-1}$. To apply the Beals criterion and show that the resolvent operator is a PDO, we need good invertibility properties. 
For $a\in S^m$ with $m>0$, we assume that $a$ is Hermitian-valued and {\em elliptic}. The latter means that if $a_{\rm min}(x,\xi)$ denotes the smallest singular value of $a(x,\xi)$, then we assume that 
\begin{equation}\label{eq:ellipticity}
|a_{\rm min}(x,\xi)| \geq C \aver{\xi}^m -1
\end{equation}
for some constant $C>0$. In other words, $a$ is invertible as soon as $|\xi|$ is sufficiently large and all its singular values are of order $\aver{\xi}^m$.

Then \cite{bony2013characterization} shows that $\ow a$ is a {\em self-adjoint} operator with domain $H^m(\Rm^d)$, the Sobolev space of order $m$, and that moreover, the {resolvent} $R_z(a)= (z-\ow a)^{-1}$ defined for $z\in \Cm$ with $\Im z\not=0$ is itself a PDO with symbol in $S^{-m}$, i.e., $R_z(a)=\ow r_z$ with $r_z\in S^{-m}$.  We may then use the Helffer-Sj\"ostrand formula \eqref{eq:HS} for $f\in C^\infty_c(\Rm)$ to show that $f(\ow a)\in \ow S^{-\infty}$; see  \cite{bony2013characterization}. 
\\[2mm]
{\bf Trace-class criterion.}
A useful result for us is the following trace-class criterion \cite[Chapter 9]{dimassi1999spectral}. Assume that $\fm\in L^1(\Rm^{2d})$ and that $|\partial^\alpha a(x,\xi)|\leq C_\alpha \fm(x,\xi)$ for all $|\alpha|\leq 2d+1$. Then $\ow a$ is a trace-class operator and 
\begin{equation}\label{eq:symboltrace}
   \Tr\ \ow a = \dfrac{1}{(2\pi)^d} \dint_{\Rm^{2d}}  \trr \,a(x,\xi) dx d\xi,\qquad \|\ow a\|_1 \leq C \max_{|\alpha|\leq 2d+1} C_\alpha \  \|\fm\|_{L^1(\Rm^{2d})}.
\end{equation}
In other words, all symbols in $S(\fm)$ with $\fm$ integrable generate trace-class operators. 

\paragraph*{Example of Dirac operator in two space dimensions.}
Consider the Dirac operator $H=D\cdot\sigma + m(y)\sigma_3 + V(x,y)$ for $m$ and $V$ smooth and bounded. Then $H=\ow a$ with $a=\xi\cdot\sigma + m(y)\sigma_3 + V(x,y)$ as $2\times2$ Hermitian matrices and $H\in \ow S^1$ while for $f\in C^\infty_c(\Rm)$, we have  $f(H)\in \ow S^{-\infty}$ since $H$ is clearly elliptic. Note that this does not require any specific form of a domain wall for $m(y)$. 

The above result does not imply any trace-class property. If we use the order function $\fm(X)=\aver{\xi}$, then we deduce that $f(H)\in S^{-\infty}(\aver{\xi})$, i.e., an operator with symbol that decays faster than algebraically as $|\xi|\to\infty$. However, we do not have any information regarding decay of the symbol in the spatial variables. The role of $m(y)$ is to provide such a decay in the spatial variable $y$ while commutators of the form $[H,P]$ for $P=P(x)$ provide decay in the spatial variable $x$.

\subsection{Classification by interface current observable.}
\label{sec:classIC}
We describe classes of operators $H=\ow a$ for which the interface current observable $\sigma_I$ and operators of the form $P(x)U(H)P(x)$ are defined and quantized \cite{bal2022topological,quinn2024approximations}.

We focus on the two-dimensional setting and denote spatial variables by $(x,y)$ and dual variables by $(\xi,\zeta)$. Let $H=\ow a$ be an elliptic operator for $a\in S^m$. Let $P(x)\in C^\infty \fS$ be a smooth switch function. We present sufficient conditions on $H$ ensuring that $[\psi(H),P]\phi(H)$ is a trace-class operator using \eqref{eq:symboltrace}. Since $H$ is elliptic, we have by spectral calculus that $\phi(H)\in S^{-\infty}$ with a symbol of thus of order $\aver{\xi,\zeta}^{-\infty}$ but no decay in $(x,y)$. We therefore need an assumption ensuring that energies in the support of $\phi(H)$ are in the bulk band gap both when $y>L\gg1$ and $y<-L\ll=-1$. This ensures that $\phi(H)\in S(\aver{y,\xi,\zeta}^{-\infty})$ has symbol decaying rapidly in all variables $(y,\xi,\zeta)$.  It will then remain to show that $[\psi(H),P]$ has symbol in $S(\aver{x}^{-\infty}\aver{y,\xi,\zeta}^{n})$ for some $n\in\Zm$, which is a straightforward consequence of the compact support of $P'(x)$. Composition calculus then ensures that $[\psi(H),P]\phi(H)$ has symbol in $S(\aver{x,y\xi,\zeta}^{-\infty})$ and is trace-class by \eqref{eq:symboltrace}.
\\[2mm]
{\bf Operators with domain walls.} 
Decay in $y$ is obtained by means of a {\em domain wall} generalizing the role of $m(y)$ transitioning from $m(y)\geq m_0>0$ for $y\geq L\ll1$ to $m(y)\leq -m_0$ for $y\leq -L$. As in the definition of $\sigma_I$, we will then need to choose $\phi(H)$ with $\phi$ supported inside the bulk gap generated by the domain wall.  Here is a sufficient hypothesis:
\\[1mm]
%%%%%%%%%%%%%%%
{\bf [H1].} $\ $ Let $H=\ow \sigma \in \ow S^m$ with $m>0$ be an {\em elliptic} differential operator, i.e., \eqref{eq:ellipticity} holds. We assume the existence of $H_\ns=\ow \sigma_\ns \in \ow S^m$ elliptic and with symbols $\sigma_\ns(\xi,\zeta)$ independent of $(x,y)$. Moreover, we assume the existence of $L>0$ such that $\sigma(x,y,\xi,\zeta)=\sigma_N(\xi,\zeta)$ when $y>L$ and $\sigma(x,y,\xi,\zeta)=\sigma_S(\xi,\zeta)$ when $y<-L$. Finally, we assume the existence of a {\em bulk} spectral gap $[E_1,E_2]$ in the sense that $E-\sigma_\ns$ is invertible for each $E\in [E_1,E_2]$. 
%%%%%%%%%%%%%%%%%%
\\[1mm]
Here, $\ns$ stands for North/South. We then have the following first result:
\begin{proposition} \label{prop:traces}
  Assume $H$ satisfies {\bf [H1]}. Let $\phi \in C^\infty_c(\Rm)$ with compact support in $(E_1,E_2)$ and $\psi_{j}$ be either a polynomial or a bounded function in $C^\infty(\Rm)$ for $j=1,2$. Finally, let $P\in C^\infty \fS$ be a smooth switch function. Then 
     $\phi(H) \in \ow S(\aver{y,\xi,\zeta}^{-\infty}) $  and $
     \psi_1(H) [\psi_2(H),P] \phi(H)  \in \ow S(\aver{x,y,\xi,\zeta}^{-\infty})$. Moreover, $[\phi(H),P]$ and $\psi_1(H) [ \psi_2(H),P] \phi(H)$ are trace-class operators with traces given by \eqref{eq:symboltrace}. In particular, $\Tr\ [ \phi(H),P]=0$.
\end{proposition}
This proposition detailed in \cite{quinn2024approximations} is a consequence of the ellipticity assumption and the resulting functional calculus, directly showing for instance that $[\psi_2(H),P]\in \ow S(\aver{\xi,\zeta}^s\aver{x}^{-\infty})$ for $s\in\Rm$, as well as the Helffer-Sj\"ostrand formula with the following trick. 

For $t\geq0$, define $\phi_t(H)=(1+H^2)^t \phi(H)$. By assumption {\bf [H1]}, we observe that $\phi(H_\ns)=\phi_t(H_\ns)=0$ since $\phi$ is supported inside the bulk band gaps of $H_\ns$. Choosing $L$ large enough that $P(y)$ is constant on each connected component of $|y|>L$, we have
\[
  \phi(H)=  P(y) (I+H^2)^{-t} (\phi_t(H)-\phi_t(H_N)) + (1-P(y)) (I+H^2)^{-t}(\phi_t(H)-\phi_t(H_S)).
\]
The first term
\[
   T_N= -\frac{1}\pi \dint_Z \bar\partial \tilde\phi (z)  P(y) (I+H^2)^{-t} (z-H)^{-1} (H-H_N) (z-H_N)^{-1} d^2z
\]
then belongs to $\ow S(\aver{\xi,\zeta}^{-2mt} \aver{y}^{-\infty})$ for every $t>0$ implying that $\phi(H)\in \ow S(\aver{y,\xi,\zeta}^{-\infty})$. The rest of the derivation follows from composition calculus.

Let $H$ satisfy {\bf [H1]}.  Define the switch function $\varphi(E)\in C^\infty(\Rm)$ in $\fS[0,1;E_1,E_2]$. Define $U(h)=e^{i 2\pi \varphi(h)}$ so that $U=I+W$ with $W$ compactly supported in $(E_1,E_2)$. 
We apply the above proposition to obtain that $i[H,P]\varphi'(H)$ and $[U(H),P]U^*(H)$ are both trace-class. This implies that $\sigma_I$ is indeed defined. This also seems to indicate that $P(x)U(H)P(x)$ is a Fredholm operator. The only remaining obstruction is that $P(x)$ is assumed to be smooth in the above proposition while $P(x)$ needs to be a projector to construct an odd Fredholm module. This is a minor technical difficulty, and we have in fact the following result \cite{quinn2024approximations}:
\begin{theorem}\label{thm:tracePDO}
  Let $H$ be an operator satisfying {\bf [H1]}. Let $P_1(x)$ be a smooth switch function in $\fS$ and $P(x)$ be a projector in $\fS$. Let $\varphi$ be a smooth non-decreasing switch function in $\fS[0,1;E_1,E_2]$. Let $U(h)=e^{i 2\pi \varphi(h)}=I+W(h)$. Then $P(x) U(H) P(x)_{|\Ran P}$ is a Fredholm operator and we have:
  \begin{equation} \label{eq:equalityedge}
     2\pi \sigma_I = \Tr 2\pi i[H,P_1]\varphi'(H) = \Tr [U(H),P_1] U^*(H) = \Tr [U(H),P] U^*(H)  = \ipup.
  \end{equation}
\end{theorem} 
Pseudo-differential calculus cannot be used for the non-smooth function $P(x)$. However, $P(x)=P_1(x)+P(x)-P_1(x)$ and the third equality in \eqref{eq:equalityedge} results from introducing an operator of the form $(P-P_1)T$ which is trace-class when  $T$ is trace-class since $P-P_1$ is bounded, and to show that the trace of $[U,P-P_1]U^*$ vanishes.  Also used is the non-commutative differentiation of the commutator $[AB,P]=A[B,P]+[A,P]B$ implying for $\phi\in C^\infty_c(\Rm)$ and cyclicity of the trace that 
\[
  \Tr [H^n,P] \phi(H) =  \Tr [H,P] nH^{n-1} \phi(H).
\]
Approximating $W(H)$ by polynomials leads to the second equality, which has been observed in many other contexts. The first equality in \eqref{eq:equalityedge}  is the definition of $\sigma_I$ while the last equality is the Fedosov formula.
\\[2mm]
{\bf Stability of the edge invariant.} The above implies that $2\pi \sigma_I\in \Zm$, which is therefore immune to a large class of perturbations. We mention that $\pup$ is Fredholm as soon as $P-UPU^*$ is compact and therefore enjoys stronger stability than $\sigma_I$. 
However, when $i[H,P]\varphi'(H)$ is trace-class, both indices are defined and agree. We may therefore use the stability of the index of $PUP$ to obtain that of the edge invariant $2\pi\sigma_I$.  

Let $a_0$ and $a_1$ be two elliptic symbols in $S^m$. For $t\in [0,1]$, we define $a_t=(1-t)a_0+ta_1=a_0+t(a_1-a_0)$ and $H_t=\ow a_t$.  We thus have $H_t-H_s = \ow (a_t-a_s)=(t-s) \ow(a_1-a_0)$. 
\begin{proposition}[Stability of $\sigma_I$] 
 Let $H_t$ be defined as above for $t\in [0,1]$. Assume that {\bf [H1]} is satisfied for each $H_t$, $t\in [0,1]$ with $(E_1,E_2)$ chosen uniformly in $t$. Assume $\varphi$ in Theorem \ref{thm:tracePDO} chosen with support in $[E_1,E_2]$. Then, $2\pi \sigma_I[H_t]=\ind PU(H_t)P$ is {\em independent} of  $t\in [0,1]$.
\end{proposition}
This is a direct consequence of the Helffer Sj\"ostrand functional calculus formula and of the Calder\'on Vaillancourt theorem stating that operators with symbols in $S(1)$ are bounded on $L^2$ spaces \cite{dimassi1999spectral}.

As an application of the preceding result, we have the following corollary. Let $a$ be an elliptic symbol in $S^m$ such that {\bf [H1]} holds. Then we have \cite{quinn2024approximations}
\begin{corollary}\label{cor:deformsigmaI}
  Let $a\in S^m$ be elliptic as above and $b\in S^{m-1}$ have compact support in $(x,y)$. Let $a_t=a+tb$. Then {\bf [H1]} holds for $H_t=\ow(a+tb)$ and $\sigma_I[H_1]=\sigma_I[H_0]$. 
  
  Let $a$ be as above and $a_0(x,y,\xi,\zeta)=a(0,y,\xi,\zeta)$. Then  $\sigma_I[H_1]=\sigma_I[H_0]$ where $H_0=\ow a_0$ satisfies {\bf [H1]}.
  
  Let $0<h_1\leq1$ and $0< h_2\leq 1$ and define $a_h(x,y,\xi,\zeta)=a(x,y,h_1\xi,h_2\zeta)$ while $H_h=\ow a_h$. Then for every multi-index $h$ as above, $H_h$ satisfies {\bf [H1]} and $\sigma_I[H_h]$ is independent of $h$.
\end{corollary}

(i) For a family of operators $H_t=H+tV$ with $t\in [0,1]$ with $V$ a perturbation that does not modify the domain walls for $|y|\geq L$, we directly obtain that $\sigma_I[H]=\sigma_I[H+V]$. 

(ii) The second result shows that variations of the coefficients in $x$ are irrelevant so long as the operator with $x-$dependent coefficients belongs to the class $S^m$. Evaluating the symbol at any point, say $x=0$, yields the same classification for inhomogeneous and homogeneous operators. 

(iii) The third result states that the invariant is independent of a rescaling of the elliptic operator of the form $\xi \to h\xi$. This implies that the computation of the index may be performed in the semi-classical regime $h\ll1$. This invariance is at the core of the derivation of the bulk-edge correspondence carried out in the next section. 

We note that the last result is simply false for the Landau operator: rescaling by $h$ rescales the (topologically non-trivial) Landau levels as well so that the interface current strongly depends on $h$. This shows that the ellipticity condition in {\bf [H1]} is crucial to obtain the above invariance. 
\\[2mm]
{\bf Stability for non-elliptic problems.}
While these problems are not elliptic, we already mentioned that similar stability results applied for Landau and magnetic Dirac operators; see \cite{combes2005edge,dombrowski2011quantization} and \cite{quinn2024asymmetric}. The reason is that a Weyl symbol of the form $\xi_1^2+ (\xi_2-Bx)^2$ still displays reasonable positivity properties when $x$ is restricted to the compact support of a localized perturbation. A very similar calculus to the one presented above then applies with little modification. 

The shallow water problem introduced in \eqref{eq:H} is more complicated to analyze. The presence of a flat band when $f$ is constant still generates essential spectrum at $0$ when $f=f(y)$ varies. Since the Coriolis force parameter ceases to be an insulator at that same value $f=0$, the topological change occurs at a frequency level where essential spectrum is present. The problem is then very far from being elliptic and the above stability results are not known to hold. The only available stability result we are aware of has been proved as an application of \cite[Proposition 4.3]{bal2022topological} and shows that for $H=D_x\gamma_1+D_y\gamma_4 - f(\eps y) \gamma_7 + V(\eps x , \eps y)$ for $V=V_{ij}$ a Hermitian $3\times3$ matrix-valued compact perturbation, then $\sigma_I$ is defined and stable against perturbations provided $\eps$ is sufficiently small (as an application of a semiclassical G{\aa}rding inequality) and coefficients $V_{22}$, $V_{33}$, $\Im V_{12}$, $\Im V_{13}$ and $\Re V_{23}$ are sufficiently small.  The other coefficients in $V$ may be arbitrarily large so long as $\eps$ is sufficiently small. Such partial stability was confirmed by numerical simulations in \cite{quinn2024approximations}.

\subsection{Classification by domain walls}
\label{sec:classDW}
This section proposes a simple and general classification of possibly non-Hermitian, {\em elliptic} (pseudo) differential operators.
\\[2mm]
{\bf One dimensional example.} Consider $H=D$. We saw in section \ref{sec:IH} that for $\varphi\in C^\infty_c(\Rm)$ with $\int \varphi'=1$, then $P(x)e^{i2\pi \varphi(H)}P(x)$ was a Fredholm operator with index equal to $1$ by spectral flow.  

We introduce another classification by domain wall. In one dimension, the coordinate $x$ acts as a domain wall and we may introduce the following operator
\[
  F = H-ix = -i \fa = \ow{a}, \qquad \fa=\partial_x+x,\quad a(x,\xi)=\xi-ix.
\]
In $\fa$ we recognize an annihilation operator, with $\fa^*=-\partial_x+x$ the corresponding creation operator. $\fa$ is a Fredholm operator from $H^1(\Rm)$ to $L^2(\Rm)$ with index equal to $1$. This index, or equivalently that of $F$, may be written explicitly in terms of the symbol $a(x,\xi)$ as 
\[
  \ind F = \frac{1}{2\pi i} \dint_{\gamma} a^{-1}da =1
\]
where $\gamma$ is an arbitrary sufficiently smooth curve winding around the origin $(0,0)$ once and the one-form $da=\partial_x a dx + \partial_\xi a d\xi$. Note that $a^{-1}$ is defined on $\Rm^2\backslash (0,0)$ and that $d(a^{-1}da)=0$ so that by the Stokes theorem, the above integral is independent of the choice of $\gamma$ with unit winding number. The index is captured by the non-trivial $L^2-$kernel $e^{-\frac12 x^2}\Cm$ of $\fa$ while the kernel of $\fa^*$ is trivial. Thus a non-trivial index is characterized by a bound zero mode $e^{-\frac12 x^2}$.
\\[2mm]
{\bf Two-dimensional example.} Consider the mass-less Dirac operator $H_0=D\cdot\sigma$. We saw a bulk classification for the operator $H_0$ perturbed by $(m+\eta\Delta)\sigma_3$. We also saw a classification for the interface Hamiltonian $H_1=H_0+m(y)\sigma_3$, where $m(y)$ acts as a domain wall in the direction $y$. The effect is to localize a range of energies to the vicinity of the $x$-axis and $\sigma_I[H_1]$ then tests the topology of $H_1$ by computing its asymmetric current.  As we did in one dimension, we can further confine matters in the vicinity of $x=0$ by introducing 
\[
  F= H-ix = H_0 + m(y) \sigma_3 -ix = \ow {a} ,\qquad a(x,y,\xi,\zeta) = \xi\sigma_1+\zeta\sigma_2+m(y)\sigma_3 -ix.
\]
We may interpret the above expression as introducing two confining domain walls to an unconfined operator $H_0$ or adding one domain wall to further confine an operator $H_1$ already confined in one variable. The operator $F$ is then Fredholm and its index admits the following simple expression:
\begin{equation}\label{eq:FH2d}
   \ind F = \frac{1}{24\pi^2} \dint_{ \Sm^3_R }  \trr (a^{-1}da)^{\wedge\,3}.
\end{equation}
Here, $\Sm^3_R$ is the sphere of radius $R$ in phase space $(x,y,\xi,\zeta)\in \Rm^4$ (oriented with $d\xi \wedge dx \wedge d\zeta \wedge dy>0$) and $R$ is chosen large enough that $a^{-1}$ is defined for $|(x,y,\xi,\zeta)|\geq R$. We again verify that $ d(\trr (a^{-1}da)^{\wedge\ 3})=0$ as a  volume form to show by the Stokes theorem  that the above integral is independent of the contour $\Sm^3_R$ so long as it encircles all phase-space points where $a^{-1}$ is not defined. 

The formula \eqref{eq:FH2d} will be referred to as a Fedosov-H\"ormander formula \cite[Chapter 19]{H-III-SP-94}. It may be seen as an Atiyah-Singer index result for Fredholm operators acting of functions of Euclidean spaces. The main point is that as complicated as computing this integral may be in practice, it is conceptually significantly simpler than computing the index of $PUP$. Only the symbol of $F$, easily related to the symbol of $H_j$, appears in \eqref{eq:FH2d}. In fact, the right-hand side in \eqref{eq:FH2d} defines a classification of symbols, whether it corresponds to the index of a Fredholm operator or not.
\\[2mm]
{\bf General classification by domain walls.} Let $d\geq1$ be spatial dimension and let $H_k=\ow{a_k}$ be an operator with elliptic symbol of order $m$ heuristically confined in the first $k$ spatial dimensions for $0\leq k\leq d-1$ and acting on $\Cm^{n_k}$-valued functions of the Euclidean space $\Rm^d$. For instance, we have $d=2$, $k=1$, and $n_1=2$ for $H_1=D\cdot\sigma+m(y)\sigma_3$. 

When the {\em effective} dimension $d-k$ is even, we assume that $H_k$ satisfies the following {\em chiral} symmetry in an appropriate basis:
\begin{equation}\label{eq:chsym}
  H_k =  \begin{pmatrix} 0 & F_k \\ F_k^* & 0 \end{pmatrix} = \sigma_- \otimes F_k^* + \sigma_+\otimes F_k,\qquad \sigma_\pm=\frac12(\sigma_1\pm i\sigma_2).
\end{equation}
No additional symmetry is assumed when $d-k$ is odd.
We next introduce the {\em domain walls} 
\begin{equation}\label{eq:dwall}
m_{j}(x):=\aver{x_{j}}^{m-1} x_{j}, \quad 1\leq j\leq d.
\end{equation}
They are constructed to have the same homogeneity of order $m$ as the Hamiltonian $H_k$. 

Assume $d-k$ even with $k\leq d=2$. We then define the new dimension $n_{k+1}=n_k$  and the augmented Hamiltonian acting on $\Cm^{n_{k+1}}$-valued functions
\begin{equation}\label{eq:augeven}
  H_{k+1} := H_k + m_{k+1} \sigma_3 \otimes I.
\end{equation}
This implements a domain wall in the variable $x_{k+1}$.

Assume now $d-k$ odd.  We define the new dimension $n_{k+1}=2n_k$ and 
\begin{equation}\label{eq:augodd}
 H_{k+1} := \sigma_1\otimes H_k+m_{k+1} \sigma_2 \otimes I  = \sigma_-\otimes F^*_{k+1} + \sigma_+\otimes F_{k+1},\quad
  F_{k+1} = H_{k} -i m_{k+1}.
\end{equation}
The operator $H_{k+1}$ now acting on $\Cm^{n_{k+1}}$-valued functions satisfies a chiral symmetry of the form \eqref{eq:chsym}, as requested since $d+k+1$ is now even.

We denote by $a_{k+1}$ the symbol of $H_{k+1}=\ow a_{k+1}$ and observe that $a_{k+1}=a_k+m_{k+1}\sigma_3\otimes I$ when $d+k$ is even and $a_{k+1}= \sigma_1\otimes a_k+m_{k+1}\sigma_2\otimes I$ when $d+k$ is odd. The procedure is iterated until $H_d$ is constructed. Note that $a_{k+2}(X)\in \Mm(2n_k)$ with dimension of the space on which the matrices act doubling every time $k$ is raised to $k+2$. 
The intermediate Hamiltonians for $0 < l \leq d-k$ all have the form
\begin{equation}\label{eq:Hinterm}
  H_{k+l} =  \sigma_1^{\otimes p} \otimes H_k + \mu\cdot\gamma\otimes I_{n_k} ,\qquad \mu=(m_{k+1},\ldots, m_{k+l}),\qquad \gamma=(\gamma_{1},\ldots,\gamma_{l} ),
\end{equation}
for some integer $p=p(l,k)$ and matrices $\gamma_j$ such that $\{\gamma_i,\gamma_j\}:=\gamma_i\gamma_j+\gamma_j\gamma_i=2\delta_{ij}I$.

Since $2d$ is even, $H_{d}=\sigma_-\otimes F^* + \sigma_+ \otimes F$ for an operator $F=F_{d}=H_{d-1}-im_d =: \ow a$, or equivalently $a=a_{d-1}-im_d$. The proposed topological classification of $H_k$ is then obtained as 
\begin{equation}\label{eq:FH}
   \ind H_k \equiv \ind F  = \rF_d[a] :=  - \dfrac{(d-1)!}{(2\pi i)^d (2d-1)!}\dint_{\Sm_R^{2d-1}} {\rm tr}\,(a^{-1} da)^{\wedge(2d-1)} .
\end{equation}
$R$ is a sufficiently large constant so that $a$ is invertible outside of the ball of radius $R$ and the orientation of $\Rm^{2d}$ and that induced on $\Sm_R^{2d-1}$ is chosen so that $d\xi_1 \wedge dx_1 \wedge \ldots \wedge d\xi_d \wedge dx_d >0$.  The right-hand side is the Fedosov-H\"ormander formula. Here are some remarks:
\\[2mm]
(i) The `Index' of $H_k$ and of $F$ is formally defined as $\rF_d[a]$ provided the above integral is defined. For elliptic operators $H_k$, $F$ is indeed a Fredholm operator with index given by the above formula. 
\\[1mm]
(ii) For operators $H_1$ in two dimensions satisfying {\bf [H1]}, then $\rF_2[a]$ may be well-defined even though $F$ may not be Fredholm.
\\[1mm]
(iii) The formula is independent of the contour of integration $\Sm_R^{2d-1}$ so long as it encircles the {\em topological charge} of $H_k$ where the matrix inverse $a^{-1}$ is not defined. This is because $d \trr (a^{-1} da)^{\wedge(2d-1)}=0$ so that the Stokes theorem may be used.
\\[1mm]
(iv) The classification applies to so-called Higher-Order Topological Insulators (HOTI), involving confinement in more than one spatial dimension. The above classification does not differentiate between HOTI and more standard topological insulators.
\\[1mm]
(v) In dimension $d=2$, the formula applies equally to operators $H_1$ that are not Hermitian. For instance $H_1=D\cdot\sigma+m(y)\sigma_3+V(x,y)$ with $V(x,y)$ an arbitrary (non-Hermitian) $2\times2$ matrix-valued compactly supported function follows a classification independent of the perturbation $V$. This is in contrast to classifications based on $\sigma_I[H]$ or $PUP$, where spectral calculus is required.
\\[1mm]
(vi) Self adjoint operators satisfying the chiral symmetry \eqref{eq:chsym} belong to the complex symmetry class AIII whereas self-adjoint operators without additional symmetry belong to class A. The chiral symmetry is necessary to obtain a nontrivial classification when $d-k$ is even as continuous deformations to a trivial operator along a path of non-chiral operators should exist.
\\[1mm]
(vii) The above classification may feel arbitrary. We show in the next section that the index of $F$ in fact equals $2\pi\sigma_I[H_{d-1}]$. This generalization of the bulk-edge correspondence, justifies the introduction of this simple invariant as a straightforward way to obtain the computationally significantly more challenging invariant $2\pi\sigma_I[H_{d-1}]$. 
\medskip

As an application, consider in $d=3$ the Weyl operator $H_0=D_1\sigma_1+D_2\sigma_2+D_3\sigma_3$. No perturbation or mass term may open a spectral gap for this operator acting on spinors in $\Cm^2$. Define the operator $H_1=\sigma_1\otimes H_0+ \sigma_2\otimes I_2 x_1$ with a domain wall $m_1(x_1)=x_1$ in the first direction but now acting on spinors in $\Cm^4$. This operator has effective dimension $2$ and is in the same class as $D_2\sigma_2+D_3\sigma_3$. Its topology is then characterized by asymmetric transport in the third dimension after a second domain wall in the $x_2$ direction is introduced: $H_2=H_1+\sigma_3\otimes I_2 x_2$.  A final confinement is achieved by introducing 
\begin{equation}\label{eq:F3d}
   F=H_2-ix_3= H_1+\sigma_3\otimes I_2 x_2 - i x_3 =  \sigma_1 \otimes H_0 + \sigma_2\otimes I_2 x_1+\sigma_3\otimes I_2 x_2 - i x_3.
\end{equation}
The above construction generalizes to arbitrary dimension. Here, $F$ is a Fredholm operator on $L^2(\Rm^3)\otimes \Cm^4$ with $\ind F=-1$. This sign reflects the choice of orientation of the Clifford matrices used to construct the operators $H_j$ as well as the orientation of the domain walls. The kernel of $F^*$ has for eigenfunction the spinor $e^{-\frac12 |x|^2}(1,-1,-1,-1)^t$. The {\em topological charge / index} of $H_0$, of $H_1$, and of $H_2$ above is {\em defined} as $\rF_3[a]$ in \eqref{eq:FH}.

The operator $H_2$ is the prototypical (continuous) example of a HOTI. Confinement in both directions $x_1$ and $x_2$ leads to an asymmetric transport in the remaining variables $x_3$, i.e., along a hinge rather than an interface.
\\[2mm]
{\bf Remarks on Fedosov-H\"ormander formula.} Formulas of the form \eqref{eq:FH} have a long history in the analysis of topological properties of physical systems. In topological insulators, the `topology' is found in the dual Fourier variable representation. The formula \eqref{eq:FH} allows one to test such a topology by domain walls as we showed above. In the analysis of Yang Mills problems, the non-trivial topology of a potential described in the physical variables is the main object of interest. The dual Fourier variables then serve as a test of such a nontrivial topology. This gives rise to the Callias formula; see \cite{bott1978some,callias1978axial,gesztesy2016callias} for details of the theory and the relationship to the formula \eqref{eq:FH}. 

The formula \eqref{eq:FH} may be interpreted as a generalized winding number and appears in many instances in the physics literature. Related to the above classification are the classifications based on imaginary-frequency Green's functions described in \cite{volovik2009universe} and analyzed in detail in \cite{essin2011bulk,gurarie2011single}. We note that the notion of Green's functions also allows one to classify systems in the presence of interactions \cite{gurarie2011single}. All results presented in this paper apply solely to single particle Hamiltonians.
\\[2mm]
{\bf Justification for elliptic operators.} We consider the setting of \cite{bal2023topological}.  Let $x\in\Rm^d$ be spatial coordinates with $\xi\in\Rm^d$ the dual (Fourier) variable and $X=(x,\xi)\in\Rm^{2d}$ the phase space variable.  For $\aver{\alpha}=\sqrt{1+|\alpha|^2}$, we define the weights $w_k(X)=\aver{x_k',\xi}$.  Let $m$ be the operator order of interest. Then $S^m_k=S^m_k[n_k]$ denotes the class of symbols $a_k$ such that for each $d-$dimensional multi-indices $\alpha$ and $\beta$, there is a constant $C_{\alpha,\beta}$ such that for each component $b$ of $a_k\in \Mm(n_k)$, we have
\begin{equation}\label{eq:Sk}
   \aver{x}^{|\alpha|} \aver{\xi}^{|\beta|} |\partial^\alpha_x\partial^\beta_\xi b(X)| \leq C_{\alpha,\beta} w_k^m(X),\qquad \forall X\in\Rm^{2d}.
\end{equation}
Here, $w_k^m=(w_k)^m$. We define the space of symbols $\tilde S^m$ as $S^m_d$ but acting on vectors of lower dimension $n_{d-1}$ instead of $n_d$.  We say that an operator $H_k=\ow {a_k}$ is {\em elliptic} when
\begin{equation}\label{eq:ellip}
  | \det a_k(X) | ^{\frac 1{n_k}} \geq C_1w_k^m(X) - C_2,\quad \forall X\in\Rm^{2d}.
\end{equation}
This means that each singular value of $a_k$ is bounded by and grows at least as $w_k^m(X)$ as $X\to\infty$.  This homogeneity in all phase space variables ensures that we can apply the theory developed in \cite[Chapter 19]{H-III-SP-94}. Let $\tilde H^0= L^2(\Rm^d)\otimes \Mm(n_{d-1})$ and $\tilde H^m \subset L^2(\Rm^d)\otimes \Mm(n_{d-1})$ be a Hilbert space that depends on $H_k$ and whose definition is given in \cite[(A.5)]{bal2023topological}. Then we have the following result, essentially a corollary of \cite[Theorem 19.3.1']{H-III-SP-94}:
\begin{theorem} [{\cite[Theorem 2.3]{bal2023topological}}] \label{thm:FH}
  Let $H_k$ and $F$ be constructed as above. Then $F$ is a Fredholm operator from $\tilde H^m$ to $\tilde H^0$. Moreover, its index is given by the Fedosov-H\"ormander formula \eqref{eq:FH}.
\end{theorem}
For Dirac operator, the above construction requires the domain walls to be linear in the spatial variables $x_j$ as indicated in the construction of $F$ in \eqref{eq:F3d}.  \cite[Theorem 19.3.1']{H-III-SP-94} comes from the index theorem \cite[Theorem 19.3.1]{H-III-SP-94} proved for symbols satisfying (for $k=d$)
\begin{equation}\label{eq:homSk}
   \aver{X}^{|\alpha|}|\partial^\alpha_X b(X)| \leq C_{\alpha} w_k^m(X),\qquad \forall X\in\Rm^{2d}
\end{equation}
and the approximation described in \cite[Lemma 19.3.3]{H-III-SP-94} of symbols satisfying \eqref{eq:Sk} (with $k=d$) by its subclass satisfying \eqref{eq:homSk}. This approximation of the larger class of symbols \eqref{eq:Sk} by fully isotropic symbols (in the phase space variables) in \eqref{eq:homSk} will also prove useful in the derivation of the bulk-edge correspondence.  The proof that $F$ is Fredholm with index given by \eqref{eq:FH} has been proved for the larger class of so-called slowly varying symbols in \cite{schrohe1992spectral}.
\subsection{Bulk-difference invariant (BDI)}
\label{sec:BDI}
Consider the Fedosov-H\"ormander formula \eqref{eq:FH2d} in dimension $d=2$. Assume that $F=H_1-ix$ where $H_1$ satisfies hypothesis {\bf [H1]}, and assume that $0\in(E_1,E_2)$. (When $0\not\in(E_1,E_2)$, find $\alpha\in (E_1,E_2)$ and apply the result to $H-\alpha$.) The symbol $a(x,y,\xi,\zeta)$ of $F$ is then equal to $\sigma_\ns(\xi,\zeta)-ix$ for $\pm y\geq L$. In other words, it depends only on the symbols of bulk Hamiltonians for $\pm y\geq L$. It turns out that the contour in \eqref{eq:FH2d} may also be continuously deformed to an integral over the hyperplanes $y=\pm L$. Indeed, as indicated below \eqref{eq:FH2d}, we have
$\rF_2[a] = \frac{1}{24\pi^2} \int_{ \mC_{L,R} }  \trr (a^{-1}da)^{\wedge\,3}$,
where $\mC_{L,R}$ is the boundary of the cylinder defined $|y|\leq L$ and $|(x,\xi,\zeta)|\leq R$. Assume $m=1$ for concreteness in {\bf [H1]}. Then $|a^{-1}|$ is bounded by $R^{-1}$ on $|(x,\xi,\zeta)|= R$ so that $|\trr (a^{-1}da)^{\wedge\,3}|$ is bounded by $R^{-3}$ for a surface integral on $|y|\leq L$ and $|(x,\xi,\zeta)|=R$ equal to $LR^2$. Thus, in the limit $R\to\infty$ at $L$ fixed, 
\begin{equation}\label{eq:indFL}
   \rF_2[a] = \frac{1}{24\pi^2} \dint_{ \mC_{L} }  \trr (a^{-1}da)^{\wedge\,3},
\end{equation}
where $\mC_{L} =\{y=L\} \cup \{y=-L\}$. As advertised, $\ind F$ only depends on $\sigma_\ns(\xi,\zeta)-ix$.

The integrals over $\mC_{L}^N =\{y=L\}$ and $\mC_{L}^S =\{y=-L\}$ separately are not necessarily integer-valued. For the Dirac operator $H_1=D\cdot\sigma+y\sigma_3$, these integrals are in fact of the form $\pm\frac12$. However, the above difference of integrals (the orientations of $\trr (a^{-1}da)^{\wedge\,3}$ are opposite on $\mC_{L}^N =\{y=L\}$ and $\mC_{L}^S =\{y=-L\}$) is indeed integer-valued as the index of a Fredholm operator $F$ or as a generalized winding number of the map $a$.
We refer to $\rF_2[a]$ as a {\em bulk-difference} invariant (BDI) rather than a difference of ill-defined bulk invariants. 

\begin{figure}[ht!]
 \begin{center}     \includegraphics[width=7.5cm]{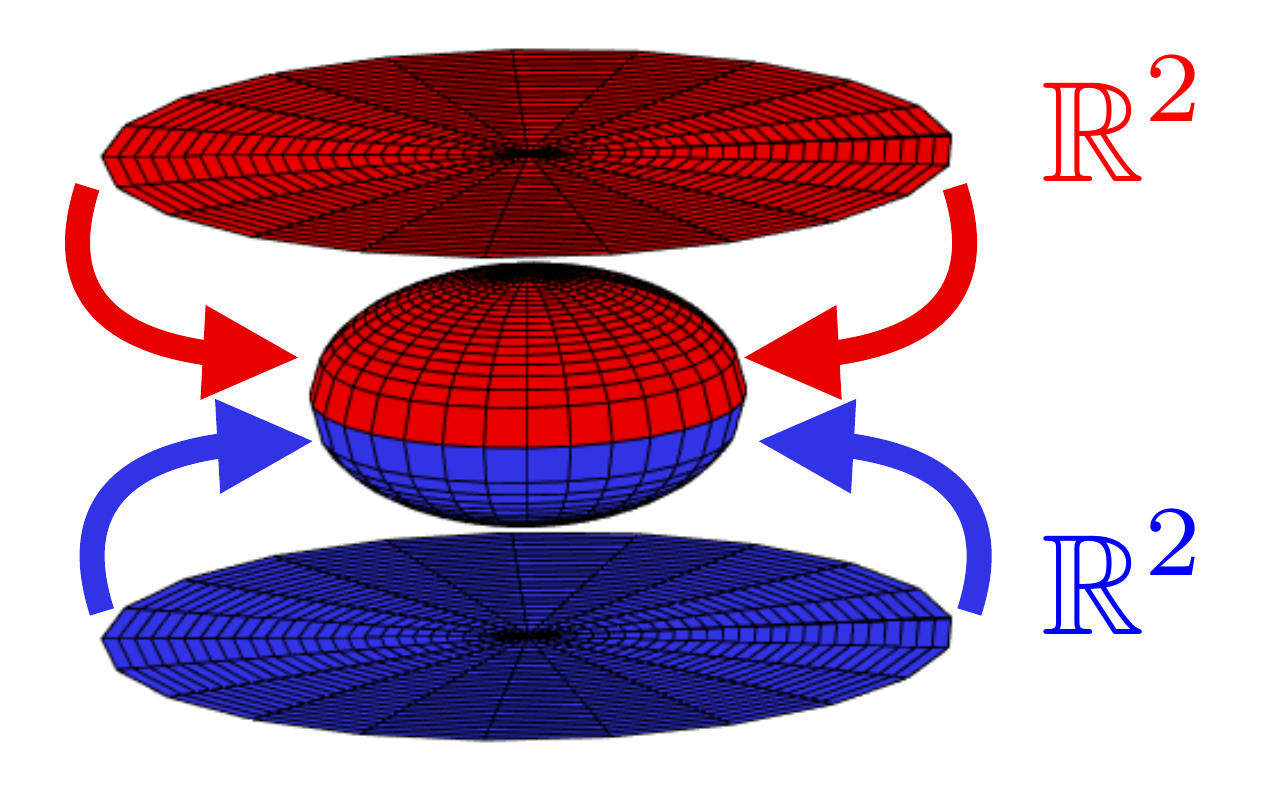} \end{center}
 \caption{Radial compactification of two Euclidean planes onto the unit sphere.}
 \label{fig:sphere}
 \end{figure}

\paragraph*{Bulk-difference invariant, winding number, and Chern number \cite{bal2022topological}.} Let $H^\ns=\mF^{-1}\hat H^\ns(k)\mF$ be two two-dimensional bulk Hamiltonians with $k=(\xi,\zeta)$ and $\hat H^\ns(k)$ given for instance by $\sigma_\ns(\xi,\zeta)$ above. Consider the diagonalization
\begin{equation}\label{eq:Hpm}
   \hat H^\ns(k) = \dsum_{i=1}^n h^\ns_i(k) \Pi^\ns_i(k)
\end{equation}
with $n$ the dimension of the square matrices $\hat H^\ns(k)$ and $\Pi^\ns_i(k)$ rank-one projectors. Let $J\subset\{1,\ldots,n\}$ and $\Pi^\ns(k)= \sum_{i\in J} \Pi^\ns_i(k)$ be two smooth families of projectors. Defining $k=|k|\theta$ in polar coordinates, let us {\em assume} the continuous gluing condition of the projectors in all directions at infinity:
\begin{equation}\label{eq:sphere}
  \lim_{|k|\to\infty} \Pi^N(|k|\theta) =  \lim_{|k|\to\infty} \Pi^S(|k|\theta) \qquad \mbox{ for all } \theta\in \Sm^1.
\end{equation}
Assume that these limits exist and are continuous in $\theta$. We then {\em define} a new projector $\Pi(k)$ for $k$ an element in the union of two planes $P_\ns\simeq\Rm^2$ that are wrapped around the unit sphere $\Sm^2\simeq(P_N\sqcup P_S)\slash \sim $ by radial compactification so that the circles at infinity are glued along the sphere's equator; see Fig. \ref{fig:sphere}. 
For $k\in P_\ns$, we define $\Pi^\BD(k)=\Pi^\ns(k)$. For a point $\phi$ on the sphere, a form of stereographic projection $\pi$ maps $\phi$ in the upper half sphere to $k\in P_N$ and $\phi$ in the lower half sphere to $k\in P_S$. More precisely, with $\phi\in \Sm^2$ parametrized by $(x,y,z)$, we have
\[
   (x,y) = \frac{k}{\sqrt{1+|k|^2}},\quad z=\frac{\pm1}{\sqrt{1+|k|^2}}, \qquad k\in P_\ns,
\] 
with $\pi$ the inverse map, i.e., $k=\pi(\phi)$. We then define the {\em bulk-difference} projector $\pi^*\Pi^\BD(\phi)=\Pi^\BD(\pi(\phi))$ as the pullback by $\pi$ (still called $\Pi^\BD(\phi)\equiv\pi^*\Pi^\BD(\phi)$ to simplify notation) a projector that is now continuous on $\Sm^2$ thanks to the continuity assumption \eqref{eq:sphere}. 
We then have the Chern number:
\begin{equation}\label{eq:cbd}
  c[\Pi^\BD] =  \dfrac{i}{2\pi} \dint_{\Sm^2} {\rm tr}\, \Pi^\BD d\Pi^\BD \wedge d\Pi^\BD =  \dfrac{i}{2\pi} \dint_{\Rm^2} {\rm tr}\, \Big( \Pi^S[\partial_1\Pi^S,\partial_2\Pi^S] - \Pi^N[\partial_1\Pi^N,\partial_2\Pi^N]\Big)  dk,
\end{equation}
where the $-$ sign above is necessary to ensure that $\Sm^2$ has a given orientation, here inherited from that of the lower plane $P_S$ and opposite that of the upper plane $P_N$. 

The BDI in \eqref{eq:cbd} may thus be seen as the classification of the vector bundle with fibers the range of $\pi^*\Pi^\BD(\phi)$ over the unit sphere $\Sm^2$. The continuity of the projectors as a function of $\phi$ ensures that $c[\Pi^\BD] \in\Zm$. We find in fact that for $\Pi^\ns(k)= \sum_{i\in J} \Pi^\ns_i(k)$, then $c[\Pi^\BD]=\sum_j c[\Pi_j^\BD]$ by additivity of Chern numbers. 

Finally, the Chern numbers are related to \eqref{eq:indFL} as follows. Let $c_i=c[\Pi_i]$ with $\Pi_i$ the gluing of $\Pi^\ns_i$ for $1\leq i\leq n$. Let $\alpha\in\Rm$. We define a bulk-difference invariant based on the notion of resolvent or Green's function \cite{gurarie2011single,volovik2009universe} and sharing similarities with the Kubo formula \cite{bernevig2013topological}. It is given for a family of Hamiltonians $\hat H(k)$ by
\[
   G=G_\alpha(\omega,k) = (z-\hat H(k))^{-1} = \dsum_{j=1}^n (z-h_j(k))^{-1} \Pi_j(k)
\]
for $k\in\Rm^2$ and $z=\alpha+i\omega$ for $\omega\in\Rm$. We {\em assume} that $\alpha$ is a fixed real number in a {\em global} spectral gap, i.e., $\alpha\not=h_j(k)$ for all $1\leq j\leq n$ and $k\in K$. Thus, $G$ and $G^{-1}$ are well-defined with obviously $G^{-1}(k,\omega)=z-\hat H(k)$. We define $i_0=i_0(\alpha)$ as the index such that (possibly after reordering of the eigenvalues $h_j$) $h_{i\leq i_0}(k)<\alpha<h_{i>i_0}(k)$. The crux of the derivation is the result showing that for a fixed value of $k$, then
\[
  \dint_{\Rm} \trr\, G[\partial_1 G^{-1}G,\partial_2 G^{-1} G] d\omega = -4\pi \dsum_{i\leq i_0}  {\rm tr}\,\Pi_i [\partial_1\Pi_i,\partial_2\Pi_i].
\]
This algebraic manipulation is proved in \cite[Lemma 3.2]{bal2022topological} and in a slightly different form in \cite{bernevig2013topological}. A consequence (since $\partial_\omega G=i$) is that 
\[
  \frac{1}{24\pi^3} \dint_{\Rm^3} \trr (G^{-1}dG)^{\wedge 3} = \sum_{i\leq i_0}  \frac{i}{2\pi} \dint_{\Rm^2} \trr\, \Pi_i d\Pi_i \wedge d\Pi_i.
\]
Identifying $x$ with $\omega$, we realize that $a(\omega,k)-\alpha=-G^{-1}(\omega,k)$ at $y=\pm L$. 

After projection onto the sphere by radial compactification, and assuming $\alpha=0$ after shifting the global bulk gap so that in includes $0$ (or replacing $a$ by $a-\alpha$ below), we deduce that 
\begin{equation}\label{eq:BDI}
   \rF_2[a] =  c[\Pi_-^\BD] 
\end{equation}
where $\Pi^\ns_-=\sum_{i<i_0} \Pi^\ns_i$ and the Chern number corresponding to projection of the bulk Hamiltonians $H^\ns$ (gapped at $\alpha=0$) onto the negative part of their spectrum $\Pi_-^\BD \equiv \pi^*\Pi_-^\BD$ is defined on the sphere by the above gluing procedure. 

Provided that we may glue the projectors as $k\to\infty$ as stipulated in \eqref{eq:sphere}, we thus obtain that the index defined in \eqref{eq:FH2d} is in fact a BDI given by a Chern number over the unit sphere.  For the Dirac operator $H=D\cdot\sigma + m(y)\sigma_3$ with $m(y)=m_+$ for $y\geq L$ and $m(y)=m_-$ for $y\leq-L$, we find that the Chern number is given by $\frac12(\sgn{m_-}-\sgn{m_+})$. This bona fide integral-valued invariant is obtained without the need to regularize the Dirac operator ($\eta=0$). We also observe that the invariant is the same as that given by $2\pi \sigma_I[H_1]$. This is not a coincidence and in fact the result of a far-reaching general principle, the bulk-edge correspondence, to which we now turn.
\\[2mm]
{\bf One-point versus radial compactification.} Lattice models have a natural small scale that provides a bound on the domain of definition of dual variables. Second-order elliptic equations with a periodic microstructure likewise have a small scale allowing one to define a compact Brillouin zone. No such natural truncation exists for (macroscopic) partial differential Hamiltonians, where the dual Fourier variables belong to $\Rm^2$ in the two-dimensional setting. Identifying $\xi=|\xi|e^{i\theta}$, the difficulty in assigning topological invariants comes from the behavior in $\theta$ of the Hamiltonian as $|\xi|\to\infty$. The Landau Hamiltonian with Weyl symbol $\xi_1^2+ (\xi_2-Bx)^2$ does not display any $\theta-$dependence for $(x,y)$ bounded and we saw that bulk invariants could be defined. 

The situation is different from the Dirac operator $H=D\cdot\sigma+m\sigma_3$ with leading symbol $|\xi| (\cos\theta \sigma_1+\sin\theta \sigma_2)$ that depends on the direction $\theta$. We saw in section \ref{sec:topoDirac} how to modify the operator so that $(m+\eta|\xi|^2)\sigma_3$ dominates as $|\xi|\to\infty$ and is independent of $\theta$.  This is a well-known regularization \cite{shen2012topological} that has been analyzed in detail in, e.g., \cite{silveirinha2015chern} for its multiple applications in topological photonics. The one-point compactification resulting in a well-defined continuous Hamiltonian on the Riemann sphere is necessary to define bulk invariants.

The radial compactification is based on a gluing assumption indicating that while the Hamiltonian depends on $\theta$ as $|\xi|\to\infty$, it does so irrespectively of the insulating mechanism. For the Dirac operator $|\xi| (\cos\theta \sigma_1+\sin\theta \sigma_2)+m\sigma_2$, we indeed observe that the $\theta-$dependence is independent of the sign of the confining parameter $m$. This is a mechanism that is expected to apply to a large class of partial differential models where the insulating mechanism is a zero-th order term; see \cite{rossi2024topology}.

The radial compactification also has the following natural interpretation: it is easier to define {\em phase differences} than {\em absolute phases}.

\section{Bulk-edge correspondence}
\label{sec:BEC}
The bulk-edge correspondence is a central principle in the analysis of topological systems heuristically stating that the interface separating two insulators inherits a topological characterization given by the difference of bulk topologies of these insulators.  For (effectively) one dimensional Hamiltonians, this implies the existence of bound states at the effectively 0-dimensional (or compact) interface. For an effective one-dimensional Hamiltonian $H_{d-1}$, then $F=H_{d-1}-ix$ describes a system with a minimum of $\ind F={\rm dim\,Ker}\,F-{\rm dim\,Ker}\,F^*$ zero modes for either $F$ or $F^*$. 

We are interested here in an interface topology characterized by $\sigma_I[H_{d-1}]$. Assuming $d=2$, we wish to show that $2\pi\sigma_I[H_1]$ may be written in terms of the bulk properties for $\pm y\geq L$. We saw in the preceding section that the bulk-difference invariant $\ind F$ only depended on the bulk properties for $\pm y\geq L$. The main objective of this section is to show that $2\pi\sigma_I[H_1]=\ind F$ holds for elliptic pseudo-differential operators.

\subsection{Bulk-edge correspondences}
\label{sec:BECS}
Due to its central role in the understanding of topological phases of matter, the bulk-edge correspondence has been the object of numerous analyses. 

The first notions of bulk edge correspondence were obtained for discrete \cite{hatsugai1993chern} or continuous \cite{halperin1982quantized} Landau operators, elucidating the relationship between the number of occupied Landau levels and the existence of extended states localized near interfaces. A number of correspondences for differential models using a semiclassical setting \cite{essin2011bulk} and index theory \cite{fukui2012bulk} is proposed in the physics literature; see also the monograph \cite{bernevig2013topological}.

There is a large number of works in the mathematical literature addressing aspects of the bulk-edge correspondence. K-theoretic approaches are used in the analysis of bulk phases in the integer quantum Hall effect in \cite{bellissard1994noncommutative}. Correspondences between K-groups of the bulk and edge algebras are constructed to relate edge invariants with bulk invariants; see \cite{prodan2016bulk} and \cite{SBindex2016}. See also \cite{bourne2017k,bourne2018chern,kellendonk2008topological} for the bulk-edge correspondence for discrete and continuous Hamiltonians and the role of Dirac operators in physical space of the form $\sum_j X_j \gamma^j$, and \cite{Avila2013} for an analysis by the method of transfer matrices also present in \cite{hatsugai1993chern}.

An analytic approach to the bulk-boundary correspondence for general tight-binding Hamiltonians, i.e., Hamiltonians acting on (vector-valued) functions of the lattice $\Zm^2$, is analyzed in a series of papers \cite{elbau2002equality,elgart2005equality,graf2007aspects}. The edge Hamiltonians are also defined as truncations of bulk Hamiltonians on $\Zm\times \Nm$. An edge conductivity similar to above construction of $\sigma_I$ and a bulk conductivity similar to that defined in section \ref{sec:bulk} are shown to be equal.

The macroscopic models we consider in this paper may be derived asymptotically from the analysis of partial differential operators with microstructure. When these microstructures are allowed to vary at the macroscopic scale, various two-scale techniques allow one to obtain macroscopic operators such as the Dirac model. We refer to \cite{fefferman2012honeycomb,Fefferman2016} for the analysis of Dirac points in periodic Schr\"odinger models and to \cite{gerard1991mathematical} for the derivation of effective Hamiltonians using semiclassical techniques. These two works were combined in \cite{drouot2021microlocal}, see also \cite{drouot2019bulk,drouot2020edge}, to obtain a general bulk-edge correspondence for second-order differential operators with periodic underlying structure. In these works, the existence of a periodic structure naturally leads to a compact Brillouin zone so that notions such as the above bulk-difference invariants are not necessary.  
%%%

%
\subsection{Bulk-edge correspondence in dimension $d=2$.}
Let $H=H_1=\ow{a_1}$ be an operator in $\ow S^m$ satisfying {\bf [H1]}. Let $\alpha\in (E_1,E_2)$ and define $F=H-\alpha-ix =\ow{a}$ where $a=a_1-\alpha-ix$. Let $\sigma_I[H]$ be the interface current observable defined in \eqref{eq:sigmaI}. Then we have the following result \cite{bal2022topological,quinn2024approximations}
\begin{theorem}[Bulk-Edge correspondence]\label{thm:TCC}
   Under the above hypothesis, we have
    \begin{equation}\label{eq:TCC}
   2\pi \sigma_I[H] = \frac{1}{24\pi^2} \dint_{ \Sm^3_R }  \trr (a^{-1}da)^{\wedge \,3} \in \Zm.
  \end{equation}
  Here, $R$ is large enough that $a^{-1}$ is defined for $|(x,y,\xi,\zeta)|\geq R$.
\end{theorem} 
We sketch the main steps of the derivation referring to the above references for the tedious details. 
The first step uses Corollary \ref{cor:deformsigmaI} implying that $a_1(x,\cdot)$ may be replaced by $a_1(0,\cdot)$ by showing that the continuous deformation linking them changes neither $\sigma_I$ nor the Fedosov-H\"ormander formula.

Since $H$ is invariant by translations in $x$, we have (with $\mF$ partial Fourier transform in $x$) that $H=\mF^{-1} \hat H(\xi) \mF$ with $\hat H(\xi) = \ow a_1(y,\xi,\zeta)$. We know from Proposition \ref{prop:traces} that $2\pi \sigma_I[H]$ may be written as an integral of the Schwartz kernel of $2\pi i[H,P]\varphi'(H)$ along the diagonal. Denoting by $\Tr_y$ the integration in the $(y,y')$ variables, 
\begin{align*}
  2\pi \sigma_I & = 2\pi \Tr_y \dint_{\Rm^2} i[H,P] (x,x') \varphi'(H)(x-x') dx' dx = \Tr_y \dint_{\Rm} \partial_\xi \hat H(\xi) \varphi'(\hat H(\xi)) d\xi.
\end{align*}

Define the symbol $s=s(Y,\xi)$ such that $\ow(s)=\partial_\xi\hat H \varphi'(\hat H)$. Here $Y=(y,\zeta)$. We now wish to use the crucial invariance of $2\pi\sigma_I$ with respect to the rescaling $\zeta\to h\zeta$ as recalled in Corollary \ref{cor:deformsigmaI}. This invariance is also central in the proof of the index theorem in \cite[Theorem 19.3.1]{H-III-SP-94} and for the same reason as here: non-commutating operator {\em almost} commute in the semiclassical regime and do so in an asymptotically tractable fashion since $i[hD_x,x]=h$ is small when $h$ is. 

We then define $H_h$ and $\hat H_h$ so that $\zeta$ is replaced by $h\zeta$ in the symbols; i.e., $\hat H=\ow a_1(\zeta)$ while $\hat H_h=\ow a_1(h\zeta)=\ow_h a_1(\zeta)$. With this notation, we thus have $\ow_h(s)=\partial_\xi\hat H_h \varphi'(\hat H_h)$ with now $s(Y,\xi;h)$. We observe that the symbol of $\partial_\xi\hat H_h \varphi'(\hat H_h)$ depends non-trivially on $h$. In this setting, 
\[
 2\pi\sigma_I = \frac1{2\pi h} \dint_{\Rm^3} \trr \ s (Y,\xi;h) dYd\xi.
\]
We know from Corollary \ref{cor:deformsigmaI} that $2\pi \sigma_I$ is {\em independent} of $h>0$. Introducing $\sigma_z=z-a_1(y,\xi,\zeta)$ and defining the symbol $\varsigma(Y,\xi;h)$ so that  $\varphi'(\hat H_h)=\ow_h\varsigma$, we use the Helffer-Sj\"ostrand formula to get
$
  \varsigma = -\frac 1\pi \int_Z \bar\partial \tilde \varphi'(z) \trr \, r_z dY d\xi d^2z
$
where $\ow_h r_z=(z-\hat H_h)^{-1}$ is the symbol of the resolvent operator. From semiclassical calculus \cite{dimassi1999spectral}, we find $r_z = \sigma_z^{-1} + \frac{ih}{2} \{\sigma_z^{-1} , \sigma_z\} \sigma_z^{-1} + O(h^2)$ with $\{\cdot,\cdot\}$ the Poisson bracket, which gives keeping only terms of order $h^0$ on both sides:
\begin{equation} \label{eq:sI1}
 2\pi \sigma_{I}=\frac{i}{4\pi^2 } \dint_{\Rm^3\times Z} \bar\partial\tilde\varphi'(z) \,\trr\,  \big(\partial_\xi \sigma_z  \{\sigma_z^{-1},\sigma_z\}\sigma_z^{-1} -  \{\partial_\xi \sigma_z,\sigma_z^{-1}\} \big)\, dY d\xi d^2z.
 \end{equation}
 An application of the Stokes theorem to write an integral along the real axis combined with some amount of differential calculus and the normalization $\int_{\Rm} \varphi'(\lambda)d\lambda=1$ yields \eqref{eq:TCC}.

\subsection{Generalized bulk-edge correspondence in higher dimensions}

The above analysis of the BEC extends to arbitrary dimensions \cite{bal2023topological}. Let $d\geq2$ be spatial dimension and $0\leq k\leq d-1$. Compared to the dimension $d=2$, there are several technical differences that we briefly sketch.
We come back here to the setting of Hamiltonians $H_k=\ow a_k$ for $a_k\in S^m_k$ elliptic, i.e., \eqref{eq:Sk} and \eqref{eq:ellip} hold. We recall the construction of $H:=H_{d-1}$ in \eqref{eq:Hinterm} obtained by twisting $H_k$ by $d-1-k\geq0$ domain walls. 

The interface current observable $\sigma_I[H_{d-1}]$ is then defined as follows. The functions $M=w_k^m$ are admissible weight functions in the sense of \cite[Definition 2.3]{bony2013characterization}. This allows one to define a functional calculus showing for instance that for $H$ elliptic in $\ow S^m_{d-1}$, then $\phi(H)\in \ow S_{d-1}^{-\infty}$ for $\phi\in C^\infty_c(\Rm)$, i.e., a pseudo-differential operator with coefficients decaying faster than any $w_{d-1}^{-N}(X)$. This leads to the result:
\begin{theorem}[{\cite[Lemma 3.1 \& Theorem 3.2]{bal2023topological}}]
  Let $\phi\in C^\infty_c(\Rm)$ and  $p,q\in\Nm$. Then $[P,\phi(H)]$ and $H^p[P,H^q]\phi(H)$  are trace-class operators with symbols in $S_{d}^{-\infty}$. When $\tilde P\in \fS[0,1]$ is an orthogonal projector, then  $T:=\tilde PU(H) \tilde P _{{\rm Ran} \tilde P}$ is a Fredholm operator on ${\rm Ran} \tilde P$. Moreover,
 \[2\pi \sigma_I  = {\rm Tr}\,[U(H),P] U^*(H) = {\rm Tr}\,[U(H),\tilde P] U^*(H) = \ind\,\tilde PU(H) \tilde P _{{\rm Ran} \tilde P}.\]
\end{theorem}

The proof of the bulk-edge correspondence, relating $2\pi\sigma_I$ to the Fedosov-H\"ormander formula, requires the following approximation step. Let us denote by $S^m_k(g^s)$ the class of symbols satisfying \eqref{eq:Sk} and by $S^m_k(g^i)$ the class of symbols satisfying \eqref{eq:homSk}.
\begin{lemma}[{\cite[Lemma 3.3]{bal2023topological}}]
Let $T=\tilde PU(H) \tilde P _{{\rm Ran} \tilde P}$ with $H=H_{d-1}\in \ow S^m_{d-1}(g^s)$ elliptic. There is a sequence of elliptic operators $H_\eps$ for $0\leq\eps\leq1$ with symbol in $S^m_{d-1}(g^{i})$ for all $\eps>0$ and such that the corresponding $[0,1]\ni\eps\to T_\eps=\tilde P U(H_\eps) \tilde P_{{\rm Ran} \tilde P}$ is continuous in the uniform sense and $T_0=T$. Thus $\ind T_\eps$ is defined, independent of $\eps$ and equal to $\ind T$. Moreover, the symbols $a_\eps$ are chosen so that for any compact domain in $X=(x,\xi)$, $a_\eps=a_{d-1}$ on that domain for $\eps$ sufficiently small. 
\end{lemma}
This result, mimicking \cite[Lemma 19.3.3]{H-III-SP-94}, allows us to compute the index of an operator in the larger class $S^m_k(g^s)$ by operators in $S^m_k(g^i)$. As a simple example, consider $a(x,\xi)=b(x)\xi$ for $x\in\Rm$ and $\xi\in\Rm$ with $b(x)$ smooth, positive, and equal to $1$ outside a compact domain. We verify that \eqref{eq:Sk} holds so that $a$ is elliptic in $S^1_1(g^s)$ whereas $\aver{x,\xi} |b'\xi| \leq C \aver{x,\xi}$ is false and $a$ does not belong to $S^1_1(g^i)$. The smaller class of symbols has the following desirable property:
\begin{lemma}[{\cite[Lemma 3.4]{bal2023topological}}]\label{lem:conttf}
  Let $[0,1]\ni t \to L_t$ be a continuous family of linear invertible transformations in $GL(2d-2,\Rm)$ in the $(x_1,\ldots,x_{d-1},\xi_1,\ldots \xi_{d-1})$ variables leaving the variables $(x_d,\xi_d)$ fixed. This includes dilations and rotations.  
  Let $a(X)\in S^m_{d-1}(g^i)$ elliptic. Then $a(t,X)=a(L_t X)\in S^m_{d-1}(g^i)$. Let $T_t$ be the corresponding Toeplitz operator. Then $T_t$ is Fredholm with index independent of $t\in [0,1]$.
\end{lemma}
With this, we can state the:
\begin{theorem}[Generalized Bulk-Edge correspondence {\cite[Theorem 4.1]{bal2023topological}}] \label{thm:tcc}
  Let $F=\ow a$ with $a_k \in S^m_k$ elliptic. Then we have 
\begin{equation}\label{eq:tcc}
  2\pi \sigma_I  =  - \dfrac{(d-1)!}{(2\pi i)^d (2d-1)!} \dint_{\Sm_R^{2d-1}}  
  {\rm tr} \,\big( (a^{-1} da)^{\wedge(2d-1)} \big) = \ind\,F.
\end{equation}
\end{theorem}
The proof of this result essentially follows that of Theorem \ref{thm:TCC} until we reach the equivalent of \eqref{eq:sI1}:
\begin{equation}\label{eq:sigmaI3}
 2\pi \sigma_I = \frac{-ic_{d-1}}{(2\pi)^d} \dint_{\Rm^{2d-1}\times \Rm} \varphi'(\lambda)  {\rm tr}\,  \sigma_z^{-1}\partial_\xi \sigma_z \{ \sigma_z^{-1},\sigma_z\} ^{d-1}_f  \Big|_{\lambda-i0}^{\lambda+i0} dY d\xi d\lambda.
\end{equation}
Here, $\{ \sigma_z^{-1},\sigma_z\} ^{d-1}_f  = \sum_{\rho \in \mS_{d-1}}  \{ \sigma_z^{-1},\sigma_z\}_{\rho_1} \ldots \{ \sigma_z^{-1},\sigma_z\}_{\rho_{d-1}}$ where the Poisson brackets act in variables $(x_k,\xi_k)$  for $1\leq k\leq d-1$ and the sum runs over permutations of such pairs of variables fixing $(x_d,\xi_d)$. The term ${\rm tr}\, (\sigma_z^{-1}d\sigma_z)^{2d-1}$ has a similar structure that may be written as a sum of products of Poisson brackets. However, it is more symmetrical by involving Poisson brackets in variables $(x_1,x_2)$, say, that are not present in \eqref{eq:sigmaI3}. This is where Lemma \ref{lem:conttf} is used to show that $\sigma_I$ is invariant under permutation of the variables $(x_1,\ldots,x_{d-1},\xi_1,\ldots \xi_{d-1})$ up to a sign given by the order of the permutation.  Summing over all permutations leads to:
\begin{equation}\label{eq:sigmaI4}
  \sigma_I = \frac{1}{(2d-2)!} \dsum_{\rho\in\mS_{2d-2}} (-1)^\rho \sigma_I(\rho (Y))
\end{equation}
where summation is over all permutations of ${1,\ldots,2d-2}$ and $(-1)^\rho$ is the signature of the permutation. Combined with \eqref{eq:sigmaI3}, this provides
\begin{equation*}%\label{eq:sigmaI5}
 2\pi \sigma_I = \dfrac{(-1)^{d-1}}{(2\pi i)^d} \dfrac{(d-1)!}{(2d-1)!}\dint_{\Rm^{2d-1}\times \Rm} \varphi'(\lambda)  {\rm tr}\, (\sigma_z^{-1}d\sigma_z)^{2d-1} \Big|_{\lambda-i0}^{\lambda+i0} dY d\xi d\lambda,
\end{equation*}
which after an application of the Stokes theorem gives the result \eqref{eq:tcc}.

\subsection{Applications of the above bulk-edge correspondence}
The bulk edge correspondence (BEC) plays an essential role in topological phases of matter in that it allows one to replace the difficult computation of a physically observable quantity $2\pi\sigma_I[H_I]$ of an interface Hamiltonian by the much simpler calculation of the integral in \eqref{eq:FH}, which in many application further simplifies to the computation of a Chern number as in \eqref{eq:cbd}.

Direct applications of \eqref{eq:TCC} and \eqref{eq:cbd} or \eqref{eq:tcc} in higher dimensions may be found in \cite{quinn2024asymmetric} and \cite{bal2023topological}. These results retrieve examples scattered throughout the physics literature with applications to p-wave and d-wave models of superconductors \cite{bernevig2013topological,volovik2009universe}, as well as to a regularized version of the shallow water model \cite{quinn2024approximations}.

A non-trivial application is in the analysis of bilayer graphene models \cite{bistritzer2011moire,watson2023bistritzer} (and the mathematically similar model of Floquet topological insulators based on Dirac systems of equations \cite{bal2022multiscale}). We follow the presentation in \cite{bal2023mathematical}. The macroscopic model for bilayer graphene is the following system of coupled Dirac equations
\begin{equation}\label{eq:He}
    H_I := \begin{pmatrix} \Omega I + D \cdot \sigma & \lambda U^*(y) \\ \lambda U(y) & -\Omega I + D \cdot \sigma\end{pmatrix}, \quad U(y)= \frac{1}{2}((1+m(y)) A + (1-m(y)) A^*),\quad A= \begin{pmatrix} 0 & 1 \\ 0& 0\end{pmatrix}
\end{equation}
where $2\Omega$ is a difference of potential between the two layers and $U(y)$ is a coupling term between the two layers. Transport within each layer is modeled by the massless Dirac term $D \cdot \sigma$. The phase $U=A$ corresponds to the so-called BA stacking (B sites of top layer above A sites of bottom layer) whereas $U=A^*$ corresponds to AB stacking. Twisting the two layers generates a transition between AB and BA stacking modeled above by a transition term $m(y)\in \fS[-1,1]$.
\begin{figure}[ht]
\centering
\includegraphics[width=.43\textwidth]{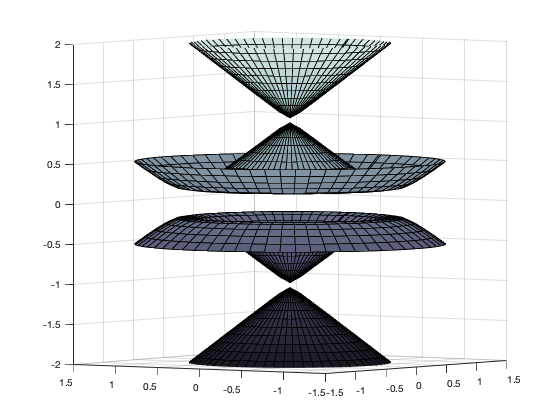}
    \centering
\includegraphics[width=.43\textwidth]{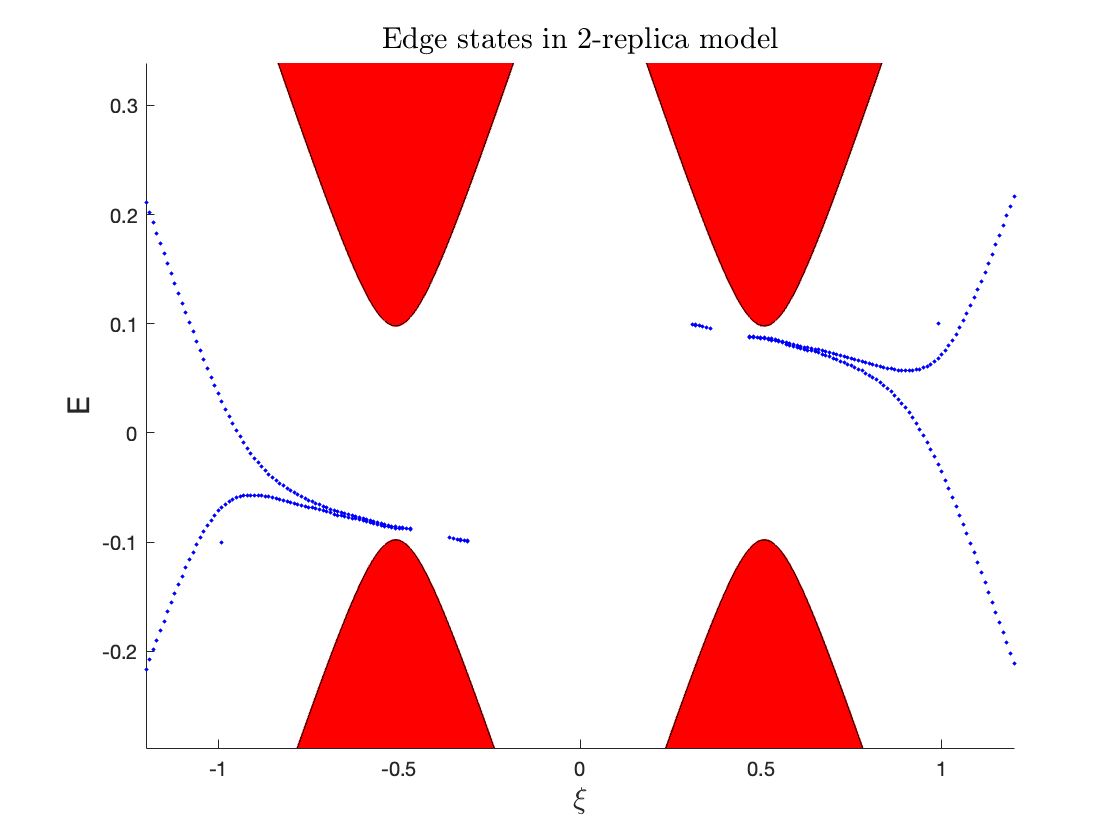}
\caption{Left: bulk spectrum for $m(y)=1$. Right: Interface spectrum with Spectral Flow=-2.}
\label{fig:tBLG}
\end{figure}
The bulk spectrum of $H_B$ is given on the left of Fig. \ref{fig:tBLG} for $m=1$. We observe a superposition of two Dirac cones with a gap opening caused by a non-vanishing coupling $\lambda\not=0$. The edge spectrum of $H_I$ is given on the right of Fig. \ref{fig:tBLG}. While we observe a spectral flow of $-2$ numerically, analytic expressions for the branches of spectrum are not known explicitly. 

The above Hamiltonian is clearly elliptic with symbol in $S^1$. The bulk Hamiltonian with BA stacking ($m=1$) has the four eigenvalues corresponding to:
$E_\pm^2 = (\Omega^2 + \frac12 \eps^2 + |\xi|^2)  \pm \sqrt{(4\Omega^2+\eps^2)|\xi|^2 + \frac 14 \eps^4}$. The `Chern' numbers corresponding to the positive branches are given after some algebra \cite{bal2023mathematical} by
\[
  c_4^N= \dfrac{\beta^2-\eps^2}{2(\eps^2+\beta^2)},\qquad c_3^N = \frac{-\eps^2-3\beta^2}{2(\eps^2+\beta^2)},  \qquad c^N=c_3^N+c_4^N = -1,
\]
where $ \beta=2\Omega+\sqrt{4\Omega^2+\eps^2}$. Overall, we thus obtain that $2\pi \sigma_I[H_I] = -2$, the bulk-difference invariant, which is consistent with the numerical simulations in Fig. \ref{fig:tBLG}. Note that $c^N_j$ the integral of the curvature $\trr \Pi_j d\Pi_j\wedge \Pi_j$ over $\Rm^2$ takes continuous values for each branch. Only after summing over branches in the computation of the BDI in \eqref{eq:BDI} does one obtain $-2\in\Zm$.

\subsection{Validity of the bulk-edge correspondence}
The notion of BEC is very robust.  While the Landau and magnetic Dirac operators are not elliptic in the presence of infinitely extended magnetic fields, their symbols are still amenable to similar mathematical analyses. For a Dirac operator with symbol of the form $\xi_1\sigma_1+(\xi_2+Bx)\sigma_2$, we recover some ellipticity for bounded values of $(x,y)$, which is sufficient to compute interface currents by spectral flows as we did in section \ref{sec:appliSF} and show that they are stable against large classes of perturbations. While not a general BEC, these calculations show that the topological edge current was indeed related to the bulk properties of the joined insulators independently of the details of the transition between them.

For the shallow water problem \eqref{eq:H}, the BEC is more problematic. We saw in Fig. \ref{fig:shallow} that the BEC was violated as soon as the Coriolis force parameter displayed discontinuities. By violation, we mean that the spectral flow and associated current observable now depend on how the Coriolis force parameter transitions from a positive to a negative value.
We already indicated in section \ref{sec:elliptic} that the presence of essential point spectrum at frequency $0$ created complications. These violations of the bulk-edge correspondence were confirmed in models of half space problems with constant $f$ but with varying boundary conditions. The interface current observable was then showed to depend on such boundary conditions, which cannot be the case for elliptic operators; see for instance \cite{graf2021topology,graf2024boundary,rossi2024topology,tauber2019anomalous,tauber2023topology}.
\\[2mm]
{\bf Extension to other geometries.}
The bulk-edge correspondence considered so far applies to problems in Cartesian geometry, for instance with confinement in the variable $y$ and propagation along the straight $x-$axis. In fact, it holds in much more general geometries with appropriate modifications. For Landau operators, see for instance \cite{buchendorfer2006scattering,frohlich2000extended,hislop2008edge}. The extension of the work of \cite{elgart2005equality} for general discrete Hamiltonians to the setting of curved and not necessarily connected interfaces is treated in \cite{drouot2024bulk,drouot2024topological}. The extension of the BEC for two-dimensional elliptic Hamiltonians in Theorem \ref{thm:TCC} also extends to more general geometries as shown in \cite[Section 4]{bal2023mathematical}. Both in the discrete and the continuous case, the interface separating the insulating phases and the transition curves of the function $P=P(x,y)$ have to be sufficiently transversal. The interface current \eqref{eq:sigmaI} needs to be modified accordingly and is then shown for instance to be immune to translations in the profile $P$. As an example, we can consider a junction topology where a mass term in a Dirac equation is positive when $xy>0$ and negative when $xy<0$. Any function $P$ transitioning from $0$ to $1$ along interfaces $x=\pm y$ for $|(x,y)|$ sufficiently large then generates a stable interface current $\sigma_I=\Tr\, i[H,P(x,y)]\varphi'(H)$ \cite[Section 4]{bal2023mathematical}.

\section*{Acknowledgements.} This work was supported in part by NSF Grant DMS-2306411.

%%%%%%%%%%%%%%%%%
{\small
%\bibliography{bibTIOP2024} 

\begin{thebibliography}{100}

\bibitem{asboth2016short}
{\sc J.~K. Asb{\'o}th, L.~Oroszl{\'a}ny, and A.~P{\'a}lyi}, {\em A short course
  on topological insulators}, Lecture notes in physics, 919 (2016).

\bibitem{atiyah1970global}
{\sc M.~F. Atiyah}, {\em Global theory of elliptic operators}, in Proc.
  Internat. Conf. on Functional Analysis and Related Topics (Tokyo, 1969),
  vol.~2130, 1970.

\bibitem{atiyah1968index}
{\sc M.~F. Atiyah and I.~M. Singer}, {\em {The index of elliptic operators.
  I.}}, Annals of Mathematics, 87 (1968), pp.~484--530.

\bibitem{atiyah1969index}
{\sc M.~F. Atiyah and I.~M. Singer}, {\em {Index theory for skew-adjoint
  Fredholm operators}}, Publications Math{\'e}matiques de l'IH{\'E}S, 37
  (1969), pp.~5--26.

\bibitem{Avila2013}
{\sc J.~C. Avila, H.~Schulz-Baldes, and C.~Villegas-Blas}, {\em Topological
  invariants of edge states for periodic two-dimensional models}, Mathematical
  Physics, Analysis and Geometry, 16 (2013), pp.~137--170.

\bibitem{avron1994}
{\sc J.~E. Avron, R.~Seiler, and B.~Simon}, {\em Charge deficiency, charge
  transport and comparison of dimensions}, Comm. Math. Phys., 159 (1994),
  pp.~399--422.

\bibitem{bal2019continuous}
{\sc G.~Bal}, {\em Continuous bulk and interface description of topological
  insulators}, Journal of Mathematical Physics, 60 (2019).

\bibitem{bal2019topological}
\leavevmode\vrule height 2pt depth -1.6pt width 23pt, {\em Topological
  protection of perturbed edge states}, Communications in Mathematical
  Sciences, 17 (2019), pp.~193--225.

\bibitem{bal2022topological}
\leavevmode\vrule height 2pt depth -1.6pt width 23pt, {\em Topological
  invariants for interface modes}, Communications in Partial Differential
  Equations, 47 (2022), pp.~1636--1679.

\bibitem{bal2023topological}
\leavevmode\vrule height 2pt depth -1.6pt width 23pt, {\em Topological charge
  conservation for continuous insulators}, Journal of Mathematical Physics, 64
  (2023), p.~031508.

\bibitem{bal2024semiclassical}
\leavevmode\vrule height 2pt depth -1.6pt width 23pt, {\em Semiclassical
  propagation along curved domain walls}, Multiscale Modeling \& Simulation, 22
  (2024), pp.~66--105.

\bibitem{bal2023magnetic}
{\sc G.~Bal, S.~Becker, and A.~Drouot}, {\em Magnetic slowdown of topological
  edge states}, Communications on Pure and Applied Mathematics, 77 (2024),
  pp.~1235--1277.

\bibitem{bal2023edge}
{\sc G.~Bal, S.~Becker, A.~Drouot, C.~F. Kammerer, J.~Lu, and A.~B. Watson},
  {\em Edge state dynamics along curved interfaces}, SIAM Journal on
  Mathematical Analysis, 55 (2023), pp.~4219--4254.

\bibitem{bal2023mathematical}
{\sc G.~Bal, P.~Cazeaux, D.~Massatt, and S.~Quinn}, {\em Mathematical models of
  topologically protected transport in twisted bilayer graphene}, Multiscale
  Modeling \& Simulation, 21 (2023), pp.~1081--1121.

\bibitem{bal2024bistopological}
{\sc G.~Bal and T.~Dang}, {\em {Topological Anderson Insulators by
  homogenization theory}}, Communications in Partial Differential Equations, 49
  (2024), pp.~989--1010.

\bibitem{bal2023integral}
{\sc G.~Bal, J.~Hoskins, S.~Quinn, and M.~Rachh}, {\em {Integral formulation of
  Dirac singular waveguides}}, arXiv preprint arXiv:2312.16701,  (2023).

\bibitem{bal2025integral}
\leavevmode\vrule height 2pt depth -1.6pt width 23pt, {\em Integral formulation
  of klein--gordon singular waveguides}, Communications on Pure and Applied
  Mathematics, 78 (2025), pp.~323--365.

\bibitem{bal2023asymmetric}
{\sc G.~Bal, J.~G. Hoskins, and Z.~Wang}, {\em {Asymmetric transport
  computations in Dirac models of topological insulators}}, Journal of
  Computational Physics, 487 (2023), p.~112151.

\bibitem{bal2022multiscale}
{\sc G.~Bal and D.~Massatt}, {\em {Multiscale invariants of Floquet topological
  insulators}}, Multiscale Modeling \& Simulation, 20 (2022), pp.~493--523.

\bibitem{bal2024mathbb}
{\sc G.~Bal and Z.~Wang}, {\em {${\mathbb{Z}_2}$ classification of FTR
  symmetric differential operators and obstruction to Anderson localization}},
  Journal of Physics A: Mathematical and Theoretical, 57 (2024), p.~285202.

\bibitem{bal2024topological}
{\sc G.~Bal and J.~Yu}, {\em Topological equatorial waves and violation (or
  not) of the bulk edge correspondence}, Journal of Physics A: Mathematical and
  Theoretical, 57 (2024), p.~405204.

\bibitem{1751-8121-41-40-405203}
{\sc J.~H. Bardarson}, {\em {A proof of the Kramers degeneracy of transmission
  eigenvalues from antisymmetry of the scattering matrix}}, Journal of Physics
  A: Mathematical and Theoretical, 41 (2008), p.~405203.

\bibitem{bellissard1994noncommutative}
{\sc J.~Bellissard, A.~van Elst, and H.~Schulz-Baldes}, {\em {The
  noncommutative geometry of the quantum Hall effect}}, Journal of Mathematical
  Physics, 35 (1994), pp.~5373--5451.

\bibitem{bernevig2013topological}
{\sc B.~A. Bernevig and T.~L. Hughes}, {\em Topological insulators and
  topological superconductors}, Princeton university press, 2013.

\bibitem{bistritzer2011moire}
{\sc R.~Bistritzer and A.~H. MacDonald}, {\em Moir{\'e} bands in twisted
  double-layer graphene}, Proceedings of the National Academy of Sciences, 108
  (2011), pp.~12233--12237.

\bibitem{bony2013characterization}
{\sc J.-M. Bony}, {\em On the characterization of pseudodifferential operators
  (old and new)}, in Studies in Phase Space Analysis with Applications to PDEs,
  Springer, 2013, pp.~21--34.

\bibitem{bott1978some}
{\sc R.~Bott and R.~Seeley}, {\em Some remarks on the paper of {Callias}},
  Communications in Mathematical Physics, 62 (1978), pp.~235--245.

\bibitem{bourne2017k}
{\sc C.~Bourne, J.~Kellendonk, and A.~Rennie}, {\em {The K-theoretic bulk--edge
  correspondence for topological insulators}}, in Annales Henri Poincar{\'e},
  vol.~18, Springer, 2017, pp.~1833--1866.

\bibitem{bourne2018chern}
{\sc C.~Bourne and A.~Rennie}, {\em Chern numbers, localisation and the
  bulk-edge correspondence for continuous models of topological phases},
  Mathematical Physics, Analysis and Geometry, 21 (2018), p.~16.

\bibitem{buchendorfer2006scattering}
{\sc C.~Buchendorfer and G.~M. Graf}, {\em Scattering of magnetic edge states},
  in Annales Henri Poincare, vol.~7, Springer, 2006, pp.~303--333.

\bibitem{callias1978axial}
{\sc C.~Callias}, {\em Axial anomalies and index theorems on open spaces},
  Communications in Mathematical Physics, 62 (1978), pp.~213--234.

\bibitem{carey2011spectral}
{\sc A.~L. Carey, J.~Phillips, and A.~Rennie}, {\em Spectral triples: examples
  and index theory}, Noncommutative geometry and physics: renormalisation,
  motives, index theory,  (2011), pp.~175--265.

\bibitem{cayssol2013floquet}
{\sc J.~Cayssol, B.~D{\'o}ra, F.~Simon, and R.~Moessner}, {\em Floquet
  topological insulators}, physica status solidi (RRL)--Rapid Research Letters,
  7 (2013), pp.~101--108.

\bibitem{chen2025scattering}
{\sc B.~Chen and G.~Bal}, {\em Scattering theory of topologically protected
  edge transport}, Pure and Applied Analysis, 7 (2025), pp.~701--731.

\bibitem{combes2005edge}
{\sc J.-M. Combes and F.~Germinet}, {\em Edge and impurity effects on
  quantization of hall currents}, Communications in mathematical physics, 256
  (2005), pp.~159--180.

\bibitem{connes1994noncommutative}
{\sc A.~Connes}, {\em Noncommutative Geometry}, Academic Press, 1994.

\bibitem{cornean2024orbital}
{\sc H.~D. Cornean, M.~Moscolari, and S.~Teufel}, {\em From orbital magnetism
  to bulk-edge correspondence}, in Annales Henri Poincar{\'e}, Springer, 2024,
  pp.~1--55.

\bibitem{davies1995spectral}
{\sc E.~B. Davies}, {\em Spectral theory and differential operators}, vol.~42,
  Cambridge University Press, 1995.

\bibitem{delplace2017topological}
{\sc P.~Delplace, J.~Marston, and A.~Venaille}, {\em Topological origin of
  equatorial waves}, Science, 358 (2017), pp.~1075--1077.

\bibitem{dimassi1999spectral}
{\sc M.~Dimassi and J.~Sj{\"o}strand}, {\em Spectral asymptotics in the
  semi-classical limit}, no.~268, Cambridge university press, 1999.

\bibitem{dombrowski2011quantization}
{\sc N.~Dombrowski, F.~Germinet, and G.~Raikov}, {\em Quantization of edge
  currents along magnetic barriers and magnetic guides}, in Annales Henri
  Poincar{\'e}, vol.~12, Springer, 2011, pp.~1169--1197.

\bibitem{drouot2019bulk}
{\sc A.~Drouot}, {\em The bulk-edge correspondence for continuous honeycomb
  lattices}, Communications in Partial Differential Equations, 44 (2019),
  pp.~1406--1430.

\bibitem{drouot2021microlocal}
\leavevmode\vrule height 2pt depth -1.6pt width 23pt, {\em Microlocal analysis
  of the bulk-edge correspondence}, Communications in Mathematical Physics, 383
  (2021), p.~2069–2112.

\bibitem{D22}
\leavevmode\vrule height 2pt depth -1.6pt width 23pt, {\em Topological
  insulators in semiclassical regime}, arXiv preprint arXiv:2206.08238,
  (2022).

\bibitem{drouot2020edge}
{\sc A.~Drouot and M.~Weinstein}, {\em Edge states and the valley hall effect},
  Advances in Mathematics, 368 (2020), p.~107142.

\bibitem{drouot2024bulk}
{\sc A.~Drouot and X.~Zhu}, {\em The bulk-edge correspondence for curved
  interfaces}, arXiv preprint arXiv:2408.07950,  (2024).

\bibitem{drouot2024topological}
\leavevmode\vrule height 2pt depth -1.6pt width 23pt, {\em Topological edge
  spectrum along curved interfaces}, International Mathematics Research
  Notices, 2024 (2024), pp.~13870--13889.

\bibitem{elbau2002equality}
{\sc P.~Elbau and G.~Graf}, {\em Equality of bulk and edge hall conductance
  revisited}, Communications in Mathematical Physics, 229 (2002), pp.~415--432.

\bibitem{elgart2005equality}
{\sc A.~Elgart, G.~Graf, and J.~Schenker}, {\em Equality of the bulk and edge
  hall conductances in a mobility gap}, Communications in mathematical physics,
  259 (2005), pp.~185--221.

\bibitem{essin2011bulk}
{\sc A.~M. Essin and V.~Gurarie}, {\em Bulk-boundary correspondence of
  topological insulators from their respective green’s functions}, Physical
  Review B, 84 (2011), p.~125132.

\bibitem{faure2023manifestation}
{\sc F.~Faure}, {\em Manifestation of the topological index formula in quantum
  waves and geophysical waves}, Annales Henri Lebesgue, 6 (2023), pp.~449--492.

\bibitem{Fefferman2016}
{\sc C.~L. Fefferman, J.~P. Lee-Thorp, and M.~I. Weinstein}, {\em {Edge States
  in Honeycomb Structures}}, Annals of PDE, 2 (2016), p.~12.

\bibitem{fefferman2012honeycomb}
{\sc C.~L. Fefferman and M.~I. Weinstein}, {\em {Honeycomb lattice potentials
  and Dirac points}}, Journal of the American Mathematical Society, 25 (2012),
  pp.~1169--1220.

\bibitem{frazier2025topological}
{\sc M.~J. Frazier and G.~Bal}, {\em Topological edge states of continuous
  hamiltonians}, New Journal of Physics, 27 (2025), p.~105001.

\bibitem{frohlich2000extended}
{\sc J.~Fr{\"o}hlich, G.~Graf, and J.~Walcher}, {\em {On the extended nature of
  edge states of quantum Hall Hamiltonians}}, in Annales Henri Poincar{\'e},
  vol.~1, Springer, 2000, pp.~405--442.

\bibitem{fu2021topological}
{\sc Y.~Fu and H.~Qin}, {\em Topological phases and bulk-edge correspondence of
  magnetized cold plasmas}, Nature Communications, 12 (2021), p.~3924.

\bibitem{fukui2012bulk}
{\sc T.~Fukui, K.~Shiozaki, T.~Fujiwara, and S.~Fujimoto}, {\em {Bulk-edge
  correspondence for Chern topological phases: A viewpoint from a generalized
  index theorem}}, Journal of the Physical Society of Japan, 81 (2012),
  p.~114602.

\bibitem{fulga2012scattering}
{\sc I.~C. Fulga, F.~Hassler, and A.~R. Akhmerov}, {\em Scattering theory of
  topological insulators and superconductors}, Physical Review B—Condensed
  Matter and Materials Physics, 85 (2012), p.~165409.

\bibitem{gerard1991mathematical}
{\sc C.~G{\'e}rard, A.~Martinez, and J.~Sj{\"o}strand}, {\em {A mathematical
  approach to the effective Hamiltonian in perturbed periodic problems}},
  Communications in mathematical physics, 142 (1991), pp.~217--244.

\bibitem{germinet2003operator}
{\sc F.~Germinet and A.~Klein}, {\em {Operator kernel estimates for functions
  of generalized Schr{\"o}dinger operators}}, Proceedings of the american
  mathematical society, 131 (2003), pp.~911--920.

\bibitem{gesztesy2016callias}
{\sc F.~Gesztesy and M.~Waurick}, {\em {The Callias index formula revisited}},
  vol.~2157, Springer, 2016.

\bibitem{GOFFENG2012357}
{\sc M.~Goffeng}, {\em Analytic formulas for the topological degree of
  non-smooth mappings: The odd-dimensional case}, Advances in Mathematics, 231
  (2012), pp.~357 -- 377.

\bibitem{gontier2023edge}
{\sc D.~Gontier}, {\em Edge states for second order elliptic operators in a
  channel}, Journal of Spectral Theory, 12 (2023), pp.~1155--1202.

\bibitem{graf2007aspects}
{\sc G.~M. Graf}, {\em {Aspects of the integer quantum Hall effect}}, in
  Proceedings of Symposia in Pure Mathematics, vol.~76, Providence, RI;
  American Mathematical Society; 1998, 2007, p.~429.

\bibitem{graf2021topology}
{\sc G.~M. Graf, H.~Jud, and C.~Tauber}, {\em Topology in shallow-water waves:
  a violation of bulk-edge correspondence}, Communications in Mathematical
  Physics, 383 (2021), pp.~731--761.

\bibitem{graf2024boundary}
{\sc G.~M. Graf and A.~Tarantola}, {\em Boundary conditions and violations of
  bulk-edge correspondence in a hydrodynamic model}, arXiv preprint
  arXiv:2410.13940,  (2024).

\bibitem{grigis1994microlocal}
{\sc A.~Grigis and J.~Sj{\"o}strand}, {\em Microlocal analysis for differential
  operators: an introduction}, vol.~196, Cambridge University Press, 1994.

\bibitem{groth2009theory}
{\sc C.~Groth, M.~Wimmer, A.~Akhmerov, J.~Tworzyd{\l}o, and C.~Beenakker}, {\em
  {Theory of the topological Anderson insulator}}, Physical review letters, 103
  (2009), p.~196805.

\bibitem{gurarie2011single}
{\sc V.~Gurarie}, {\em {Single-particle Green’s functions and interacting
  topological insulators}}, Physical Review B, 83 (2011), p.~085426.

\bibitem{hafezi2013imaging}
{\sc M.~Hafezi, S.~Mittal, J.~Fan, A.~Migdall, and J.~Taylor}, {\em Imaging
  topological edge states in silicon photonics}, Nature Photonics, 7 (2013),
  pp.~1001--1005.

\bibitem{PhysRevLett.61.2015}
{\sc F.~D.~M. Haldane}, {\em {Model for a Quantum Hall Effect without Landau
  Levels: Condensed-Matter Realization of the "Parity Anomaly"}}, Phys. Rev.
  Lett., 61 (1988), pp.~2015--2018.

\bibitem{PhysRevLett.100.013904}
{\sc F.~D.~M. Haldane and S.~Raghu}, {\em Possible realization of directional
  optical waveguides in photonic crystals with broken time-reversal symmetry},
  Phys. Rev. Lett., 100 (2008), p.~013904.

\bibitem{halperin1982quantized}
{\sc B.~I. Halperin}, {\em Quantized hall conductance, current-carrying edge
  states, and the existence of extended states in a two-dimensional disordered
  potential}, Physical review B, 25 (1982), p.~2185.

\bibitem{hatsugai1993chern}
{\sc Y.~Hatsugai}, {\em {Chern number and edge states in the integer quantum
  Hall effect}}, Physical review letters, 71 (1993), p.~3697.

\bibitem{hislop2008edge}
{\sc P.~D. Hislop and E.~Soccorsi}, {\em {Edge currents for quantum Hall
  systems I: one-edge, unbounded geometries}}, Reviews in Mathematical Physics,
  20 (2008), pp.~71--115.

\bibitem{H-II-SP-83}
{\sc L.~V. H{\"o}rmander}, {\em {The Analysis of Linear Partial Differential
  Operators II: Differential Operators with Constant Coefficients}}, Springer
  Verlag, 1983.

\bibitem{H-III-SP-94}
\leavevmode\vrule height 2pt depth -1.6pt width 23pt, {\em {The Analysis of
  Linear Partial Differential Operators III: Pseudo-Differential Operators}},
  Springer Verlag, 1994.

\bibitem{PhysRevLett.95.146802}
{\sc C.~L. Kane and E.~J. Mele}, {\em {$\Zm_2$ Topological Order and the
  Quantum Spin Hall Effect}}, Phys. Rev. Lett., 95 (2005), p.~146802.

\bibitem{kellendonk2017c}
{\sc J.~Kellendonk}, {\em {On the $C^*$-algebraic approach to topological
  phases for insulators}}, in Annales Henri Poincar{\'e}, vol.~18, Springer,
  2017, pp.~2251--2300.

\bibitem{kellendonk2008topological}
{\sc J.~Kellendonk and S.~Richard}, {\em Topological boundary maps in physics:
  General theory and applications}, Perspectives in Operator Algebras and
  Mathematical Physics,  (2008), pp.~105--121.

\bibitem{kellendonk2004quantization}
{\sc J.~Kellendonk and H.~Schulz-Baldes}, {\em Quantization of edge currents
  for continuous magnetic operators}, Journal of Functional Analysis, 209
  (2004), pp.~388--413.

\bibitem{kitaev2009periodic}
{\sc A.~Kitaev, V.~Lebedev, and M.~Feigel’man}, {\em Periodic table for
  topological insulators and superconductors}, in AIP Conference Proceedings,
  vol.~1134, AIP, 2009, pp.~22--30.

\bibitem{klitzing1980new}
{\sc K.~v. Klitzing, G.~Dorda, and M.~Pepper}, {\em {New method for
  high-accuracy determination of the fine-structure constant based on quantized
  Hall resistance}}, Physical review letters, 45 (1980), p.~494.

\bibitem{kotani2014quantization}
{\sc M.~Kotani, H.~Schulz-Baldes, and C.~Villegas-Blas}, {\em Quantization of
  interface currents}, Journal of Mathematical Physics, 55 (2014), p.~121901.

\bibitem{laughlin1981quantized}
{\sc R.~B. Laughlin}, {\em Quantized hall conductivity in two dimensions},
  Physical Review B, 23 (1981), p.~5632.

\bibitem{lee2019elliptic}
{\sc J.~P. Lee-Thorp, M.~I. Weinstein, and Y.~Zhu}, {\em Elliptic operators
  with honeycomb symmetry: Dirac points, edge states and applications to
  photonic graphene}, Archive for Rational Mechanics and Analysis, 232 (2019),
  pp.~1--63.

\bibitem{li2009topological}
{\sc J.~Li, R.-L. Chu, J.~K. Jain, and S.-Q. Shen}, {\em {Topological Anderson
  insulator}}, Physical review letters, 102 (2009), p.~136806.

\bibitem{lin2022mathematical}
{\sc J.~Lin and H.~Zhang}, {\em Mathematical theory for topological photonic
  materials in one dimension}, Journal of Physics A: Mathematical and
  Theoretical, 55 (2022), p.~495203.

\bibitem{Loring}
{\sc T.~A. Loring}, {\em {$K$-theory} and pseudospectra for topological
  insulators}, Annals of Physics, 356 (2015), pp.~383--416.

\bibitem{LS19}
{\sc T.~A. Loring and H.~Schulz-Baldes}, {\em Spectral flow argument localizing
  an odd index pairing}, Canadian Mathematical Bulletin, 62 (2019),
  pp.~373--381.

\bibitem{LS20}
\leavevmode\vrule height 2pt depth -1.6pt width 23pt, {\em The spectral
  localizer for even index pairings}, Journal of Noncommutative Geometry, 14
  (2020), pp.~1--23.

\bibitem{lu2014topological}
{\sc L.~Lu, J.~D. Joannopoulos, and M.~Solja{\v{c}}i{\'c}}, {\em Topological
  photonics}, Nature Photonics, 8 (2014), pp.~821--829.

\bibitem{matsuno1966quasi}
{\sc T.~Matsuno}, {\em Quasi-geostrophic motions in the equatorial area},
  Journal of the Meteorological Society of Japan. Ser. II, 44 (1966),
  pp.~25--43.

\bibitem{MPLW}
{\sc J.~Michala, A.~Pierson, T.~A. Loring, and A.~B. Watson}, {\em Wave-packet
  propagation in a finite topological insulator and the spectral localizer
  index}, Involve, a Journal of Mathematics, 14 (2021), pp.~209--239.

\bibitem{moessner2021topological}
{\sc R.~Moessner and J.~E. Moore}, {\em Topological phases of matter},
  Cambridge University Press, 2021.

\bibitem{parker2020topological}
{\sc J.~B. Parker, J.~Marston, S.~M. Tobias, and Z.~Zhu}, {\em Topological
  gaseous plasmon polariton in realistic plasma}, Physical Review Letters, 124
  (2020), p.~195001.

\bibitem{Prodan}
{\sc E.~Prodan}, {\em A computational non-commutative geometry program for
  disordered topological insulators}, vol.~23 of SpringerBriefs in Mathematical
  Physics, Springer, 2017.

\bibitem{prodan2016bulk}
{\sc E.~Prodan and H.~Schulz-Baldes}, {\em {Bulk and boundary invariants for
  complex topological insulators: From K-Theory to Physics}}, Springer Verlag,
  Berlin, 2016.

\bibitem{qin2023topological}
{\sc H.~Qin and Y.~Fu}, {\em {Topological Langmuir-cyclotron wave}}, Science
  Advances, 9 (2023), p.~8041.

\bibitem{quinn2024approximations}
{\sc S.~Quinn and G.~Bal}, {\em Approximations of interface topological
  invariants}, SIAM Journal on Mathematical Analysis, 56 (2024),
  pp.~5521--5582.

\bibitem{quinn2024asymmetric}
\leavevmode\vrule height 2pt depth -1.6pt width 23pt, {\em {Asymmetric
  transport for magnetic Dirac equations}}, Pure and Applied Analysis, 6
  (2024), pp.~353--377.

\bibitem{rechtsman2013photonic}
{\sc M.~C. Rechtsman, J.~M. Zeuner, Y.~Plotnik, Y.~Lumer, D.~Podolsky,
  F.~Dreisow, S.~Nolte, M.~Segev, and A.~Szameit}, {\em {Photonic Floquet
  topological insulators}}, Nature, 496 (2013), pp.~196--200.

\bibitem{RS-79-III}
{\sc M.~Reed and B.~Simon}, {\em Methods of modern mathematical physics. III.},
  Academic Press, Inc., New York, 1979.

\bibitem{rossi2024topology}
{\sc S.~Rossi and A.~Tarantola}, {\em {Topology of 2D Dirac operators with
  variable mass and an application to shallow-water waves}}, Journal of Physics
  A: Mathematical and Theoretical, 57 (2024), p.~065201.

\bibitem{russo1977hausdorff}
{\sc B.~Russo}, {\em {On the Hausdorff-Young theorem for integral operators}},
  Pacific Journal of Mathematics, 68 (1977), pp.~241--253.

\bibitem{schrohe1992spectral}
{\sc E.~Schrohe}, {\em {Spectral invariance, ellipticity, and the Fredholm
  property for pseudodifferential operators on weighted Sobolev spaces}},
  Annals of Global Analysis and Geometry, 10 (1992), pp.~237--254.

\bibitem{schulz2015z2}
{\sc H.~Schulz-Baldes}, {\em {$\Zm_2$-Indices and Factorization Properties of
  Odd Symmetric Fredholm Operators}}, Documenta Mathematica, 20 (2015),
  pp.~1481--1500.

\bibitem{SBindex2016}
\leavevmode\vrule height 2pt depth -1.6pt width 23pt, {\em {Topological}
  insulators from the perspective of non-commutative geometry and index
  theory}, Jahresbericht der deutschen Mathematiker-Vereinigung, 118 (2016),
  p.~247–273.

\bibitem{schulz2000simultaneous}
{\sc H.~Schulz-Baldes, J.~Kellendonk, and T.~Richter}, {\em {Simultaneous
  quantization of edge and bulk Hall conductivity}}, Journal of Physics A:
  Mathematical and General, 33 (2000), p.~L27.

\bibitem{shen2012topological}
{\sc S.-Q. Shen}, {\em Topological insulators}, vol.~174, Springer, 2012.

\bibitem{silveirinha2015chern}
{\sc M.~G. Silveirinha}, {\em Chern invariants for continuous media}, Physical
  Review B, 92 (2015), p.~125153.

\bibitem{slobozhanyuk2017three}
{\sc A.~Slobozhanyuk, S.~H. Mousavi, X.~Ni, D.~Smirnova, Y.~S. Kivshar, and
  A.~B. Khanikaev}, {\em Three-dimensional all-dielectric photonic topological
  insulator}, Nature Photonics, 11 (2017), p.~130.

\bibitem{souslov2019topological}
{\sc A.~Souslov, K.~Dasbiswas, M.~Fruchart, S.~Vaikuntanathan, and V.~Vitelli},
  {\em Topological waves in fluids with odd viscosity}, Physical Review
  Letters, 122 (2019), p.~128001.

\bibitem{tauber2019anomalous}
{\sc C.~Tauber, P.~Delplace, and A.~Venaille}, {\em Anomalous bulk-edge
  correspondence in continuous media}, Physical Review Research, 2 (2020),
  p.~013147.

\bibitem{tauber2023topology}
{\sc C.~Tauber and G.~C. Thiang}, {\em Topology in shallow-water waves: A
  spectral flow perspective}, Annales Henri Poincar{\'e}, 24 (2023),
  pp.~107--132.

\bibitem{thiang2016k}
{\sc G.~C. Thiang}, {\em {On the K-theoretic classification of topological
  phases of matter}}, in Annales Henri Poincar{\'e}, vol.~17, Springer, 2016,
  pp.~757--794.

\bibitem{PhysRevLett.49.405}
{\sc D.~J. Thouless, M.~Kohmoto, M.~P. Nightingale, and M.~den Nijs}, {\em
  {Quantized Hall Conductance in a Two-Dimensional Periodic Potential}}, Phys.
  Rev. Lett., 49 (1982), pp.~405--408.

\bibitem{tsui1982two}
{\sc D.~C. Tsui, H.~L. Stormer, and A.~C. Gossard}, {\em Two-dimensional
  magnetotransport in the extreme quantum limit}, Physical Review Letters, 48
  (1982), p.~1559.

\bibitem{volovik2009universe}
{\sc G.~Volovik}, {\em The Universe in a Helium Droplet}, International Series
  of Monographs on Physics, OUP Oxford, 2009.

\bibitem{watson2023bistritzer}
{\sc A.~B. Watson, T.~Kong, A.~H. MacDonald, and M.~Luskin}, {\em
  {Bistritzer--MacDonald dynamics in twisted bilayer graphene}}, Journal of
  Mathematical Physics, 64 (2023).

\bibitem{yafaev1992mathematical}
{\sc D.~R. Yafaev}, {\em Mathematical scattering theory: general theory},
  no.~105, American Mathematical Soc., 1992.

\bibitem{zworski2022semiclassical}
{\sc M.~Zworski}, {\em Semiclassical analysis}, vol.~138, American Mathematical
  Society, 2022.

\end{thebibliography}
%\bibliographystyle{siam}

}

\end{document}